\documentclass[a4paper,useAMS,usenatbib,fleqn]{mnras}
\usepackage[T1]{fontenc}
\usepackage{ae,aecompl}

\usepackage{graphicx}	
\usepackage{amsmath}	
\usepackage{amssymb}	
\usepackage{subfig}      
\usepackage{natbib}
\usepackage{booktabs}
\usepackage{cleveref}
\crefname{figure}{Fig.}{Figs.}
\crefname{table}{Table}{Tables}
\crefname{chapter}{Chapter}{Chapters}

\newcommand{\perc}[1]{$#1$ per cent}
\newcommand{\units}[1]{\, \mathrm{#1}}
\newcommand{\unitstx}[1]{\mathrm{#1}}
\newcommand{\diff}{\mathop{}\!\mathrm{d}}
\newcommand{\diffn}[1]{\mathop{}\!\mathrm{d}^#1}

\newcommand{\pderiv}[2]{\frac{\mathrm{\partial} #1}{\mathrm{\partial} {#2}}}
\newcommand{\rfsec}[1]{\mbox{\S\ref{sec:#1}}}
\newcommand{\equnp}[1]{eq.~\ref{eq:#1}}

\newcommand{\subrfig}[1]{\protect\subref{fig:#1}}

\newcommand{\mach}{\mathcal{M}}
\newcommand{\kboltz}{k_{\mathrm{B}}}
\newcommand{\zeq}[1]{\mbox{$z=#1$}}
\newcommand{\nbody}{$N$-body~}
\newcommand{\delvir}{\Delta_{\mathrm{vir}}}
\newcommand{\msun}{\units{M_\odot}}
\newcommand{\Rv}{R_{\mathrm{vir}}}
\newcommand{\Mv}{M_{\mathrm{vir}}}
\newcommand{\Vv}{V_{\mathrm{vir}}}
\newcommand{\Tv}{T_{\mathrm{vir}}}
\newcommand{\cvir}{c_{\mathrm{vir}}}

\volume{461}
\graphicspath{ {figs/} }
\DeclareGraphicsExtensions{.jpg,.pdf}

\title[Penetrating Gas Streams in Galaxy Clusters]{The Role of Penetrating Gas Streams in Setting the Dynamical State of Galaxy Clusters}

\author[Zinger et al.]{ E.\@ Zinger$^1$\thanks{E-mail: elad.zinger@mail.huji.ac.il}, A.\@ Dekel$^1$, Y.\@ Birnboim$^1$, A.\@ Kravtsov$^2$ \& D.\@ Nagai$^3$ \\
$^1$Center for Astrophysics and Planetary Science, Racah Institute of Physics, The Hebrew University, Jerusalem 91904, Israel\\
$^2$Department of Astronomy \& Astrophysics, The University of Chicago, Chicago, IL 60637 USA \\
$^3$Department of Physics, Yale University, New Haven, CT 06520, USA\\}

\date{Accepted 2016 May 25. Received 2016 May 25; in original form 2015 October 18}

\pubyear{2016}

\begin{document}
\pagerange{\pageref{firstpage}--\pageref{lastpage}}
\maketitle
\label{firstpage}

\begin{abstract}
We utilize cosmological simulations of 16 galaxy clusters at redshifts
\zeq{0} and \zeq{0.6} to study the effect of inflowing streams on the
properties of the X-ray emitting intracluster medium. We find that the
mass accretion occurs predominantly along streams that originate from
the cosmic web and consist of heated gas. Clusters that are unrelaxed
in terms of their X-ray morphology are characterized by higher mass
inflow rates and deeper penetration of the streams, typically into the
inner third of the virial radius. The penetrating streams generate
elevated random motions, bulk flows and cold fronts.  The degree of
penetration of the streams may change over time such that clusters can
switch from being unrelaxed to relaxed over a time-scale of several giga years.
\end{abstract}

\begin{keywords}
galaxies: clusters: general -- galaxies: clusters: intracluster medium.
\end{keywords}

\section{Introduction}\label{sec:intro}
The high temperatures achieved by the gas within the potential wells
of clusters of galaxies lead to strong X-ray emission, of the order of 
$10^{43}\textrm{--}10^{46} \units{erg\,s^{-1}}$
\citep{Peterson2006}. This allows a unique vista into the dynamics of
the gas within the dark matter halo, which is unavailable in less
massive systems.

Observations reveal that clusters are complex dynamic systems whose
inner structure poses challenging open questions. These include the
differences in the dynamical state between `relaxed' and `unrelaxed'
clusters \citep{Buote1996}, the origin of `cool-core' (CC) versus
`non-cool-core' (NCC) clusters \citep{Sanderson2006,Sanderson2009}, the
bulk motions and cold fronts in the intracluster medium  
\citep[ICM;][]{Burns1998}, and the wide metal enrichment of the
ICM \citep{Mushotzky1997,DeGrandi2004}. The fact that clusters form at
the nodes of the cosmic web and are fed through streams may allow us
to make progress in the understanding of these open issues concerning
the ICM.

By and large, the ICM gas is assumed to be in hydro-static equilibrium
within the potential well of the dark matter halo, achieving
temperatures of \mbox{$\sim 10^7\textrm{--}10^8 \units{K}$}. Gas
accretion into the central regions occurs gradually as the gas losses
energy to radiation, primarily in the X-ray
\citep{Sarazin1988}. However, detailed observations reveal that in
many clusters the X-ray morphology of the cluster points to an
unrelaxed (NR) dynamical state, as attested by an abundance of
substructure, irregular isophotes and prominent filamentary
structures. Clusters in which these features are absent are considered
relaxed (R).

It is commonly assumed that the dynamically unrelaxed state is the
result of one or many merger events in the cluster. In this paper we
propose that the existence of large-scale gas streams which penetrate
into the centre of cluster can also play an equally important role in
determining whether a cluster is R or NR.

The treatment of clusters up until now has been largely as
quasi-spherical systems, but it has now become clear that one must
consider the full three dimensional structure of clusters to truly
understand these systems.  A key to a better understanding of the
dynamics of clusters is taking into account the fact that clusters
form at the nodes of the cosmic web.

The simplified model of spherical accretion has in recent years been
replaced by a more realistic picture where the accretion on to and
through dark matter haloes is predominantly along streams that follow
the large-scale filaments of the cosmic web. Typically, a few, nearly
co-planar streams feed each halo \citep{Danovich2012}. The massive
haloes of rich clusters represent high-sigma peaks in the density
fluctuation field, namely they are much more massive than the
Press--Schechter mass of typical haloes at that time. 

At high redshift, as these haloes were fed by streams that were
narrower than the virial radius of the halo and denser than the
mean virial density. Due to the high density, the radiative cooling
time of the gas in these streams was shorter than the local
compression time, preventing the formation of a stable virial shock
within the streams, while such a shock is present in the rest of the
halo virial surface \citep{Birnboim2003,Dekel2006}. This allowed the
streams in massive haloes at high redshift to penetrate through the
virial radius into the inner halo
\citep{Keres2005,Ocvirk2008,Dekel2009,Keres2009}. As we shall see, the
streams can still be found in massive clusters at \zeq{0} albeit with
very different thermodynamic properties.

These penetrating streams are likely to affect the inner ICM. As we
will show, the dichotomy between R and NR clusters is strongly related
to the presence of deeply penetrating streams in the cluster. The
penetrating streams can carry large quantities of gravitational energy
into the cluster centre which drive random gas motions and result in
an unrelaxed state. Clusters in which streams do not penetrate into
the centre are found to be dynamically relaxed.

Observational studies of clusters have detected filamentary structures
which may be the streams we find in the simulations. Observations of
the Coma cluster
\citep[e.g.,][]{Finoguenov2003,Brown2011a,Bonafede2013,Simionescu2013}
reveal diffuse X-ray emissions from filamentary structures which
extend out to the cluster outskirts. \citealt{Eckert2015} report the
detection of filamentary structures of hot gas extending over 8 Mpc in
the cluster Abell 2744. The filaments of the cosmic web from which the
gas streams originate have also been detected as extended X-ray
sources between clusters \citep{Werner2008,Dietrich2012}.

The dynamical state of the cluster has been found to correlate
strongly with the presence, or absence, of a CC in the
cluster. Clusters which are dynamically relaxed typically host a CC
while NR cluster are typically classified as NCC
\citep{Smith2005,Sanderson2009a,Chen2007}. The common wisdom is that a
CC cannot easily form in a dynamically disturbed environment. It is
important to note that while the simulations analysed here result in
both R and NR clusters the simulations do \emph{not} reproduce the CC
versus NCC bimodality.

The penetrating streams in NR clusters may also impact the metal
enrichment of the ICM.  There are observational indications of high
metal enrichment throughout the ICM, even at large distances from the
BCG \citep{Rebusco2006,Simionescu2015}. This suggests that the ICM gas
is churned and mixed, thereby scattering the metals throughout the
cluster. Turbulent motions created by the inflow of gas, smoothly and
through mergers, may act as an efficient mixing process.
  
Gas motions in the ICM are observed to be highly complex.
\citet{Markevitch2000} and \citet{Churazov2003} find evidence of inner
shocks and `cold fronts' which are associated with bulk motions of gas
within the ICM (see \citealt{Markevitch2007} for a review). Cold
fronts are contact discontinuities between gas of different
temperatures and densities which maintain pressure equilibrium along
the discontinuity, as opposed to shocks in which the pressure jumps in
value across the shock front. These cold fronts are found to form as a
result of motions of bodies of gas within the ICM
\citep{Ascasibar2006,Roediger2011}. As we will see, deeply penetrating
streams in NR clusters can generate cold fronts in the cluster centre.

The paper is organized as follows: in \rfsec{sims}, we describe our
suite of simulated systems. In \rfsec{rlx_urlx}, we compare between the
R and NR clusters in the simulation suite and analyse the source of
differences between them. In \rfsec{streams}, the mass accretion along
inflowing gas streams is presented and a link between stream
penetration and relaxedness is established. \rfsec{mergers} examines
the respective roles of gas streams and merging substructure in
determining the dynamical state of the cluster. In \rfsec{discuss}, we
discuss our results and in \rfsec{summary} summarize our main
findings.

\begin{table}
\centering
  	 \begin{tabular}{@{}lccccc@{}}
	 \toprule
	 Cluster&$\Mv$&$\Rv$&$\Tv$&$\Vv$& State\\ 
	  &$(10^{14}\units{\msun})$&$(\unitstx{Mpc})$&$(10^7\units{K})$&$(\unitstx{km\,s^{-1}})$  & R/NR \\ 
        \midrule
CL101 & 22.1 & 3.37 & 10.1 &  1678 & NR \\
CL102 & 13.7 & 2.88 & 7.4  & 1433 & NR \\
CL103 & 15.7 & 3.01 & 8.1  & 1497 & NR \\
CL104 & 11.9 & 2.74 & 6.7  & 1365 & R \\
CL105 & 12.0 & 2.75 & 6.7  & 1369 & NR \\
CL106 & 9.5  & 2.54 & 5.8  & 1266 & NR \\
CL107 & 6.6  & 2.26 & 4.5  & 1125 & NR \\
CL3   & 6.3  & 2.22 & 4.4  & 1107 & R \\
CL5   & 3.1  & 1.76 & 2.8  & 875 & R \\
CL6   & 3.3  & 1.80 & 2.9  & 894 & NR \\
CL7   & 3.3  & 1.78 & 2.8  & 886 & R \\
CL9   & 1.9  & 1.48 & 1.7  & 739 & NR \\
CL10  & 1.3  & 1.32 & 1.6  & 658 & R \\
CL11  & 1.8  & 1.45 & 1.9  & 721 & NR \\
CL14  & 1.7  & 1.43 & 1.8  & 709 & R \\
CL24  & 0.86 & 1.14 & 1.2  & 569 & NR \\
 	\bottomrule 
	\end{tabular}
	\caption{Cluster properties at \zeq{0}. State refers to
          R versus NR systems. Virial quantities were
          calculated for an overdensity of $\delvir=337$ above the
          mean density of the universe.}
	 \label{tab:clusterProperties}
 \end{table}

\begin{table}
\centering
  	 \begin{tabular}{@{}lccccc@{}}
	 \toprule
	 Cluster&$\Mv$&$\Rv$&$\Tv$&$\Vv$& State\\ 
	  &$(10^{14}\units{\msun})$&$(\unitstx{Mpc})$&$(10^7\units{K})$&$(\unitstx{km\,s^{-1}})$  & R/NR \\ 
        \midrule
CL101 & 3.3  & 1.29 & 4.0 & 1053 & R \\
CL102 & 3.3  & 1.28 & 4.0 & 1051 & NR \\
CL103 & 2.9  & 1.23 & 3.7 & 1009 & NR \\
CL104 & 7.2  & 1.66 & 6.7 & 1362 & R \\
CL105 & 6.2  & 1.58 & 6.0 & 1296 & NR \\
CL106 & 2.6  & 1.18 & 3.3 & 965  & NR \\
CL107 & 2.9  & 1.22 & 3.6 & 1002 & NR \\
CL3   & 2.8  & 1.21 & 3.6 & 995  & NR \\
CL5   & 1.8  & 1.04 & 2.6 & 858  & NR \\
CL6   & 2.4  & 1.15 & 3.2 & 946  & NR \\
CL7   & 2.3  & 1.13 & 3.1 & 929  & R \\
CL9   & 1.2  & 0.91 & 2.0 & 744  & NR \\
CL10  & 1.1  & 0.89 & 1.9 & 730  & R \\
CL11  & 0.86 & 0.78 & 1.5 & 643  & NR \\
CL14  & 0.99 & 0.86 & 1.8 & 703  & NR \\
CL24  & 0.37 & 0.62 & 0.9 & 505  & NR \\
 	\bottomrule 
	\end{tabular}
	\caption{Cluster properties at \zeq{0.6}. State refers to
          R versus NR systems. Virial quantities were
          calculated for an overdensity of $\delvir=224$ above the
          mean density of the universe.}
	 \label{tab:clusterProperties06}
 \end{table}

\section{Simulations}\label{sec:sims}
The simulation suite analysed in this study is comprised of $16$
cluster-sized systems at \zeq{0} spanning a mass range of $8.6\times
10^{13} \textrm{--} 2.2\times 10^{15} \units{\msun}$, and their most massive
progenitors at \zeq{0.6}.

The systems were extracted from cosmological simulations in a flat $\Lambda$ cold dark matter 
($\Lambda$CDM) model: $\Omega_{\mathrm{m}}=1-\Omega_{\Lambda}=0.3$,
$\Omega_{\mathrm{b}}=0.04286$, $h=0.7$, and $\sigma_8=0.9$, where $\Omega_{\mathrm{m}}$
and $\Omega_{\Lambda}$ are the density parametres of mass and
cosmological constant, the Hubble constant is
$100h\units{km\,s^{-1}\,Mpc^{-1}}$, and $\sigma_8$ is the power
spectrum normalization on an $8h^{-1} \units{Mpc}$ scale at
\zeq{0}. The simulations were carried out with the Adaptive Refinement
Tree (\textsc{art}) \nbody+gas-dynamics AMR code \citep{Kravtsov1999},
an Eulerian code that uses adaptive refinement in space and time, and
(non-adaptive) refinement in mass \citep{Klypin2001} to reach the high
dynamic range required to resolve cores of haloes formed in
self-consistent cosmological simulations.

The computational boxes of the large-scale cosmological simulations
were either $120 h^{-1}\units{Mpc}$ or $80 h^{-1}\units{Mpc}$, and the
simulation grid was adaptively refined to achieve a peak spatial
resolution of $\sim 7$ and $5\, h^{-1}\units{kpc}$, respectively. These
simulations are discussed in detail in \citet{Kravtsov2006},
\citet{Nagai2007a} and \citet{Nagai2007}. 

Besides the basic dynamical processes of collision-less matter (dark
matter and stars) and gas-dynamics, several physical processes
critical for galaxy formation are incorporated: star formation, metal
enrichment and feedback due to Type II and Type Ia supernov\ae\@ and
self-consistent advection of metals. The cooling and heating rates of
the gas take into account Compton heating and cooling of plasma,
heating by the UV background \citep{Haardt1996}, and atomic and
molecular cooling, which is tabulated for the temperature range
$10^2\,\units{K} < T < 10^9\,\units{K}$, a grid of metallicities, and
UV intensities using the \textsc{cloudy} code (version 96b4;
\citealt{Ferland1998}). The \textsc{cloudy} cooling and heating rates take into
account metallicity of the gas, which is calculated self-consistently
in the simulation, so that the local cooling rates depend on the local
metallicity of the gas. The star formation recipe incorporated in
these simulations is observationally motivated (e.g.\@
\citealt{Kennicutt1998}) and the code also accounts for the stellar
feedback on the surrounding gas, including injection of energy and
heavy elements (metals) via stellar winds, supernov\ae, and secular
mass loss.

In this paper, the virial quantities of mass, radius, velocity and
temperature ($\Mv, \Rv, \Vv, \Tv$) of the clusters are defined for an
overdensity $\delvir = 337$ at \zeq{0} and $\delvir = 224$ at
\zeq{0.6} above $\rho_{\mathrm{mean}}$, the mean mass density of the
universe \citep{Bryan1998}. The properties of the clusters for \zeq{0}
and \zeq{0.6} are summarized in
\cref{tab:clusterProperties,tab:clusterProperties06}, respectively.

The virial quantities are customarily set based on the spherical
`top-hat' model which follows the expansion and subsequent collapse of
an initial uniform density perturbation in an Einstein--de Sitter
universe. Virialization of the halo in the model occurs when the
radius of the halo is one half of its radius at maximum expansion, and
the resulting mean density in the halo is found to be $178$ times the
mean density of the universe. Extending this model for a $\Lambda$CDM
universe results in a higher value of $\delvir\simeq 337$
\citep{Bryan1998}.

The virial parameters $\Mv$ and $\Rv$ are therefore defined to be
\begin{equation}\label{eq:virialDef}
\frac{3\Mv}{4\mathrm{\pi} \Rv^3}= \Delta_i \rho_{\mathrm{ref}} ,
\end{equation}
with various choices in the literature for the overdensity parameter
($\Delta_i = 178, 200, 337$) and reference density
$\rho_{\mathrm{ref}}$ -- either $\rho_{\mathrm{crit}}$ or
$\rho_{\mathrm{mean}} = \rho_{\mathrm{crit}}\Omega_{\mathrm{m}}$. Additional
distance scales for clusters are sometimes set by overdensity factors
of $500$ and $1500$.\footnote{The dependence of the scaling parameters
  on the choice of $\Delta$ and reference density can be found for
  a given choice of halo model. As an example, for an isothermal
  sphere density profile, different choices of $\Delta$ and
  $\rho_{\mathrm{ref}}$ will result in different values of $\Mv$ and
  $\Rv$ such that
\begin{equation}\label{eq:virialDependence}
\frac{M_{\mathrm{vir,1}}}{M_{\mathrm{vir,2}}}\propto
\frac{R_{\mathrm{vir,1}}}{R_{\mathrm{vir,2}}}\propto \left( \frac{\Delta_1\rho_{\mathrm{ref,1}}}{\Delta_2\rho_{\mathrm{ref,2}}}\right)^{-1/2},
\end{equation} and thus $R_{200}\simeq 1.3 R_{337}$.}

The simulations do not include an AGN feedback mechanism, and this may
lead to unrealistic conditions in the central region of the cluster. The
AGN feedback may provide the heating source to avert the cooling flow
problem in cluster cores. In this paper we do not address the issue of
the cooling flow problem but rather focus on the division into R and
NR clusters. Excluding AGN feedback allows us to separate the role of
gas streams in determining the conditions in the inner regions of the
ICM from that of the AGN.

Since the simulations do not include the relevant cluster core
physics, such as heating by AGN and thermal conduction, the CC versus
NCC dichotomy cannot be properly recreated in our cluster
suite. Recent studies by \citealt{Rasia2015} and \citealt{Hahn2015} have shown that
the dichotomy can be recreated in simulations if the proper numerical
models are implemented.

As a result of these shortcomings, the entire suite of simulated
clusters suffers from overcooling in the central core ($\lesssim 0.01
\Rv$), which leads to a nonphysical high density cusp. However, the
simulations do capture the dynamics and structure at larger scales,
recreating the R versus NR dichotomy. Specifically, the radial range of
$0.05 \textrm{--} 0.25\Rv$, in which we study the effects of stream
penetration, is well resolved and can give a reliable picture of the
large-scale effects which determine the relaxedness of clusters.

Traditionally, $\Rv$ is used to demark the edge of a halo
\citep{Molnar2009}. However, this definition has been shown to be
inadequate for clusters since other important measures such as the
`splashback' radius \citep{Diemer2014,More2015} and the virial
accretion shock have been shown to extend to well beyond the virial
radius \citep{Lau2015}. In this work we consider the regions between
$\Rv$ and the virial shock front as part of the cluster, and for the
study at hand, we shall define the virial accretion shock as the edge
of the cluster. An in-depth analysis of the definition of the edge of
a cluster, specifically in the context of quenching of star formation
in satellite galaxies, appears in a separate paper (Zinger et al., in
preparation). 

The classification of the simulated clusters R or NR (see
\cref{tab:clusterProperties,tab:clusterProperties06}) was carried out
to emulate as close as possible the methods employed by observational
studies. The classification is described in detail in
\citet{Nagai2007a,Nagai2007}, and is based on mock \emph{Chandra}
X-ray images of the clusters which were created on a scale of
$R_{500,\mathrm{c}}$ (an overdensity of $\Delta_{\mathrm{c}}=500$ above the
critical density $\rho_{\mathrm{crit}}$) where $R_{500,\mathrm{c}}\simeq
0.45\Rv$ (\equnp{virialDependence}). Based on the mock observations,
clusters are identified as R if they possess regular X-ray morphology
with no secondary luminosity peaks and minimal deviations from
elliptical symmetry. By contrast, NR clusters are those with distinct
secondary luminosity peaks, filamentary X-ray structures, or
significant shifts of the centres of isophotes. This classification
was carried out for each cluster in three orthogonal Cartesian
projections, and only clusters which appear relaxed in all three
projections are classified as R. Conversely, clusters which appear
unrelaxed in at least one projection are classified as NR \citep[see
  also][]{Lau2011}. We revisit the issue of this classification in
relation to the gas streams in \rfsec{mergers}

Since we will be dealing with gas motions and inflowing streams, it is
only proper that we elucidate how the centre of the cluster and the
inertial frame of reference used to measure the gas velocities are
defined. The cluster centre is defined by the location of the dark
matter particle with the highest local matter density
\citep{Lau2009}. To define the inertial frame, a centre-of-mass
velocity profile was calculated for the three velocity components for each
cluster in the suite. A radius was selected where the profiles were
seen to level off (typically at $\sim 1\textrm{--}2 \Rv$). The gas
velocities are then measured with respect to the centre-of-mass
velocity as calculated for the gas within that radial limit.  Since
the profiles are relatively flat and constant over a wide range of
radii, the centre-of-mass velocity is largely insensitive to the
choice of the radial limit.

\section{Properties of R versus NR Clusters}\label{sec:rlx_urlx}
We examine the differences between the R and NR populations of
clusters in our simulation suite and attempt to link the dynamical
state of the clusters to the intensity and penetration of the gas
streams.

\subsection{Hydrodynamic Stability}\label{sec:jeans}
We begin with an examination of the hydrodynamical stability of the
ICM to gravitational collapse. To this end, we probe the dynamics of
the gas when smoothed over an intermediate scale, i.e.\@ one that is
larger than the cell-size yet smaller than the relevant cluster scales
of several $100\units{kpc}$.

To do so we make use of a generalized Jeans equation for the radial
velocity component adapted for use in an Eulerian system (rather than
a system of particles), averaged over spherical shells
(\citealt{Rasia2004} and see also \citealt{Lau2013} for a similar
derivation)
\begin{multline}  \label{eq:jeans1}
\overline{\rho} \pderiv{\overline{v}_r }{t}+
\overline{\rho}\,\overline{v}_r \pderiv{\overline{v}_r}{r}= \\
-\overline{\rho}\overline{\pderiv{\Phi}{r}}
-\pderiv{}{r}\left(\frac{\kboltz}{m} \overline{\rho}T \right)
 -\pderiv{}{r}\left(\overline{\rho}\, \overline{\sigma^2_{rr}}\right)\\
-\frac{\overline{\rho}}{r}\left( 2\overline{\sigma^2_{rr}}-\overline{\sigma^2_{\theta\theta}}
-\overline{\sigma^2_{\varphi\varphi}}\right)
+\frac{\overline{\rho}}{r}\left(\overline{v}_\theta^2 +\overline{v}_\varphi^2\right).
\end{multline}
Like the Jeans equation, this equation is derived as the first moment
of the Boltzmann equation for the distribution function averaged over
velocity space and over a spherical shell of a given radius. The
result is a motion equation for a spherical shell, in which all the
values represent averages of the form:
\begin{align}\label{eq:shell_avg}
\overline{\rho}(r)&=\frac{1}{4\mathrm{\pi}}\int f(\vec{x},\vec{v}) \diff\Omega{\diffn{3}}v \\
\overline{v}_\alpha(r)&=\frac{1}{4\mathrm{\pi} \overline{\rho}}\int v_\alpha f(\vec{x},\vec{v}) \diff\Omega \diffn{3}v,
\end{align}
where $f(\vec{x},\vec{v})$ is the distribution function in phase space
and $\overline{\sigma^2_{ij}}$ is defined as
\begin{equation}\label{eq:sigma_def}
\overline{\sigma^2_{ij}}(r)=\overline{v_i
  v_j}-\overline{v}_i\overline{v}_j .
\end{equation}

Since we wish to apply the equation to an Eulerian hydro-simulation,
in which gas motions are smoothed on the scale of the cell size, we
assume that the $\sigma^2$ terms can be separated into a contribution
from the microscopic velocities, which are isotropic and represented
by the temperature $T$, and the macroscopic cell velocities, which are
used to determine the terms in \cref{eq:sigma_def}. We therefore add a
thermal pressure term (second term on the RHS of \cref{eq:jeans1}) to
account for the contribution of the subgrid velocities. The thermal
pressure is derived for an Ideal Gas equation of state,
$\overline{\rho}\kboltz T m^{-1}$, where $\kboltz$ is the Boltzmann
constant and $m$ is the average particle mass. The temperature $T$ in
this term is the mass-weighted mean temperature of the spherical
shell.

There has been some contention as to the validity of using the Jeans
equation, which describes collisional dynamics, for the treatment of
the collisional gas in simulations. \citet{Suto2013} have argued this
point and suggested that the Euler equation is a more appropriate
choice. However, \citet{Lau2013} have shown that one can obtain the
generalized Jeans equation \emph{with} the added thermal term by a
spatial averaging of the Euler equation, and that both methods are
valid, and in fact, equivalent.

The LHS of \cref{eq:jeans1} is simply the full convective derivative
of the shell velocity. The terms on the RHS are the force terms, the
first relating to the gravitational attraction, the second is force
due to thermal pressure and the third is the force related to
`pressure' from macroscopic motions of the gas (`turbulence'). The
other terms are related to anisotropy and asphericity and vanish for a
spherically symmetric ($v_\theta=v_\varphi=0$) and isotropic
($\sigma_{rr}^2 = \sigma_{\theta\theta}^2=\sigma_{\varphi\varphi}^2$)
systems. The final three terms, the force due to random gas motions,
anisotropy and asphericity, comprise the non-thermal contribution to
the hydrodynamical stability of the system.

For the case of a system in steady state \mbox{($\mathrm{\partial} v_r/
  \mathrm{\partial} t = 0$)} and equilibrium ($v_r = 0$), the LHS of
\cref{eq:jeans1} is zero and the equation takes the form
\begin{multline}\label{eq:jeans2}
-\overline{\rho}\overline{\pderiv{\Phi}{r}}
-\pderiv{}{r}\left(\frac{\kboltz}{m} \overline{\rho}T \right)
 -\pderiv{}{r}\left(\overline{\rho}\, \overline{\sigma^2_{rr}}\right)\\
-\frac{\overline{\rho}}{r}\left( 2\overline{\sigma^2_{rr}}-\overline{\sigma^2_{\theta\theta}}
-\overline{\sigma^2_{\varphi\varphi}}\right)
+\frac{\overline{\rho}}{r}\left(\overline{v}_\theta^2 +\overline{v}_\varphi^2\right)=0.
\end{multline}  

Employing this equation allows us to examine whether a system is
indeed in steady state equilibrium and, more importantly, what is the relative
contributions of the thermal and non-thermal parts to balancing the
force of gravity. We recast \cref{eq:jeans2} by dividing it by the amplitude of the
gravitational force
\begin{equation}\label{eq:jeans3} 
\frac{F_T + F_{NT}}{|F_G|} = 1 ,
\end{equation} 
where
\begin{align}
F_G=&-\overline{\rho}\overline{\pderiv{\Phi}{r}}=-\overline{\rho}\frac{GM(<r)}{r^2} \label{eq:fGrav} \\
F_T=&-\pderiv{}{r}\left(\frac{\kboltz}{m} \overline{\rho}T \right) \label{eq:fTherm}\\
F_{NT}=&-\pderiv{}{r}\left(\overline{\rho} \overline{\sigma^2_{rr}}\right)
-\frac{\overline{\rho}}{r}\left( 2\overline{\sigma^2_{rr}}-\overline{\sigma^2_{\theta\theta}}
-\overline{\sigma^2_{\varphi\varphi}}\right)\nonumber \\
&+\frac{\overline{\rho}}{r}\left(\overline{v}_\theta^2 +\overline{v}_\varphi^2\right)  \label{eq:fNTherm} .
\end{align}

\begin{figure*}
  \subfloat[R Clusters at \zeq{0}]{\label{fig:jeans_a1_rlx}
    \includegraphics[width=8cm,keepaspectratio]{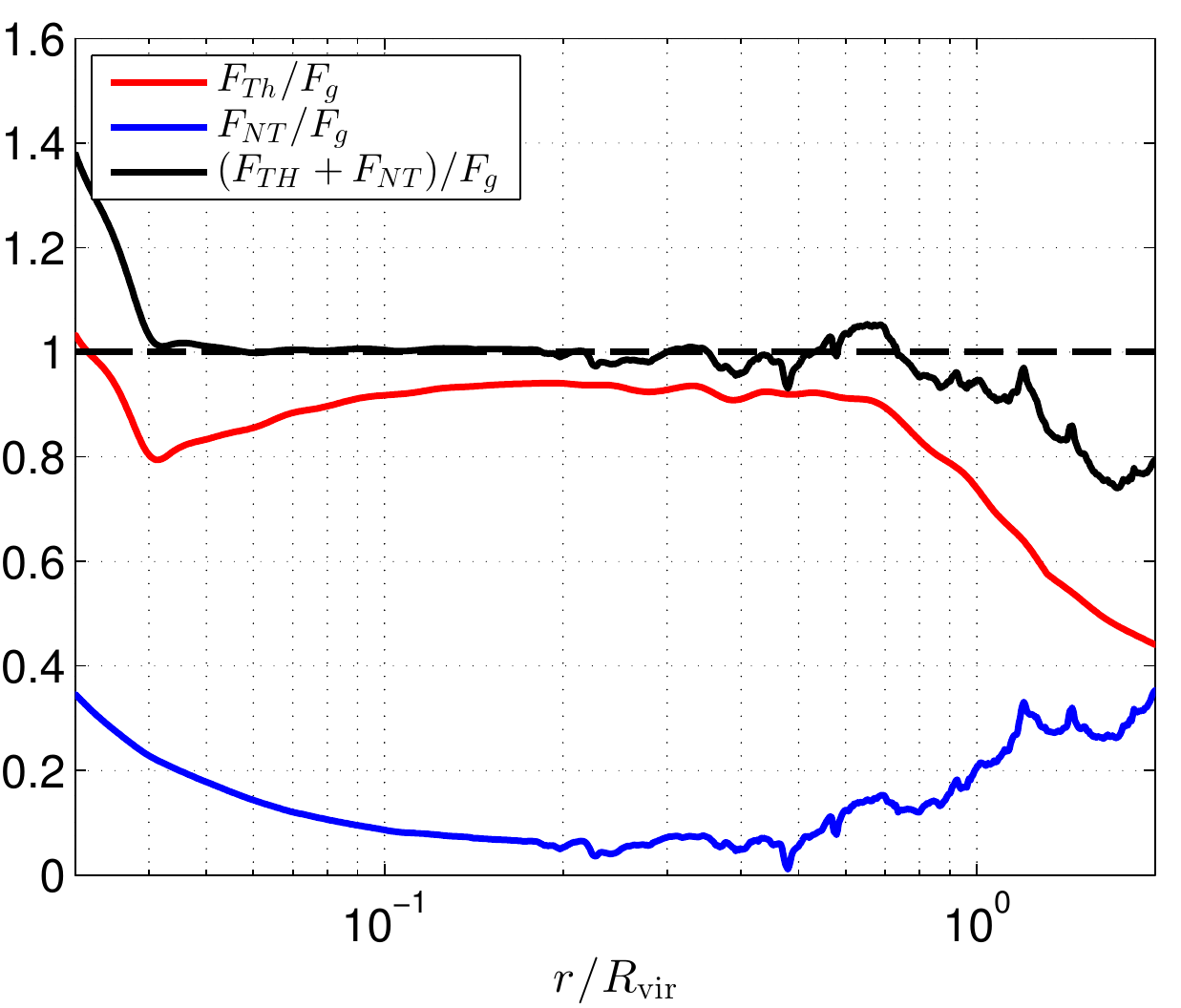}}
  \subfloat[NR Clusters at \zeq{0}]{\label{fig:jeans_a1_urlx}
    \includegraphics[width=8cm,keepaspectratio]{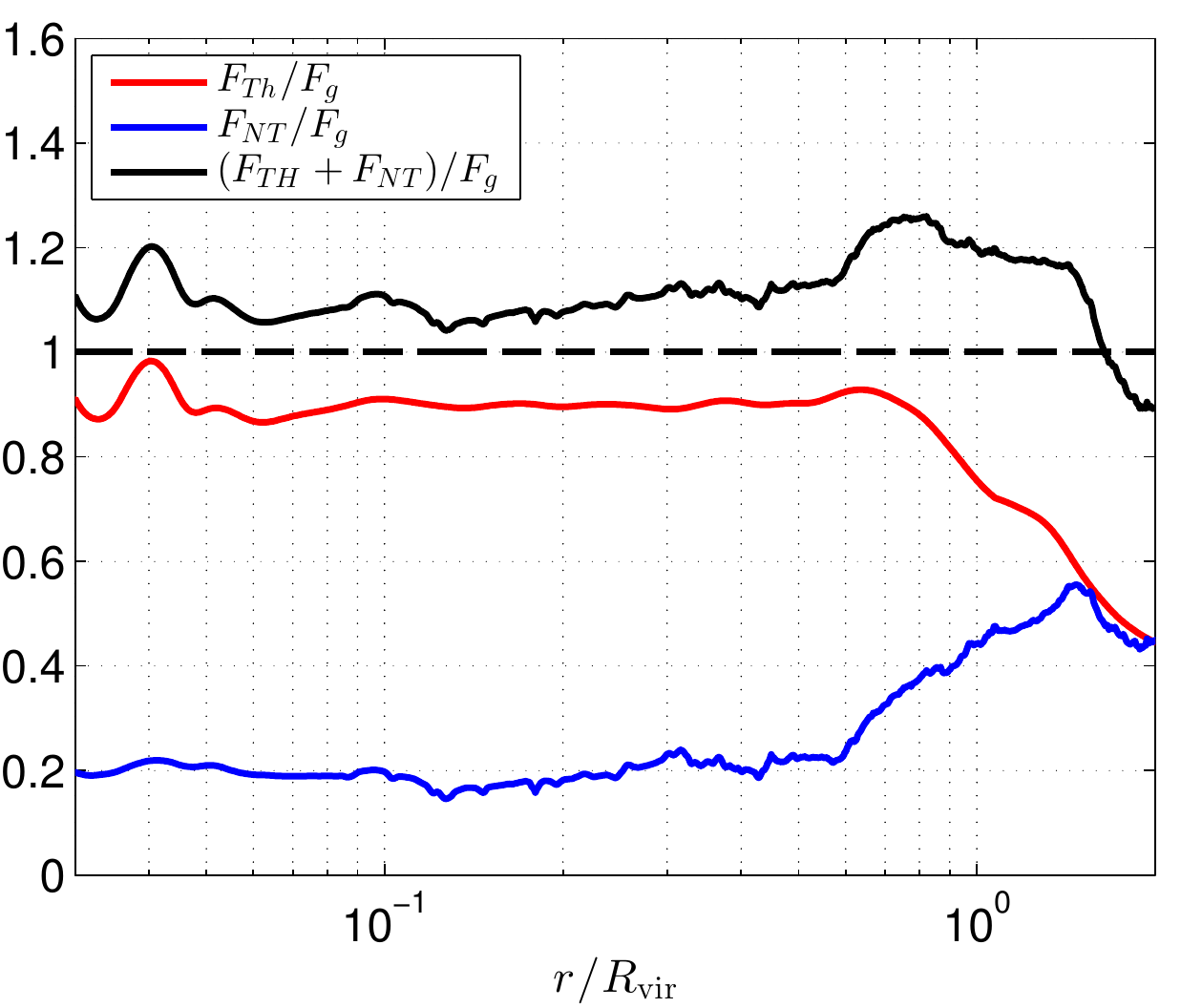}}\\
  \subfloat[R Clusters at \zeq{0.6}]{\label{fig:jeans_a06_rlx}
    \includegraphics[width=8cm,keepaspectratio]{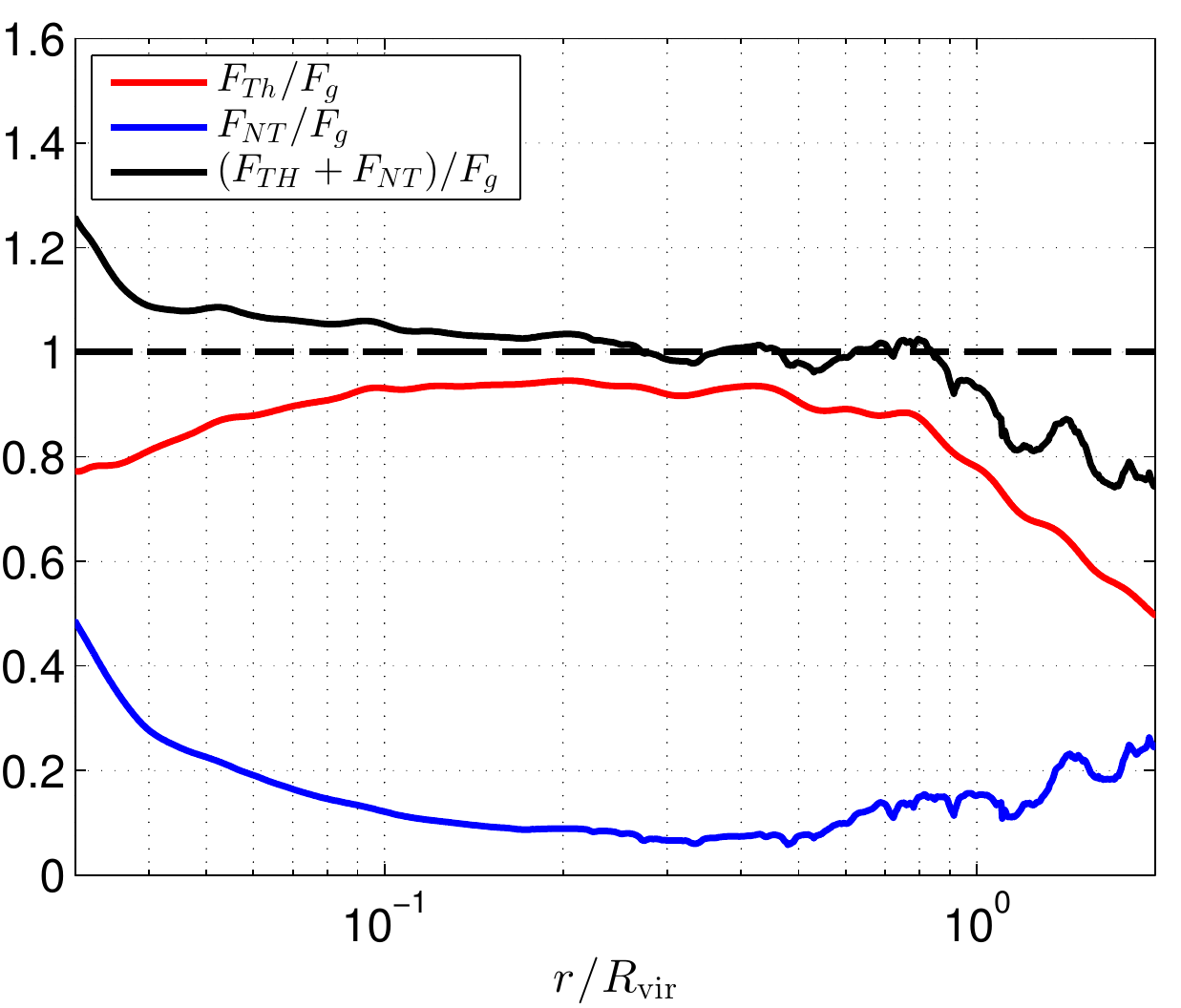}}
  \subfloat[NR Clusters at \zeq{0.6}]{\label{fig:jeans_a06_urlx}
    \includegraphics[width=8cm,keepaspectratio]{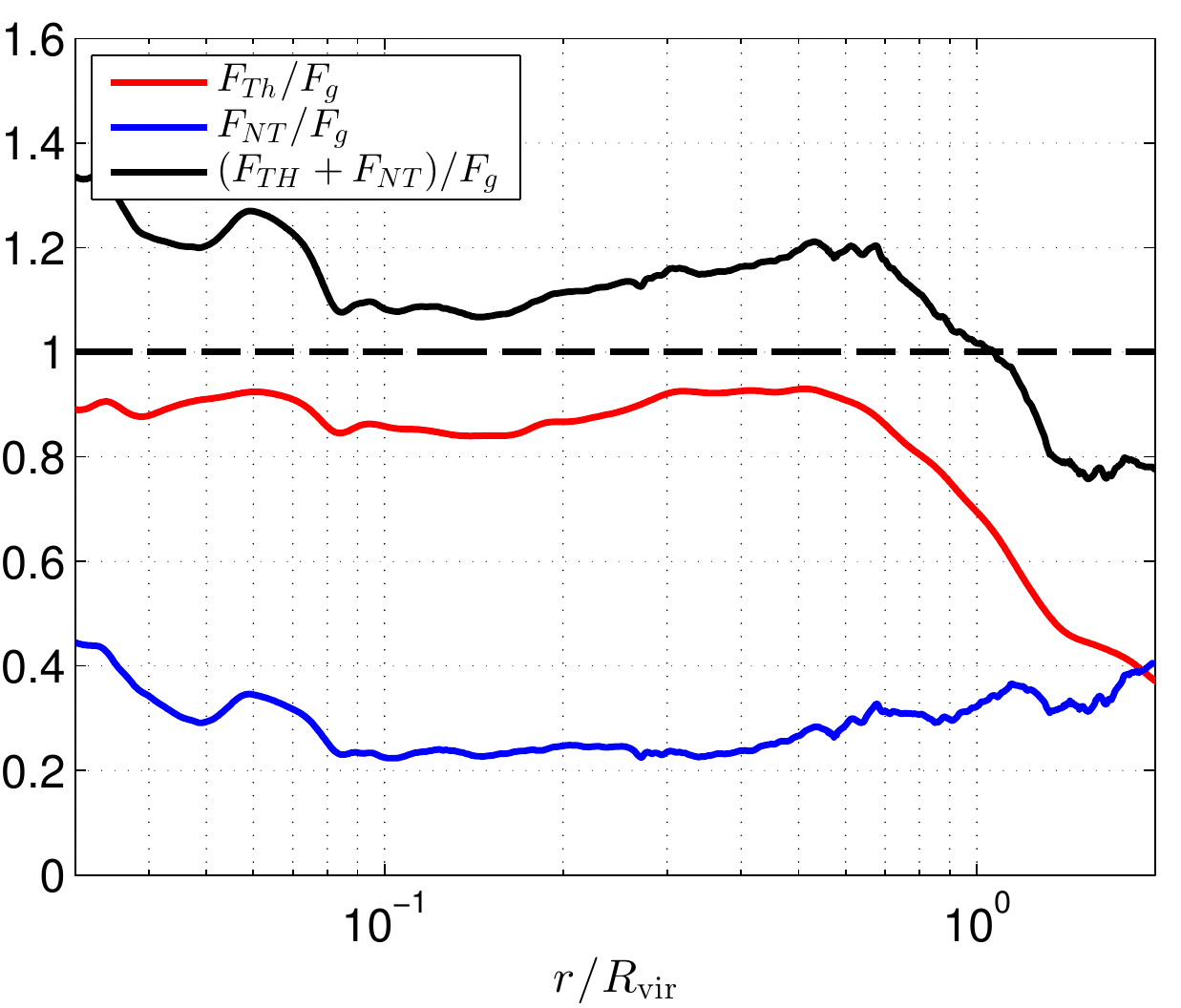}}
 \caption{The hydrodynamic stability of the clusters at \zeq{0}
   (top) and \zeq{0.6} (bottom) is explored via a
   generalized form of the Jeans equation \cref{eq:jeans3} for
   spherical shells. The ratio of the thermal to gravitational force
   (red) is shown alongside the non-thermal to gravitational
   force ratio (blue). The sum of the two ratios is shown in
   black, and a value of $1$ corresponds to a balance between
   the gravity and the thermal and non-thermal pressure support, in a
   steady-state cluster. The profiles are averaged over the R clusters
   \subrfig{jeans_a1_rlx} \& \subrfig{jeans_a06_rlx} and NR clusters
   \subrfig{jeans_a1_urlx} \& \subrfig{jeans_a06_urlx}
   separately. Gravity is balanced by the pressure in the R clusters
   but in the NR population, due to the higher contribution of
   non-thermal pressure, the systems are unbalanced and cannot be said
   to be in steady state. }
  \label{figs:jeans}
\end{figure*}

\begin{figure*}
  \subfloat[R Clusters at \zeq{0}]{\label{fig:jeansNT_a1_rlx}
    \includegraphics[width=8cm,keepaspectratio]{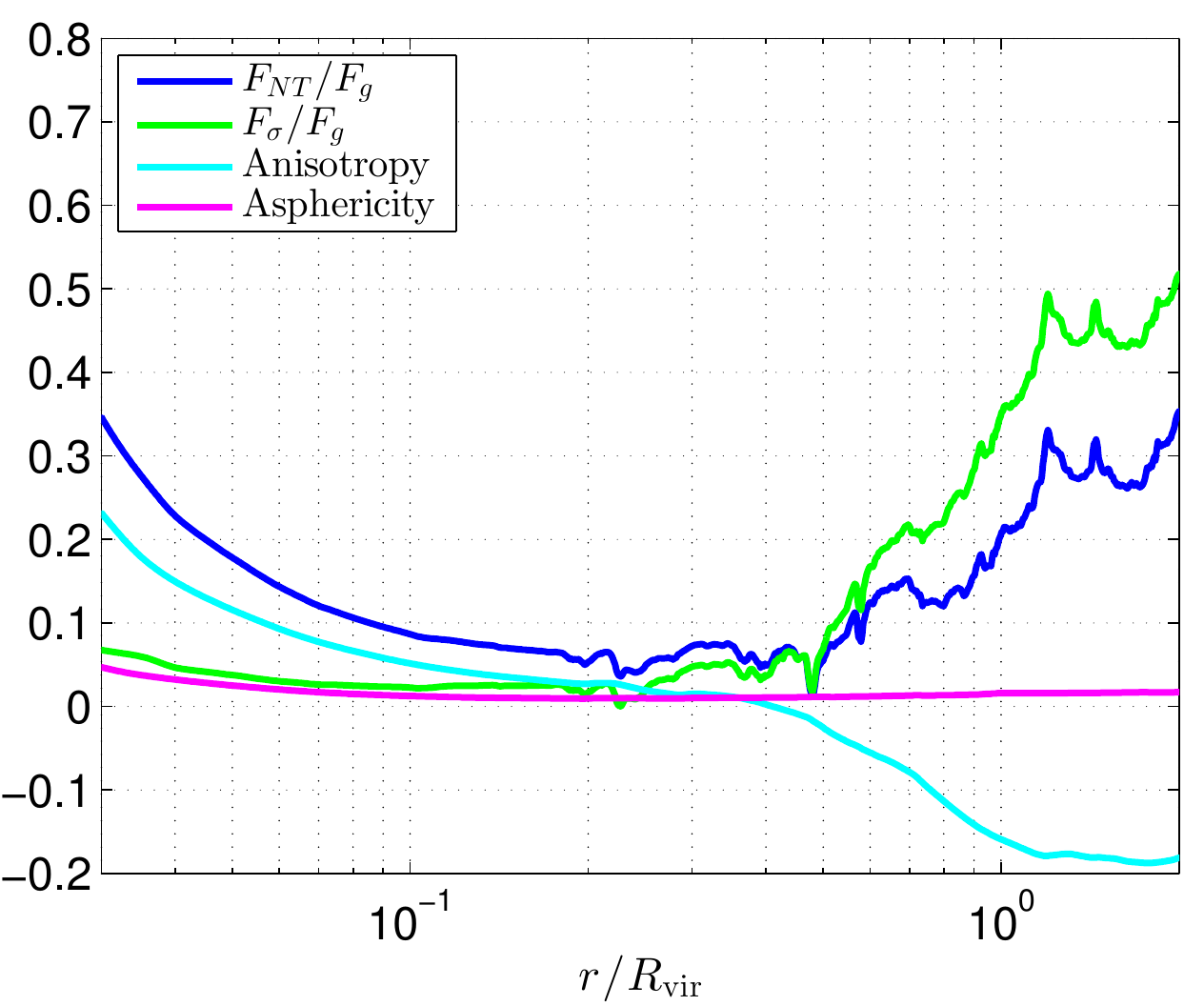}}
  \subfloat[NR Clusters at \zeq{0} ]{\label{fig:jeansNT_a1_urlx}
    \includegraphics[width=8cm,keepaspectratio]{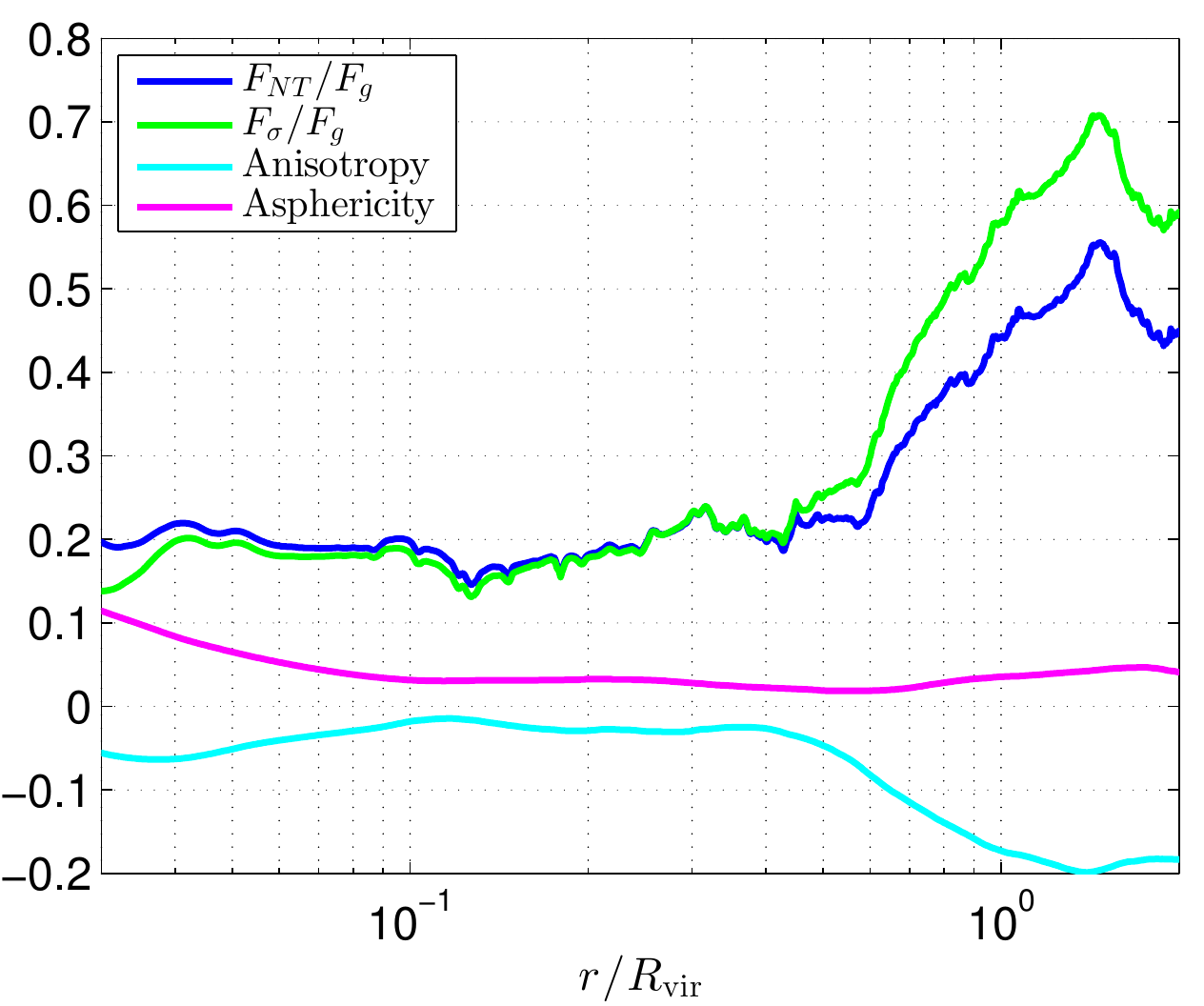}}\\
  \subfloat[R Clusters at \zeq{0.6}]{\label{fig:jeansNT_a06_rlx}
    \includegraphics[width=8cm,keepaspectratio]{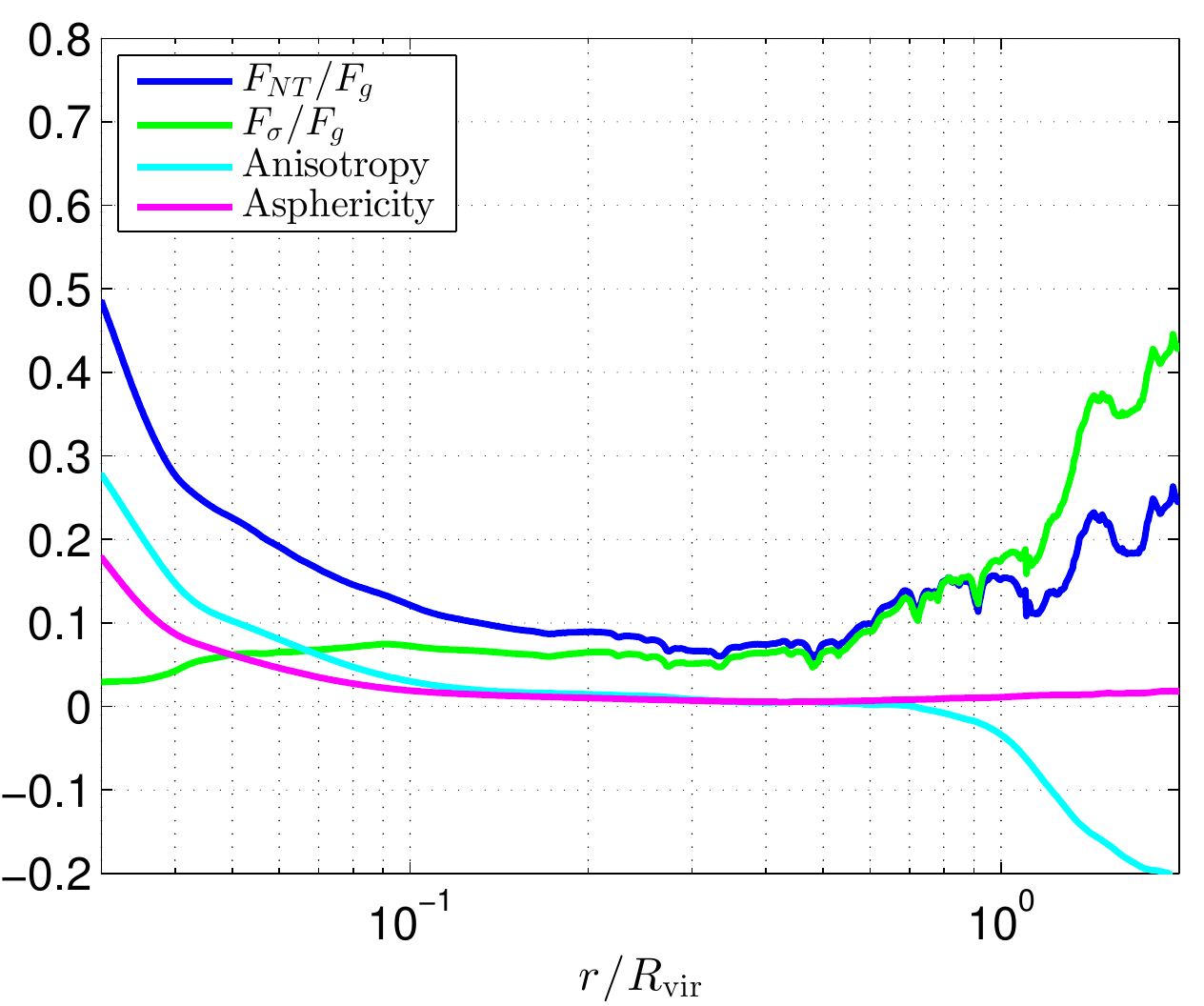}}
  \subfloat[NR Clusters at \zeq{0.6} ]{\label{fig:jeansNT_a06_urlx}
    \includegraphics[width=8cm,keepaspectratio]{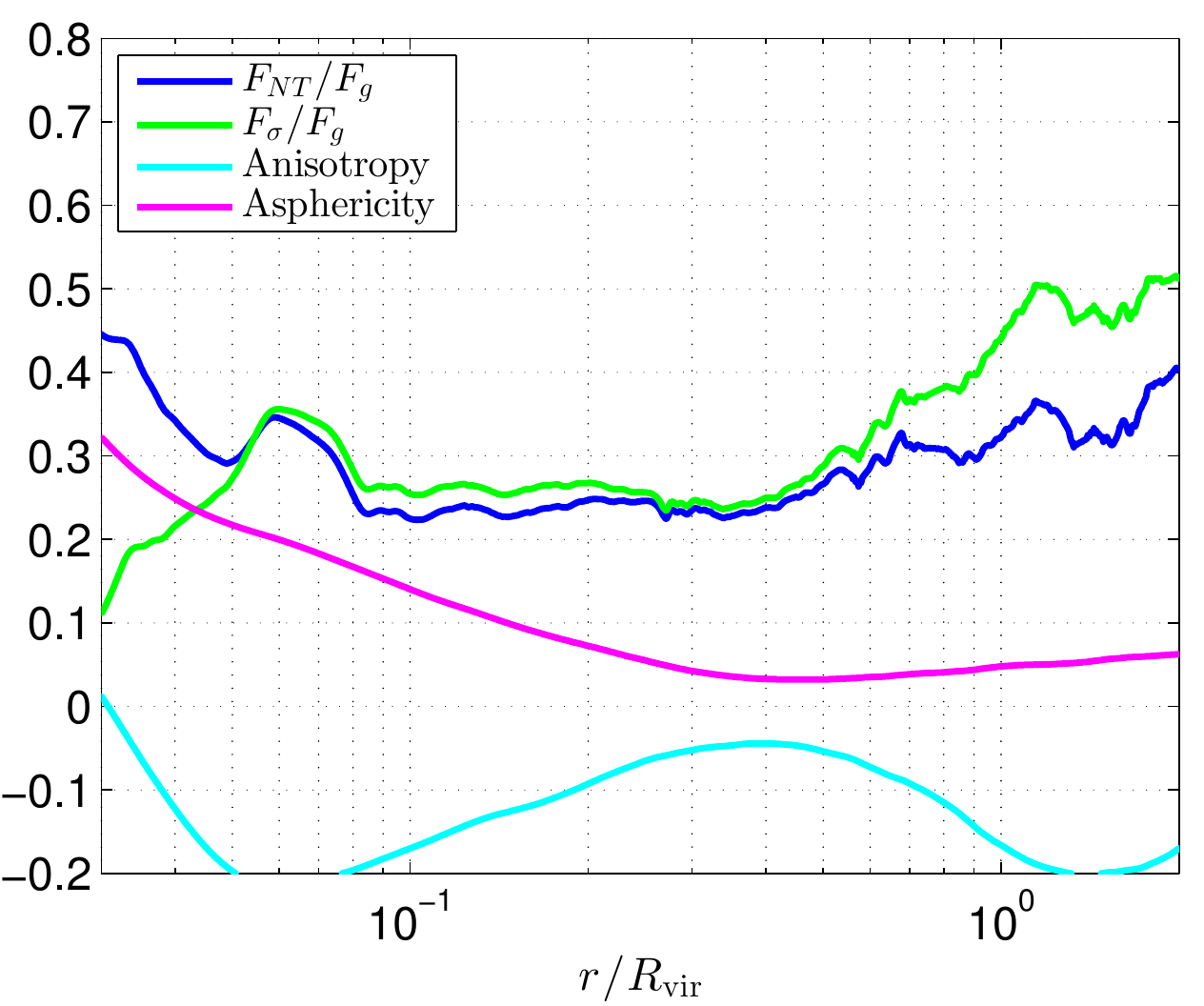}}
\caption{Analysing the non-thermal pressure support in the clusters at
  \zeq{0} (top) and \zeq{0.6} (bottom). Following
  \cref{eq:fNTherm}, the non-thermal part of the generalized Jeans equation 
  (blue) is decomposed into a `turbulent' pressure term (green) 
  an anisotropic term (cyan) and an asphericity term (magenta). 
  The decomposition is averaged over the R clusters
  \subrfig{jeansNT_a1_rlx} \& \subrfig{jeansNT_a06_rlx} and NR
  clusters \subrfig{jeansNT_a1_urlx} \& \subrfig{jeansNT_a06_urlx}
  separately. The pressure due to random motions is the dominant
  component except in the very central areas in the R clusters, where
  an ordered rotational motion is responsible for the increase in the
  contribution of the asphericity and anisotropic terms. In the NR
  clusters the terms relating to anisotropy and asphericity are more
  important, especially at \zeq{0.6}. }
  \label{fig:jeansNT}
\end{figure*}

\begin{figure*}
  \subfloat[R Clusters at \zeq{0}]{\label{fig:flux_rlx_a1}
    \includegraphics[width=8cm,keepaspectratio,trim=1cm 0 0 0, clip]{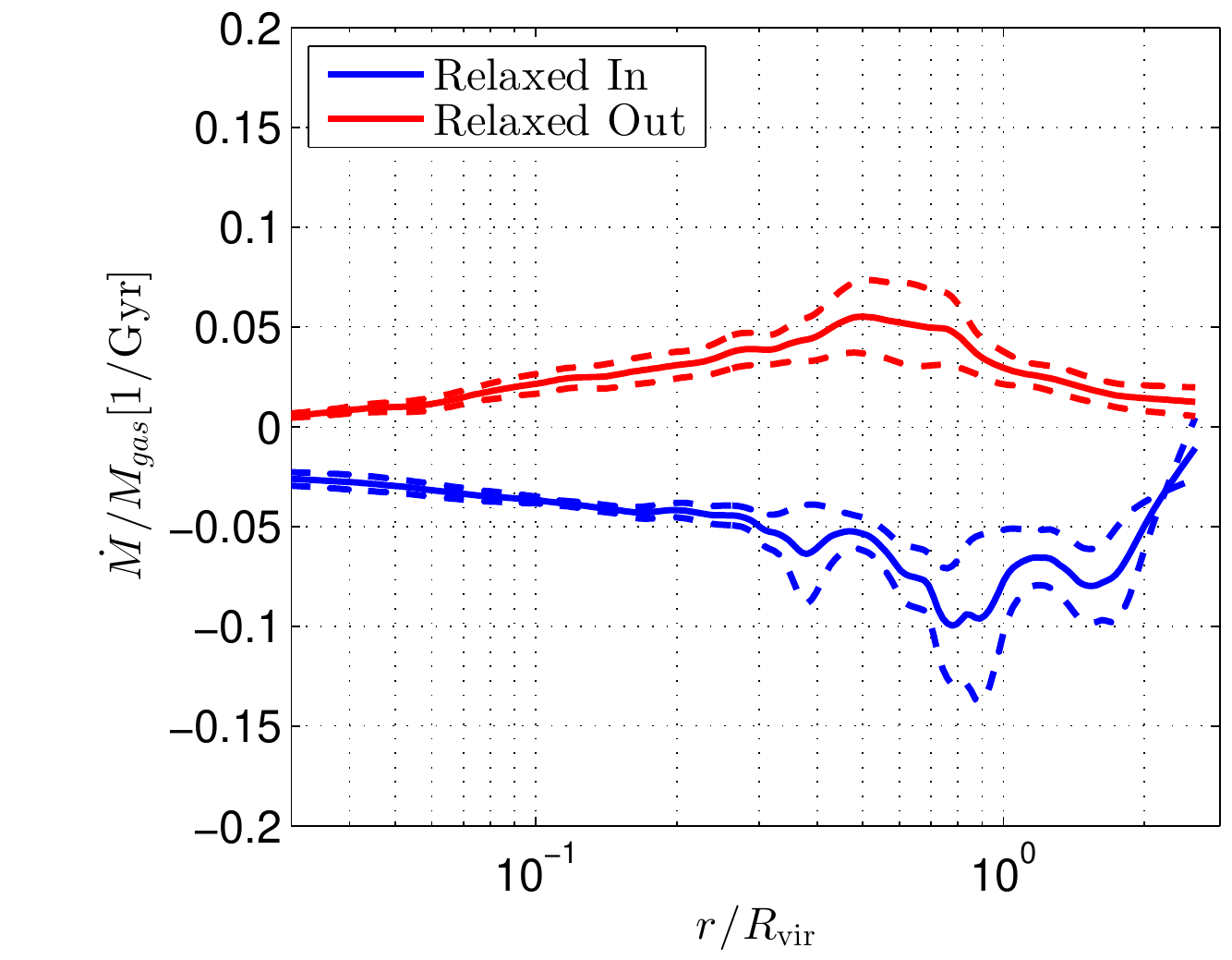}} 
  \subfloat[NR Clusters at \zeq{0}]{\label{fig:flux_urlx_a1}
    \includegraphics[width=8cm,keepaspectratio,trim=1cm 0 0 0, clip]{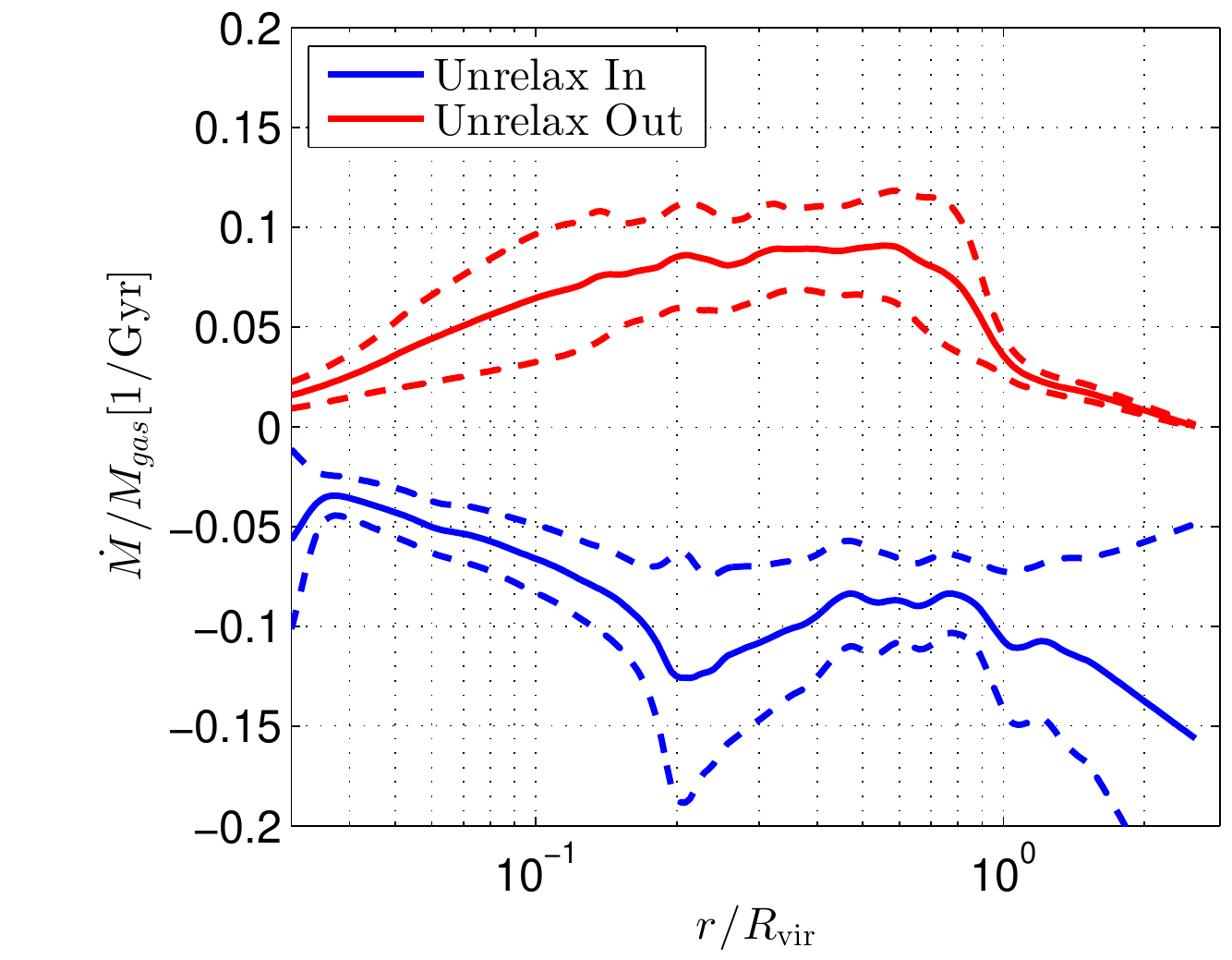}}\\
\subfloat[R Clusters at \zeq{0.6}]{\label{fig:flux_rlx_a06}
    \includegraphics[width=8cm,keepaspectratio,trim=1cm 0 0 0, clip]{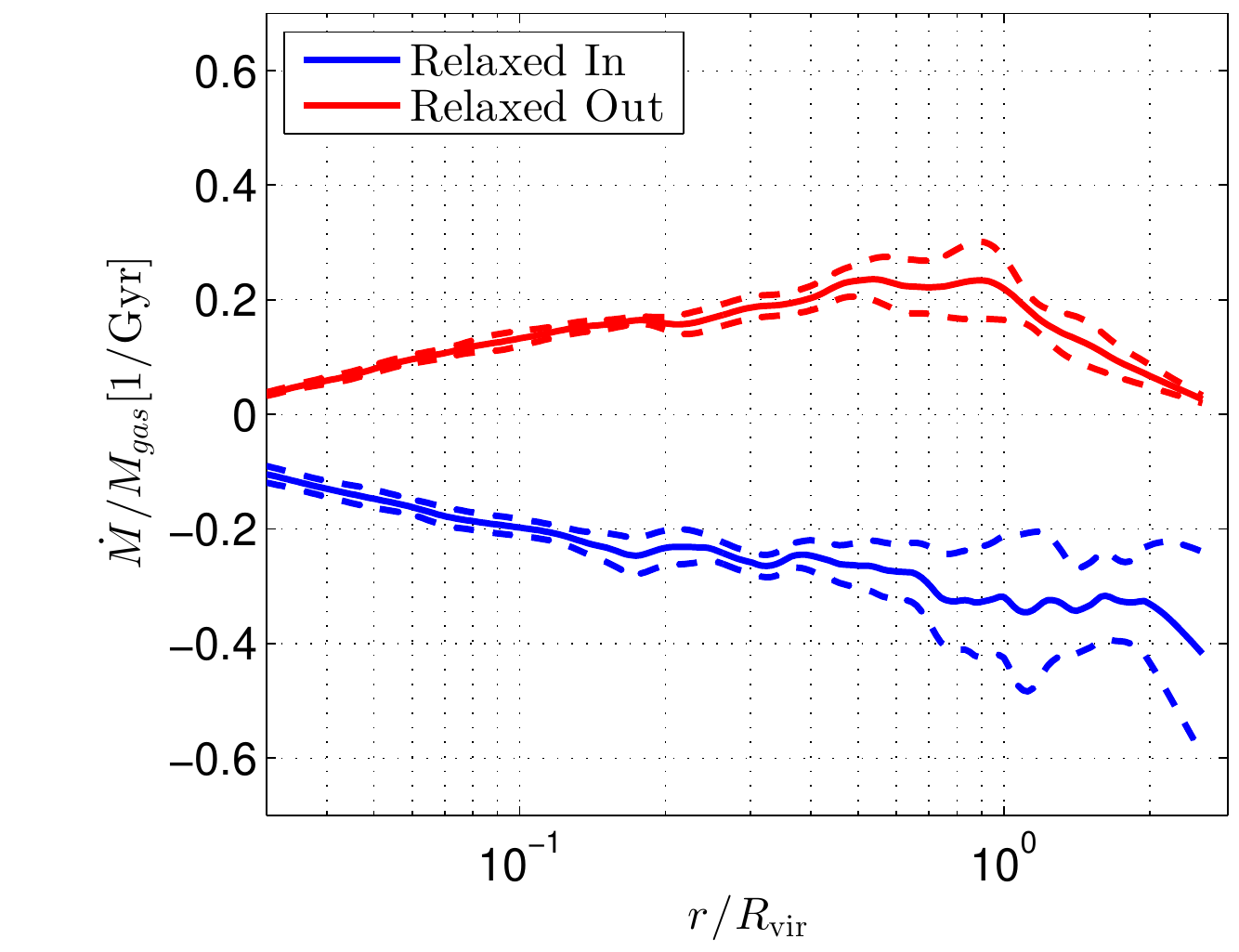}} 
  \subfloat[NR Clusters at \zeq{0.6}]{\label{fig:flux_urlx_a06}
    \includegraphics[width=8cm,keepaspectratio,trim=1cm 0 0 0, clip]{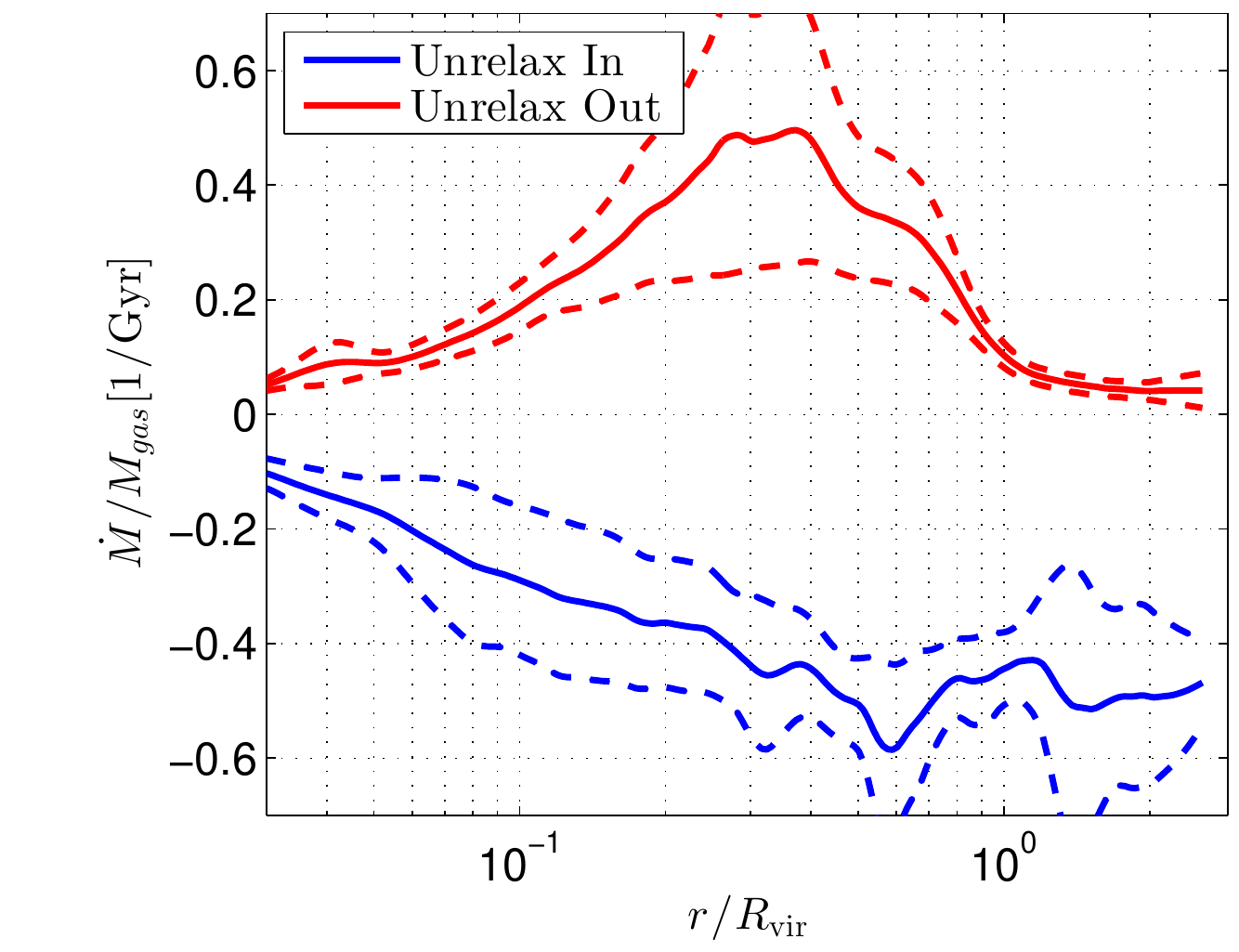}} 
  \caption{The mean mass inflow and outflow rates averaged over the
    population of R clusters \subrfig{flux_rlx_a1} \&
    \subrfig{flux_rlx_a06} and NR clusters \subrfig{flux_urlx_a1} \&
    \subrfig{flux_urlx_a06} at \zeq{0} (top) and \zeq{0.6}
    (bottom). The mean inflow rate is shown in solid blue and the mean
    outflow rate in red. Dashed lines mark a $1\sigma$ deviation about
    the mean. NR clusters are characterized by higher inflow rates at
    all radii, as well as higher outflows.}
  \label{fig:flux_a1}
\end{figure*}

In \cref{fig:jeans_a1_rlx,fig:jeans_a1_urlx}, we show the terms of
\cref{eq:jeans3} averaged separately over the R and NR clusters
respectively, at \zeq{0}. The same is shown for \zeq{0.6} in
\cref{fig:jeans_a06_rlx,fig:jeans_a06_urlx}.  A striking difference
between the two populations is that for R clusters, the thermal and
non-thermal terms balance the gravitational force within $\Rv$,
indicating that these clusters are indeed in a steady state
equilibrium. The same cannot be said of the NR clusters, where the
total contributions are greater than the gravitational pull. This
implies that the full derivative of the radial velocity (LHS of
\equnp{jeans1}) is positive, which can describe either a cluster which
is expanding ($v_r>0$) or one which is collapsing ($v_r<0$) but whose
internal pressure (thermal and non-thermal) is in the process of
balancing the collapse \citep{Nelson2014}. An examination of the
mass-accretion rate in these clusters
(\cref{fig:flux_urlx_a1,fig:flux_urlx_a06}) shows that the mass is
inflowing into the clusters, thus we are observing clusters in which
the pressure is building up to balance the collapse. In this way we
see that the NR nature of the clusters, originally defined by
observational measures, is manifest in the dynamical properties of the
cluster as well.

In both populations, the thermal component accounts for $\sim 90$ per
cent of the support against gravity, but in NR clusters the
non-thermal component is $\sim 20$ per cent of the gravitational
force, roughly twice as much as in R clusters. It is this excess which
embodies the `unrelaxedness' of the clusters.  In the outskirts of
both the R and NR clusters, beyond $\Rv$, the thermal pressure
steadily drops leaving the gas unsupported against the pull of gravity
in these regions.

Another difference between the two populations is seen in the central
regions of the clusters, within $0.2\Rv$.  In this region in the R
clusters, the contribution of the thermal pressure is seen to
gradually decline corresponding to a rise in the non-thermal
contributions \citep[see
  also][]{Lau2009,Molnar2010a,Vazza2011,Nelson2014a}. The NR clusters,
on the other hand, show no such change. To understand this issue, we
must ascertain which of the non-thermal terms is responsible for this
increase, which is at odds with observational limits on the
non-thermal support \citep{Sanders2011,Pinto2015}.

We decompose the non-thermal support into the `turbulent' pressure and
the anisotropic and aspherical components, as detailed in
\cref{eq:fNTherm}, and average the profiles separately over R and NR
clusters. This decomposition is shown in \cref{fig:jeansNT} for the
clusters at \zeq{0} and \zeq{0.6}.

The rise in non-thermal support in the very central regions of R
clusters ($\lesssim 0.1\Rv$), is due to the anisotropic component, and
to a lesser degree the aspherical component, both of which are due to
tangential motions, which both rise gradually towards the centre. The
source for this rise is a strong, ordered rotational motion in the
centres of these clusters, as was shown in \citealt{Lau2011}. The
build-up of the rotational motion is due to the overcooling in the
centre of the cluster, and is therefore an unphysical feature of the
simulation we do not expect to find in cluster observations.

The dominant non-thermal component in clusters is the pressure due to
random motions, which accounts for $\sim 50$ per cent of the
gravitational support outside $\Rv$, and drops to $\sim 20$ per cent
within $\Rv$ for NR clusters and only several per cent in R
clusters. The asphericity in all clusters can be seen to be small at
all radii. In the outskirts ($\gtrsim \Rv$) there is a significant
amount of anisotropy which dwindles to several per cent at $r\approx
0.5\Rv$.

In the NR population, there is a clear difference between the two
epochs, with the anisotropic and ashperical component playing a minor
role in the central regions at \zeq{0} in contrast to their very
significant contribution in the central parts at \zeq{0.6}.

\subsection{Mass Inflow into the Cluster}\label{sec:massFlux}
We compare the R and NR cluster populations in terms of the gas
accreted on to the systems, by studying the mass inflow rate through
radial shells,
\begin{equation}\label{eq:massflux} 
\dot M(r)=\iint_{\Omega} \rho \left[\left(\vec{v}-\vec{v}_{\mathrm{cm}}\right)
  \cdot \hat{r}\right] r^2 \diff\Omega \, ,
\end{equation} 
where $\rho$ and $\vec{v}$ are the density and velocity of a gas
element and $\vec{v}_{\mathrm{cm}}$ is the centre of mass velocity of
the system, defined on a scale of $\sim \Rv$ (see \rfsec{sims}). In
practice, the density and velocity data grids in the simulations were
sampled on a spherical shell of a given radius, while maintaining a
constant surface area for each sampling point. The sampling points
were distributed along 256 latitudinal rings, 256 points uniformly
distributed along each ring, with the spacing between the rings
adjusted according to the distance between points along the rings to
ensure a constant surface area per sampling point. Thus, the rings in
the `equatorial' regions are more densely packed than in close to the
`poles'.

We separate the inflowing $(v_r < 0)$ and outflowing $(v_r > 0)$
material for each cluster and average the mass inflow/outflow rate
over the R and NR clusters separately. In \cref{fig:flux_a1}, we show
the inflow and outflow rates for the clusters at \zeq{0} and
\zeq{0.6}, respectively.

In the outer regions of the cluster, out to the virial shock front
which extends out to $\sim 2\Rv$ \citep[][Zinger et al., in preparation]{Lau2015}, 
the inflow rate is roughly equal to analytic estimates for the average
inflow rate at $\Rv$ based on the EPS formalism \citep{Birnboim2007}:
\begin{equation}\label{eq:analytflux} 
\frac{\dot M}{M} \simeq 0.1 M_{15}^{0.15}a^{-2.25} \units{Gyr^{-1}},
\end{equation} 
where $M_{15}\equiv\Mv /10^{15}\msun$ and $a=(1+z)^{-1}$ is the
expansion parameter of the universe\footnotemark. This estimate is
expected to be largely independent of radius, is identical for both
the dark matter and baryons \citep{Dekel2013}, and is found to be
consistent with cosmological simulations
\citep{Wechsler2002,Neistein2008}. \footnotetext{The power $2.25$,
  which is chosen to fit the results from EPS and simulations at
  $z<1$, is a deviation from $5/2$, the value that can be simply
  derived analytically for the EdS regime at $z>1$
  \citep{Dekel2013}. The power $0.15$ is determined by the slope of
  the power spectrum.}

The differences between the R and NR population are striking -- in
both epochs the inflow and outflow rates in the NR systems are larger
by roughly a factor of 2 at all radii. More mass is being driven
into the NR clusters, and the gravitational energy released as a
result of the inflow into the potential well (see
\rfsec{streamEnergy}) leads to the build-up of random motions, as seen
in \cref{fig:flux_urlx_a1,fig:flux_urlx_a06}.

An interesting feature seen in \cref{fig:flux_a1} is the very tight
scatter in the inner parts of R clusters, compared with the much
larger scatter in the NR population. The main reason for this feature
is the similarity of the inflow and outflow profiles of the different
R clusters, and in particular the dearth of substructure in these
systems compared to the NR population. We discuss the role of
substructure and mergers in determining the relaxedness of a cluster
in \rfsec{mergers}.

\begin{figure*}
  \subfloat[Mass Inflow Rate]{\label{fig:cl6Map_flux_b8_a1}
    \includegraphics[height=7.5cm,keepaspectratio]{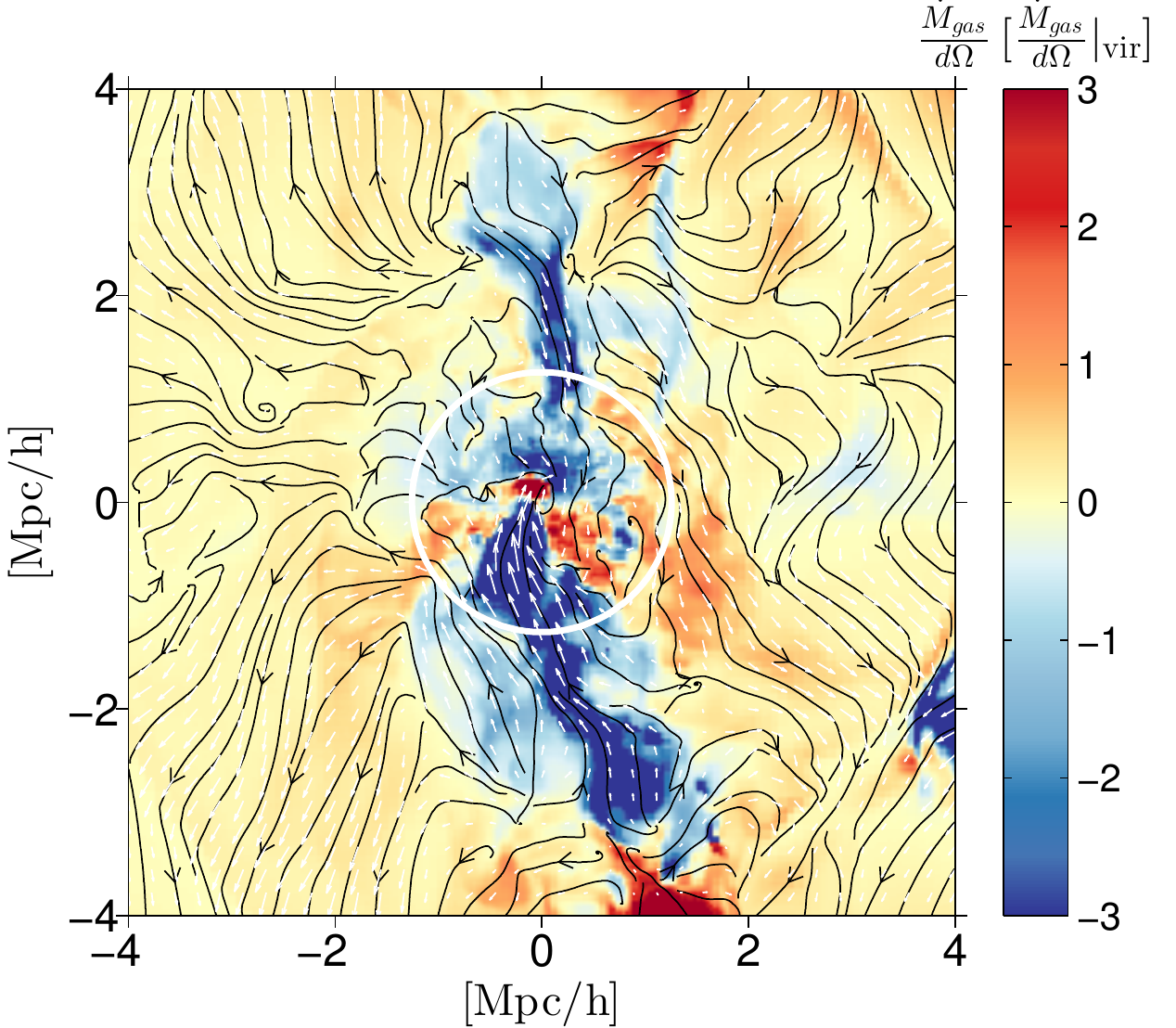}}
  \subfloat[Density]{\label{fig:cl6Map_dens_b8_a1}
    \includegraphics[height=7.5cm,keepaspectratio]{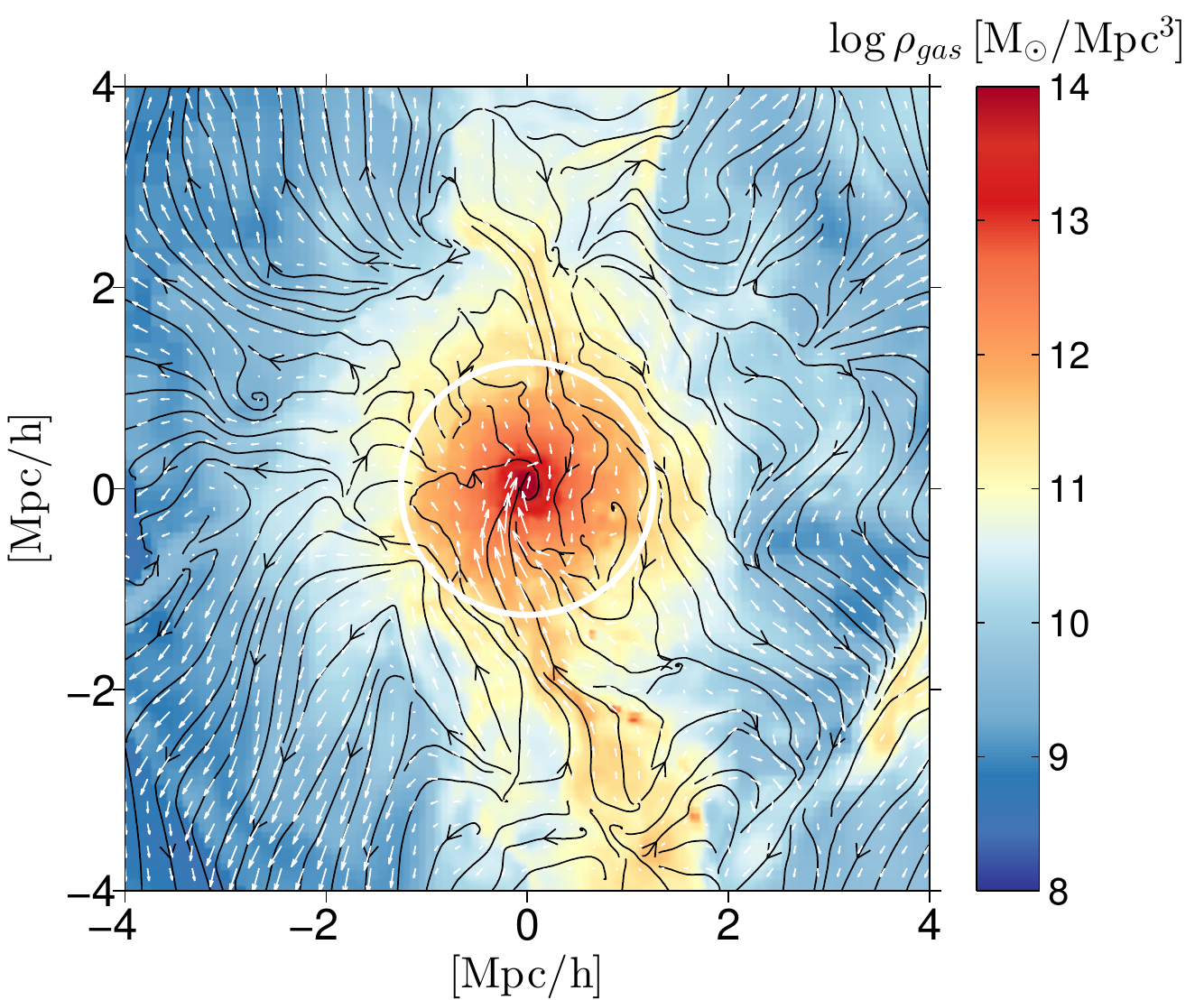}}\\
  \subfloat[Temperature]{\label{fig:cl6Map_temp_b8_a1}
    \includegraphics[height=7.5cm,keepaspectratio]{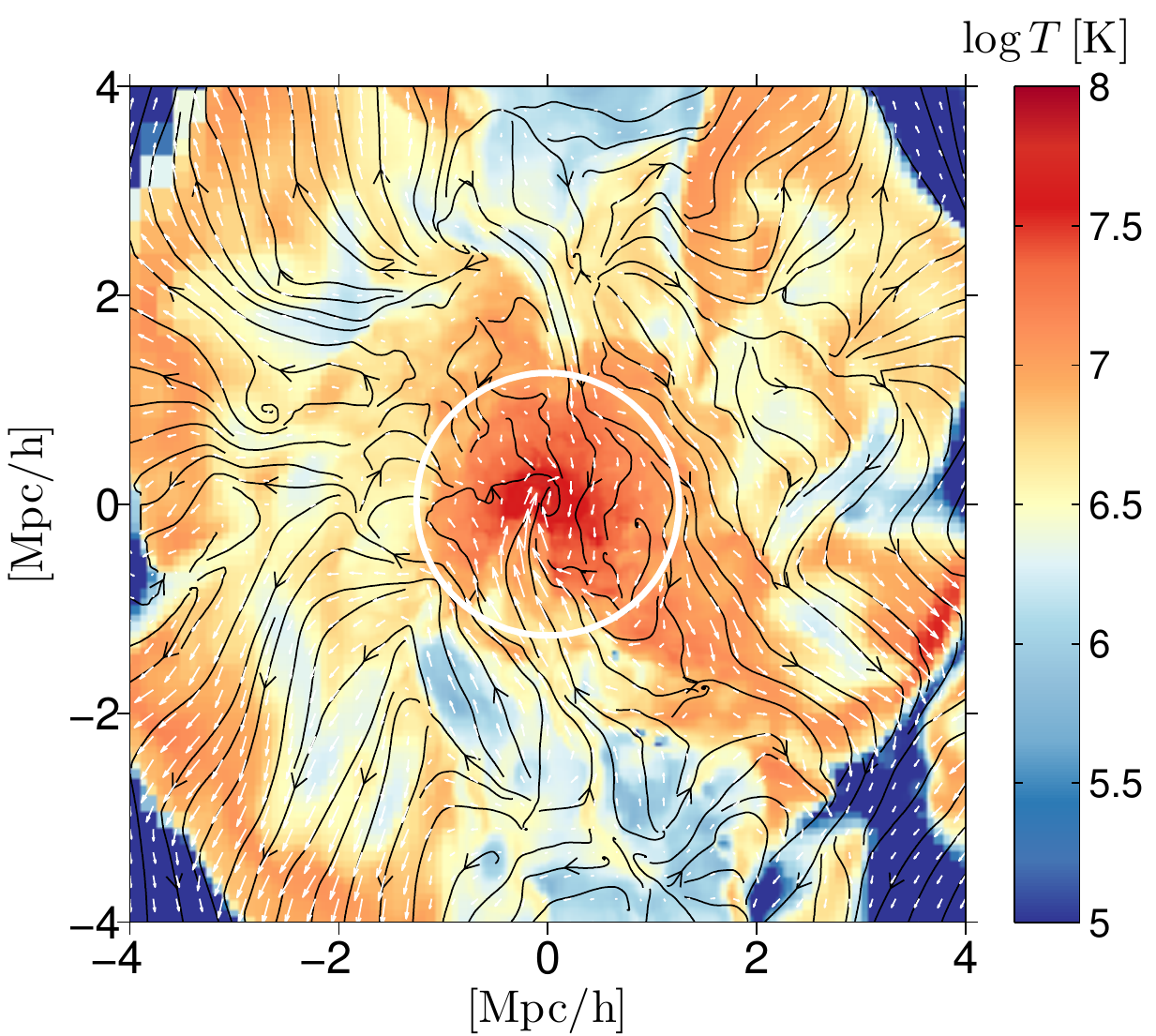}}
  \subfloat[Entropy]{\label{fig:cl6Map_ent_b8_a1}
    \includegraphics[height=7.5cm,keepaspectratio]{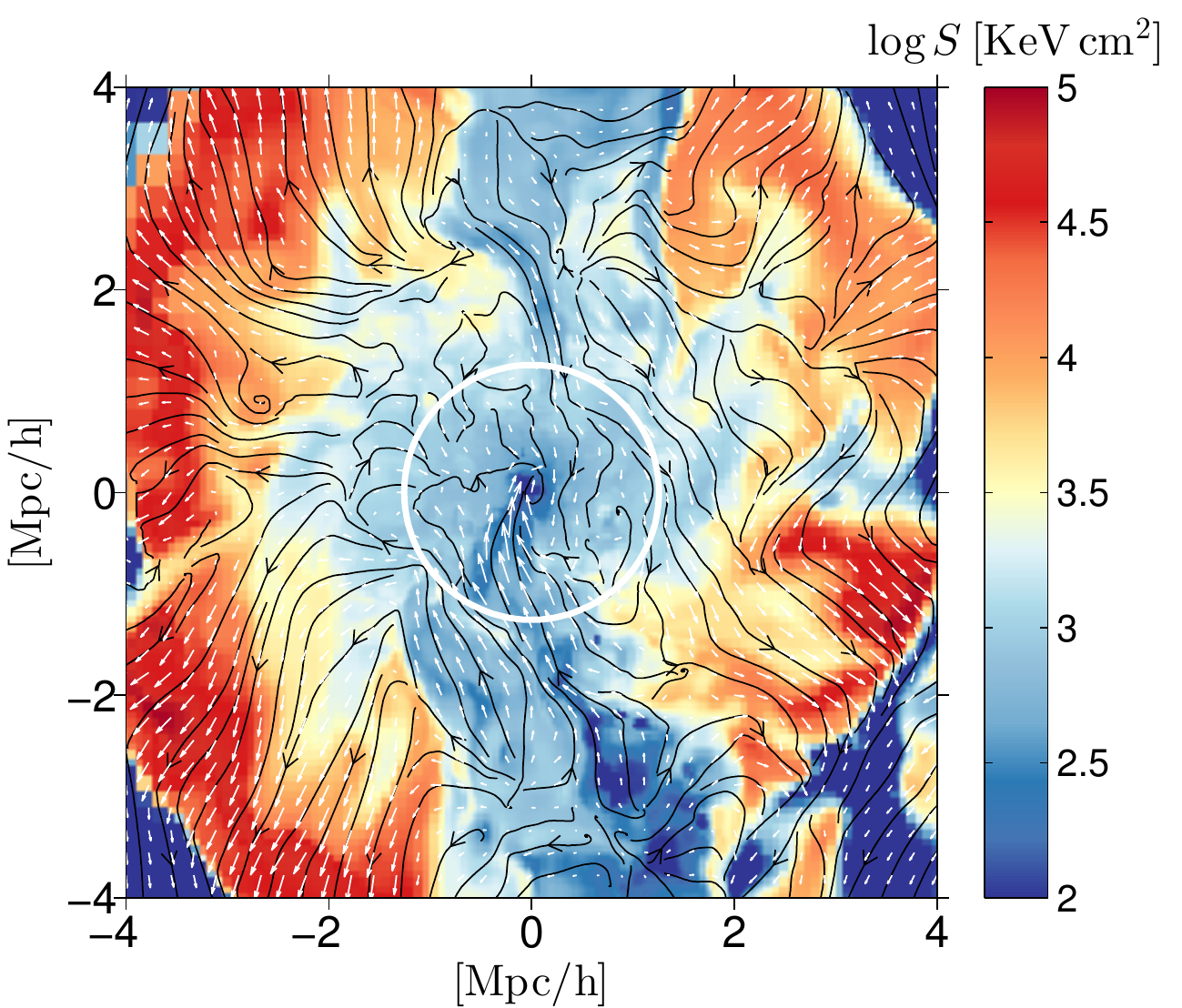}}
  \caption{Maps of mass inflow rate \subrfig{cl6Map_flux_b8_a1},
    density \subrfig{cl6Map_dens_b8_a1}, temperature
    \subrfig{cl6Map_temp_b8_a1} and entropy \subrfig{cl6Map_ent_b8_a1}
    in an equatorial slice of $180h^{-1} \units{kpc}$ of the cluster CL6
    at \zeq{0}, focusing on the areas between $\Rv$ and the accretion
    shock. The velocity field in the plane is described by white
    arrows and in addition, by black streamlines. The virial radius is
    marked by a white circle. It is clear that mass inflow rate occurs
    almost exclusively along the streams puncturing the hot medium
    which are clearly seen in the entropy map. The streams are only
    marginally cooler than the hot medium at large radii and are
    heated to $\Tv$ within $\Rv$.}
  \label{fig:cl6Maps}
\end{figure*}

\begin{figure*}
  \subfloat[Mass inflow rate on the scale of $\Rv$.]{\label{fig:cl6Map_flux_b4_a1}
    \includegraphics[height=7.69cm,keepaspectratio]{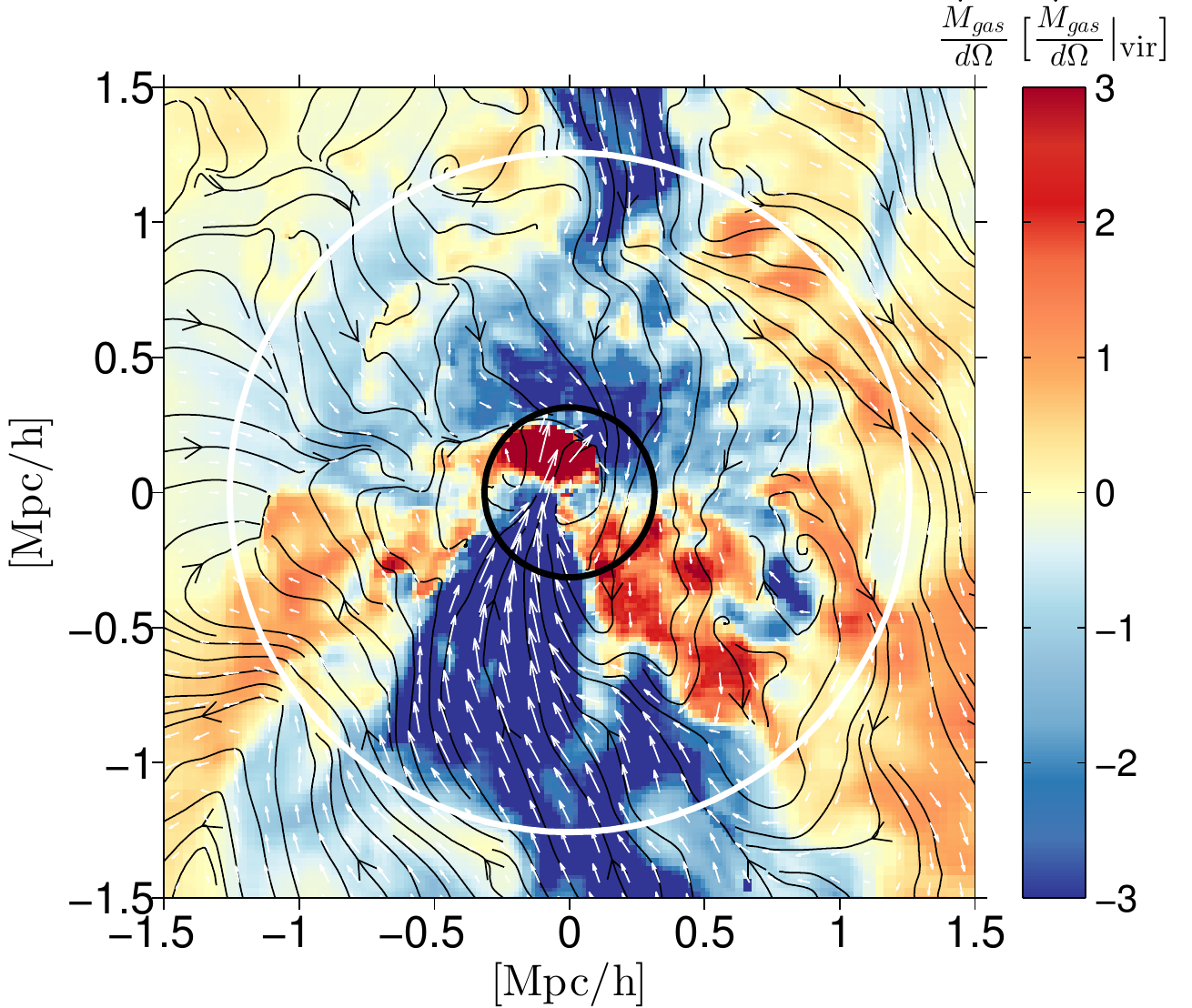}} 
  \subfloat[Mach number of the gas flow on the scale of $\Rv$.]{\label{fig:cl6Map_mach_b4_a1}
    \includegraphics[height=7.5cm,keepaspectratio]{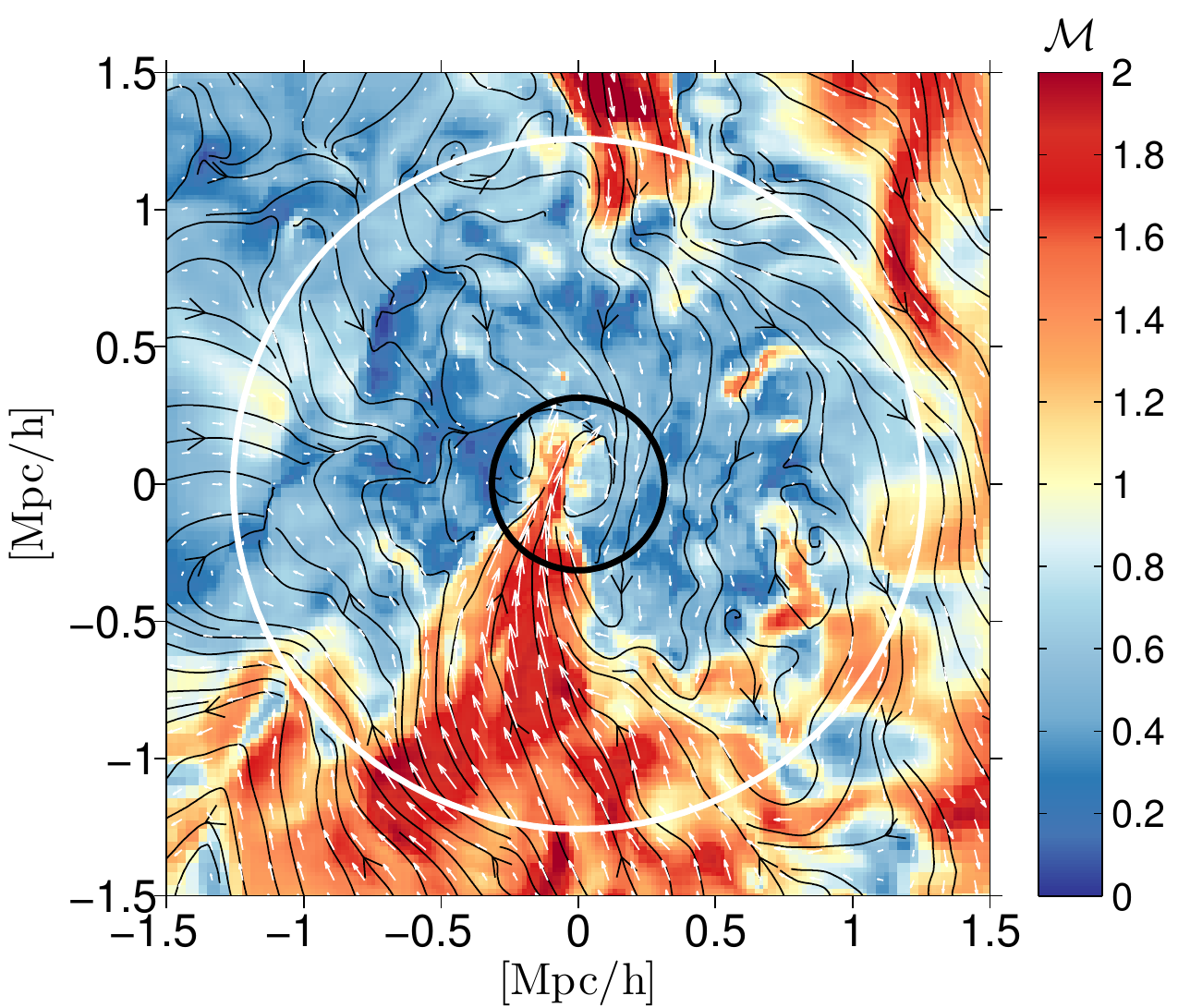}} \\
  \subfloat[Mass inflow rate in the cluster centre.]{\label{fig:cl6Map_flux_b1_a1}
    \includegraphics[height=7.69cm,keepaspectratio]{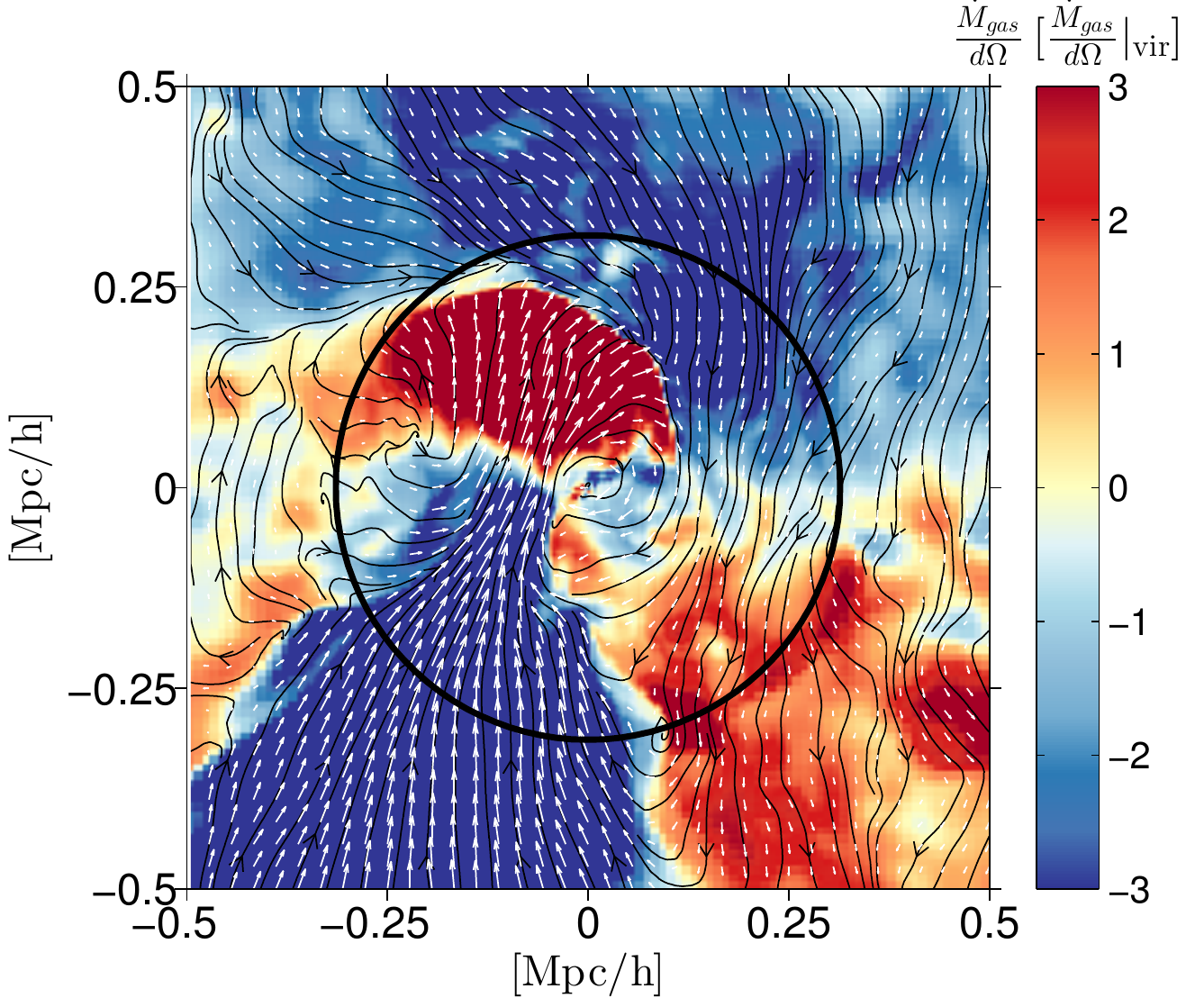}}
  \subfloat[Mach number of the gas flow in the cluster centre.]{\label{fig:cl6Map_mach_b1_a1}
    \includegraphics[height=7.5cm,keepaspectratio]{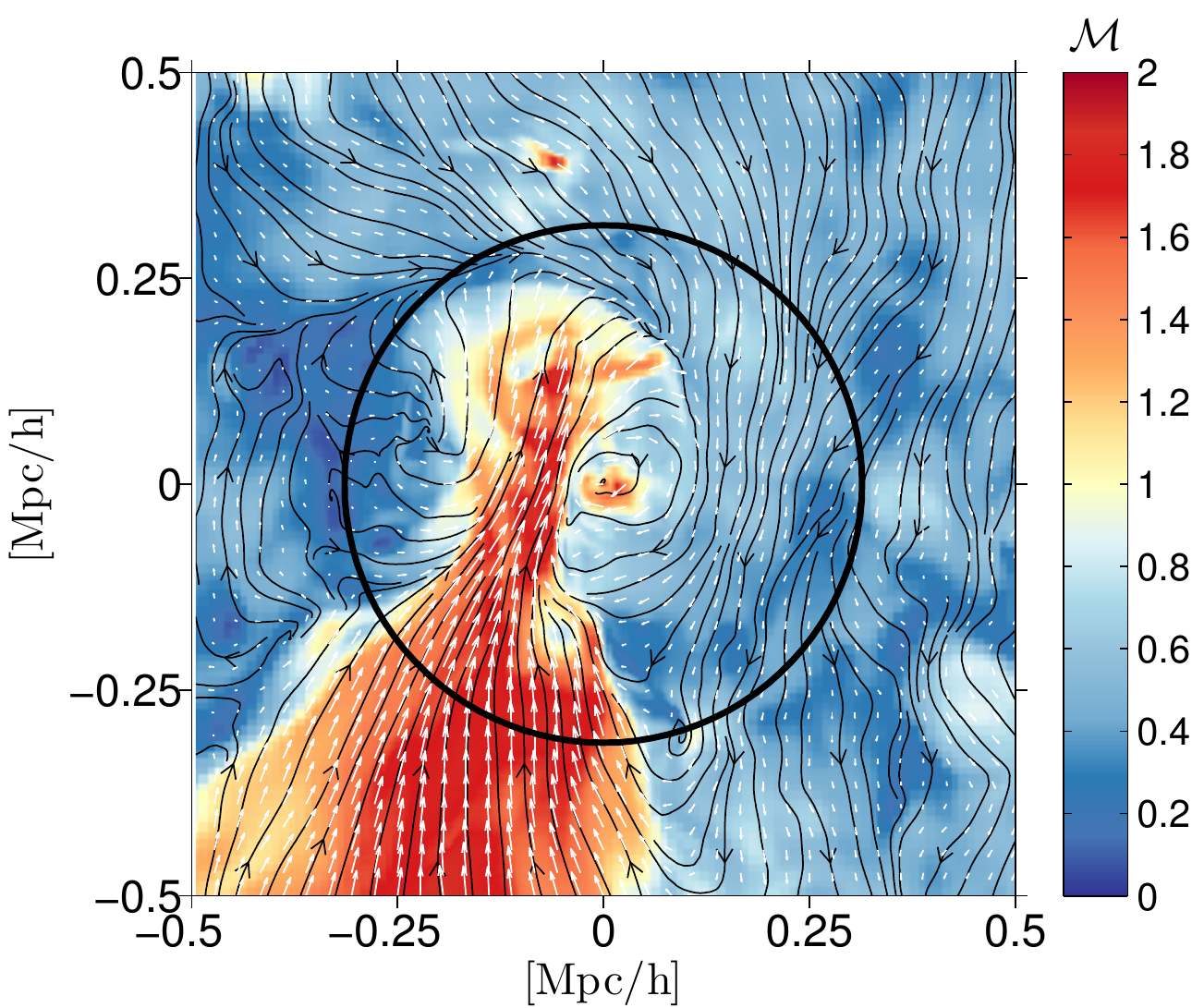}} 
  \caption{Maps of mass inflow rate (left) and Mach number
    (right) in a thin equatorial slice of the NR cluster CL6,
    also shown in \cref{fig:cl6Maps}, at \zeq{0} shown on the scale of
    $\Rv$ \subrfig{cl6Map_flux_b4_a1} \& \subrfig{cl6Map_mach_b4_a1}
    and focusing on the central regions \subrfig{cl6Map_flux_b1_a1} \&
    \subrfig{cl6Map_mach_b1_a1}. $\Rv$ is marked by a white circle and
    a $0.25\Rv$ radius black circles marks the inner regions. The Mach
    number $\mach$ is the ratio of the local velocity and the sonic
    speed derived from the shell averaged temperature. A supersonic
    $(\mach>1)$ stream enters form the bottom and penetrates to the
    very centre of the cluster. The abrupt change in the sign of the
    inflow rate in the up-wards flowing stream (transition from blue
    to red) is due to the stream overshooting the centre of the
    system. A shock front forms where the stream collides with the
    medium.}
  \label{fig:cl6FluxMach}
\end{figure*}

\begin{figure*}
  \subfloat[Temperature Map]{\label{fig:cl6CF_temp_b1_a1}
    \includegraphics[height=7.5cm,keepaspectratio]{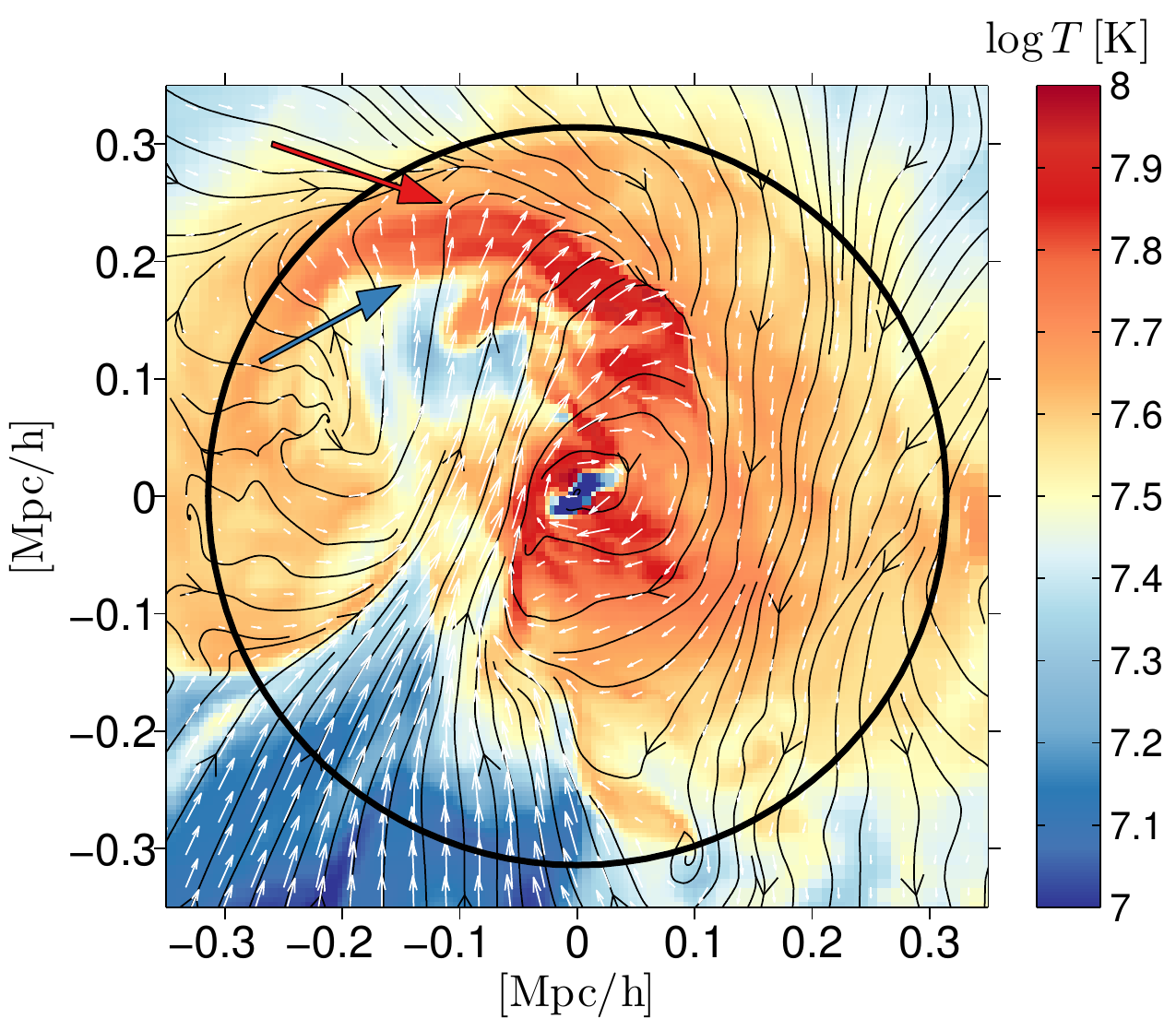}}
  \subfloat[Thermal Pressure Map]{\label{fig:cl6CF_press_b1_a1}
    \includegraphics[height=7.5cm,keepaspectratio]{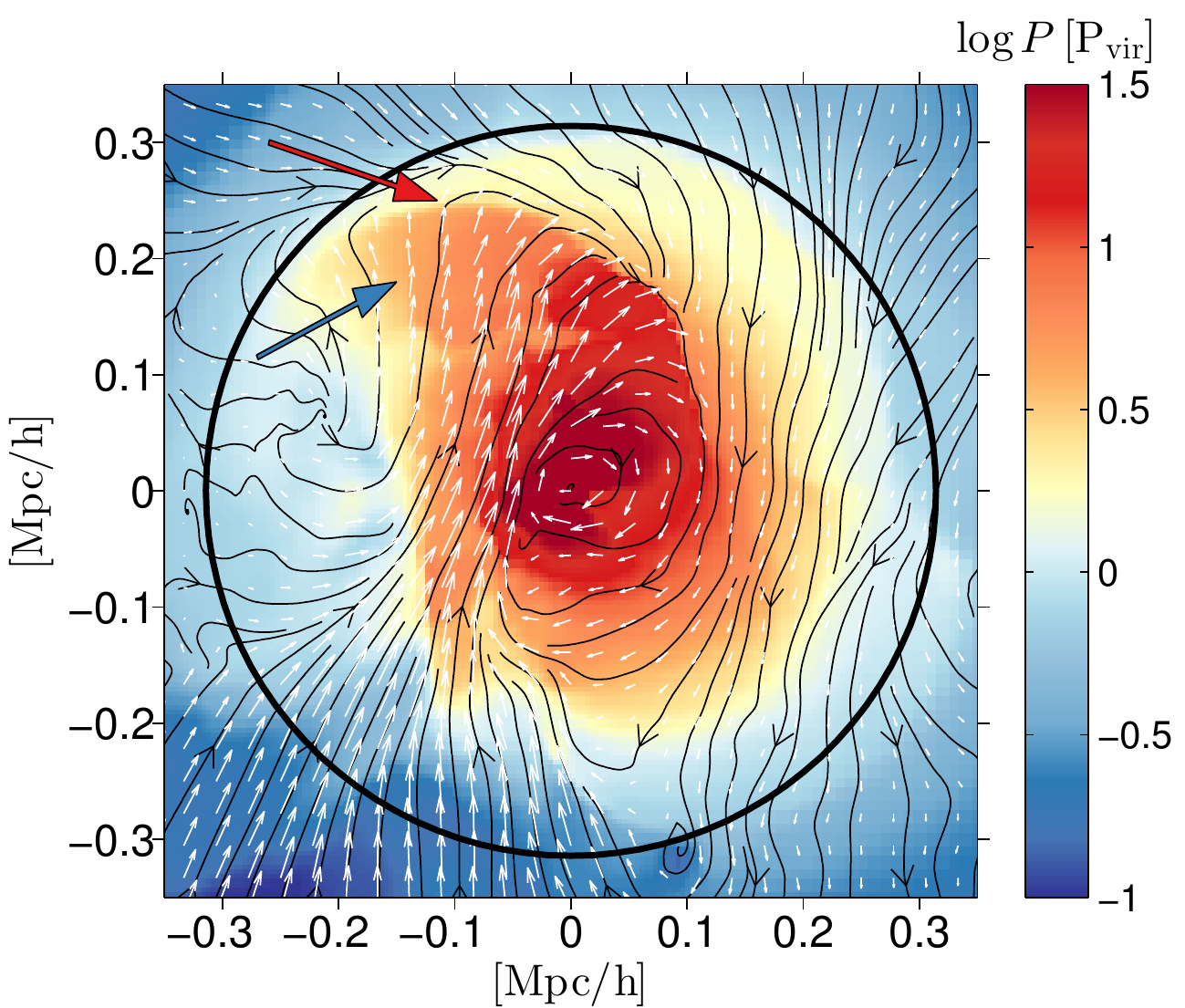}}
  \caption{Zoom in on the central region of the cluster CL6 previously
    shown in \cref{fig:cl6Maps,fig:cl6FluxMach}, in temperature
    \subrfig{cl6CF_temp_b1_a1} and pressure
    \subrfig{cl6CF_press_b1_a1}. A shock is clearly evident where the
    upward flowing stream collides with gas flowing from the top of
    the picture, marked by a red arrow). Just below the shock, a
    contact discontinuity (`cold front') can be seen, marked by a blue
    arrow.}
  \label{fig:cl6CF}
\end{figure*}

\section{Gas Streams}\label{sec:streams}
In the previous section a link was established between the relaxedness
of a cluster and the mass inflow rate into the cluster. We now turn to
examine in detail the way gas accretes on to a cluster and specifically
the filamentary nature of the accretion, along gas streams, and
examine the effect they may have on the ICM, especially in the central
regions.

\subsection{Properties of Gas Streams}\label{sec:streamsProperties}
In \cref{fig:cl6Maps}, we show the temperature, density, entropy and
mass inflow rate of the gas in the form of colour-maps for a thin
equatorial slice in the cluster CL6, focusing on the regions where the
streams enter the system beyond $\Rv$. It is clear that the mass
inflow, even in the central parts, is dominated by gas streams which
flow into the cluster and extend well beyond $\Rv$ (see also
\cref{fig:cl14Pen,fig:cl3Pen}). The accretion of baryonic mass into
haloes is known to occur predominately along the dark matter filaments
of the cosmic web \citep{Klar2012,Gheller2015}, both in the form of
clumpy accretion \citep{Vazza2013} and also, as smooth streams, as can
be seen in these simulations.

The streams are best seen in terms of mass inflow rate and entropy
since the gas flow is mostly adiabatic along the stream and the
entropy is conserved, even as the stream is compressed by the ambient
gas and its temperature and density increase as a result. In the outer
regions beyond $\Rv$, their temperature is also lower, enhancing the
entropy contrast with their surroundings.

However, this gas inflow cannot be termed a `cold flow' since the
temperature within the streams is lower only by a factor of a few than
the surrounding medium. This is due to the gas being heated first when
crossing the cylindrical shock surrounding the filaments, and
subsequently by shock-heating within the supersonic streams as they
flow into the system. As a result, the gas temperature in a stream is
already well above $10^6\units{K}$ when it enters the system
\citep{Werner2008}. At these temperatures, radiative cooling occurs
predominately through Bremsstrahlung radiation which scales as
$\propto T^{1/2}$. A drop in temperature in a radiatively cooling
parcel of gas will lead to a drop in the cooling efficiency, and thus
runaway cooling cannot develop (as it does for lower temperatures) and
the gas will remain hot. The heating of the gas, by compression and by
shocks, in the stream continues as the supersonic stream flows into
the cluster, especially within $\Rv$ as the density of the ICM
increases.

Unlike the `cold flows' seen in simulations of high redshift galaxies
\citep{Birnboim2003,Keres2005,Dekel2006,Dekel2009,Keres2009}, these
gas streams feeding the clusters are more aptly described by the term
`warm streams', due to the pre-heating in the filaments. After
crossing $\Rv$, the streams are heated to $\Tv\gtrsim 10^7\units{K}$
by the time they reach the inner regions ($r \lesssim 0.25\Rv$).

In \cref{fig:cl6FluxMach}, we examine the streams in the cluster CL6 both
at the scale of $\Rv$
(\cref{fig:cl6Map_flux_b4_a1,fig:cl6Map_mach_b4_a1}) and zooming in on
the central regions
(\cref{fig:cl6Map_flux_b1_a1,fig:cl6Map_mach_b1_a1}). The mass inflow
rate is shown, which allows easy identification of the streams. In
addition, we show the Mach number of the gas flows with respect to the
surrounding hot medium. The Mach number $\mach$ shown is the ratio
between the local gas velocity and the typical speed of sound at a
given radius
\begin{equation}\label{eq:csonic}
c_{\mathrm{s}}(r) = \sqrt{\pderiv{P}{\rho}\Bigg|_{S}  } =\sqrt{\gamma
  \frac{\kboltz\overline{T}}{\mu m_{\mathrm{p}}}},
\end{equation} 
where the second expression derives from the ideal gas equation of
state used to describe the gas. The factor $\gamma$ is the adiabatic index
($\gamma = 5/3$ for a monoatomic gas), $\kboltz$ is the
Boltzmann constant, $\mu m_{\mathrm{p}}\simeq 0.59 m_{\mathrm{p}}$ is the average particle
mass ($m_{\mathrm{p}}$ being the proton mass), and $\overline{T}$ is the
mass-weighted mean temperature of a spherical shell (including the
stream) at a given radius.

A very prominent supersonic ($\mach > 1$) stream coming from outside
$\Rv$ can be identified at the bottom of the picture and flowing into
the central region. Focusing on the centre of the cluster, we see the
stream overshooting the centre (leading to an abrupt change in sign of
the radial inflow rate, from blue to red). The stream collides with
gas flowing in the opposite direction leading to the formation of a
shock front, as can be seen in the temperature and pressure maps of
the region (\cref{fig:cl6CF_temp_b1_a1,fig:cl6CF_press_b1_a1},
respectively). The shock front, marked by the red arrow in
\cref{fig:cl6CF}, is arc-shaped and characterized by a sharp change in
the pressure and density. Shocks such as these, generated by the gas
streams, are the mechanism for transforming the gravitational energy
brought in by the gas streams into heat and random motions.

Just below the shock, a contact discontinuity, also known as a `Cold
Front', can be seen as a sharp change in the temperature map, marked
by a blue arrow \citep[see also][]{Nagai2003}. The contact discontinuity is in
pressure equilibrium and thus is not noticeable in the pressure
map. An in-depth analysis of this shock and `cold front' is presented
in a future paper (Zinger et al., in preparation). 

\subsection{Energy Deposition by the Gas Streams}\label{sec:streamEnergy}
\begin{figure*}
  \subfloat[Mass inflow rate on the scale of $\Rv$]{\label{fig:cl14Pen_b2}
    \includegraphics[height=7.5cm,keepaspectratio]{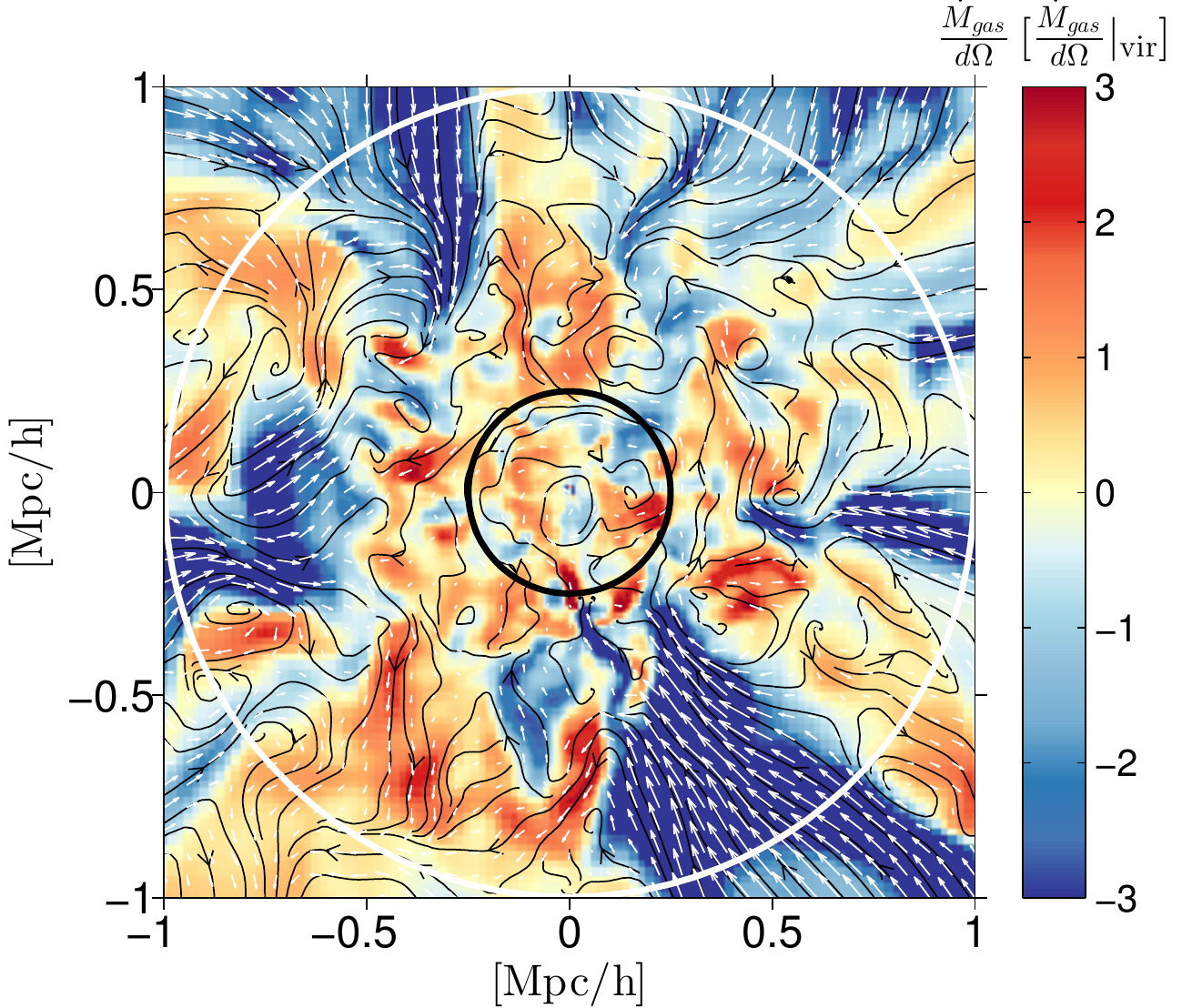}}
  \subfloat[Mass inflow rate focusing on the inner regions]{\label{fig:cl14Pen_b1}
    \includegraphics[height=7.5cm,keepaspectratio]{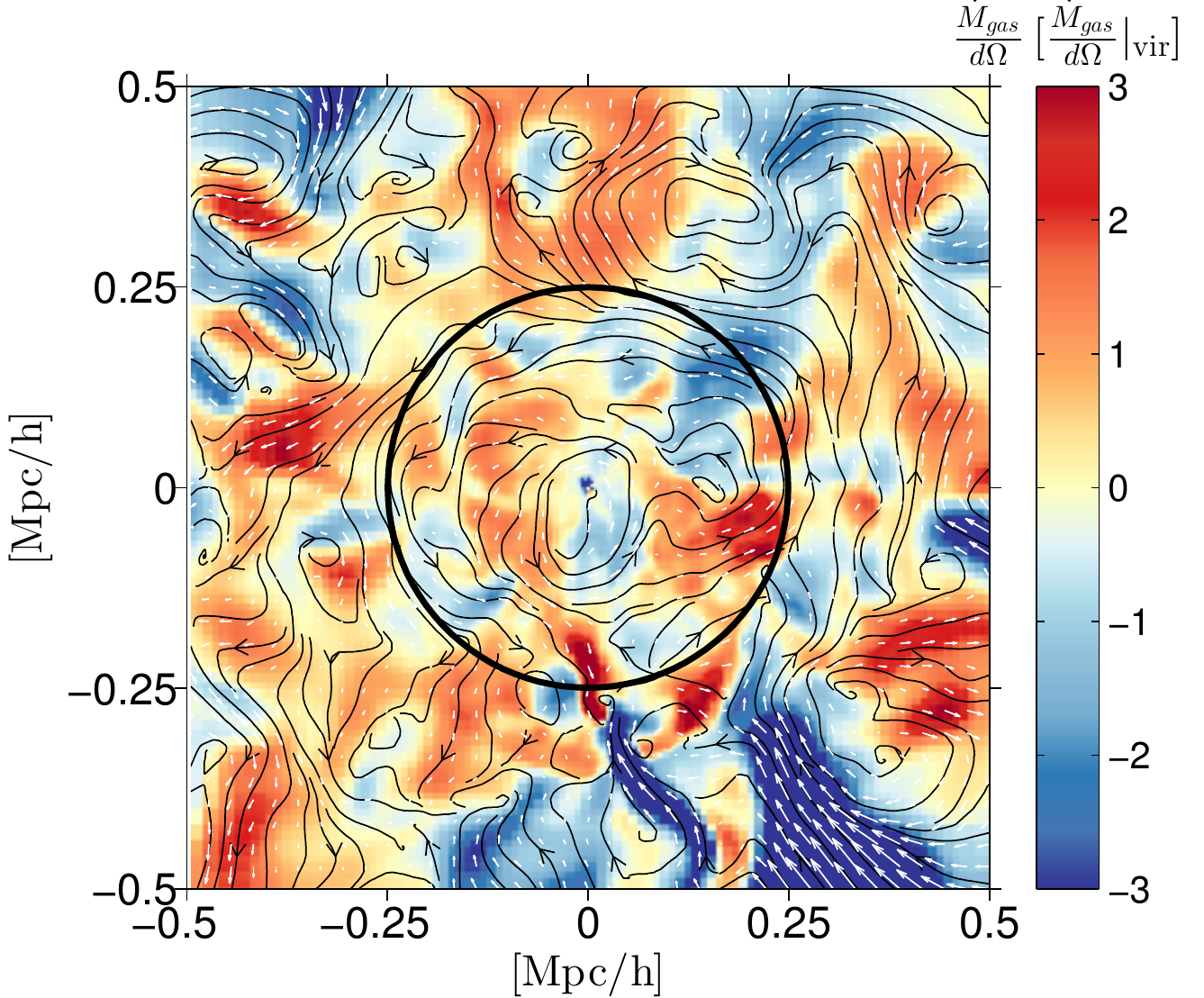}}
  \caption{Maps of mass inflow rate in a thin equatorial slice of the
    R cluster CL14 at \zeq{0}, shown on the scale of $\Rv$
    \subrfig{cl14Pen_b2} and focusing on the inner regions
    \subrfig{cl14Pen_b1}. White arrows describe the velocity field in
    addition to black streamlines. The virial radius is marked by a
    white circle and the black circle demarks the $0.25\Rv$
    radius. Several streams can be identified entering the system,
    e.g.\@ bottom-right and top-left corners in \subrfig{cl14Pen_b2},
    but they stop at $r\approx 0.5\Rv$ and do not penetrate into the
    central regions in which no coherent flows can be found.}
  \label{fig:cl14Pen}
\end{figure*}

\begin{figure*}
  \subfloat[Mass inflow rate at \zeq{0.6}]{\label{fig:cl3Pen_a06}
    \includegraphics[height=7.5cm,keepaspectratio]{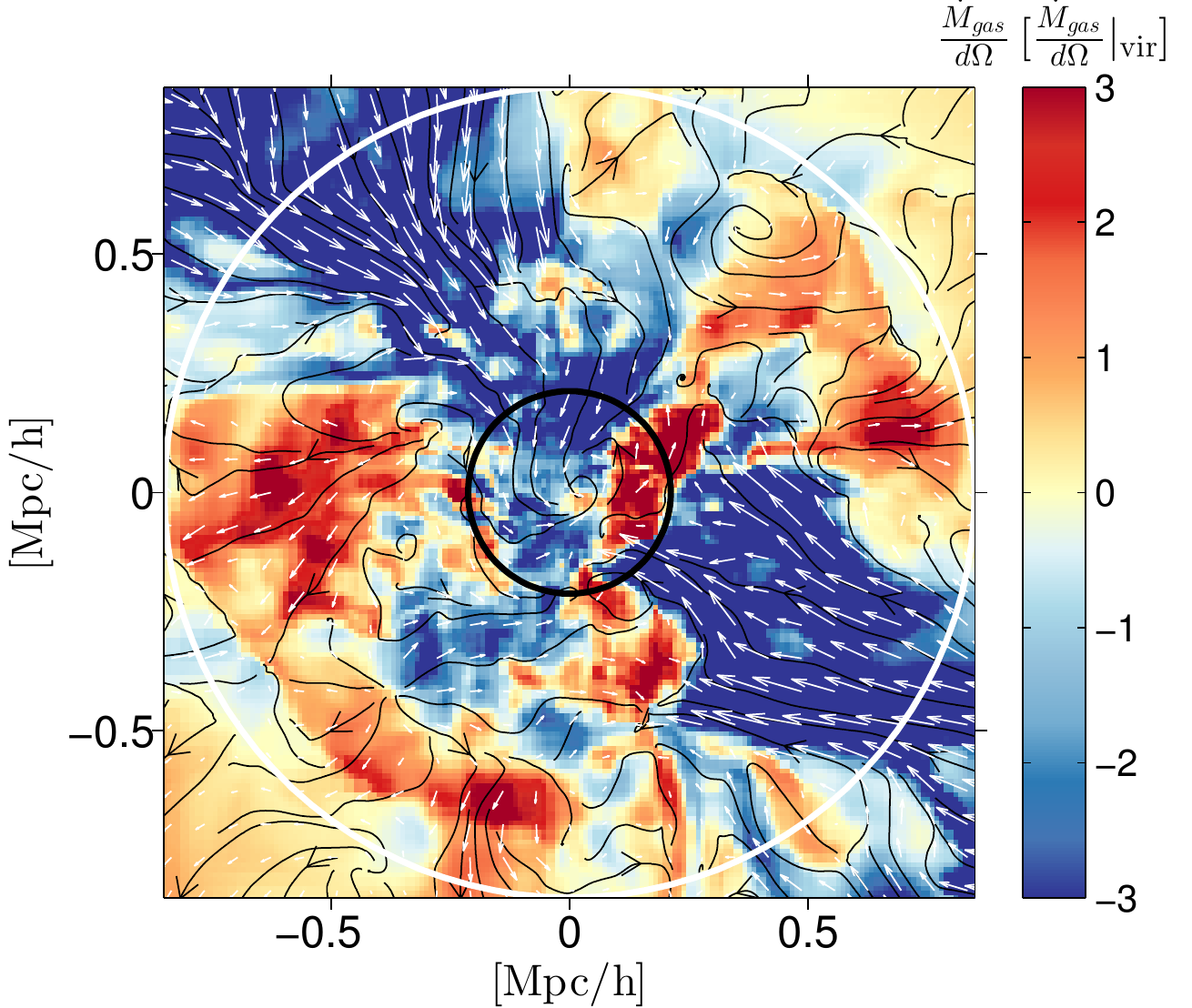}}
  \subfloat[Mass inflow rate at \zeq{0}]{\label{fig:cl3Pen_a1}
    \includegraphics[height=7.5cm,keepaspectratio]{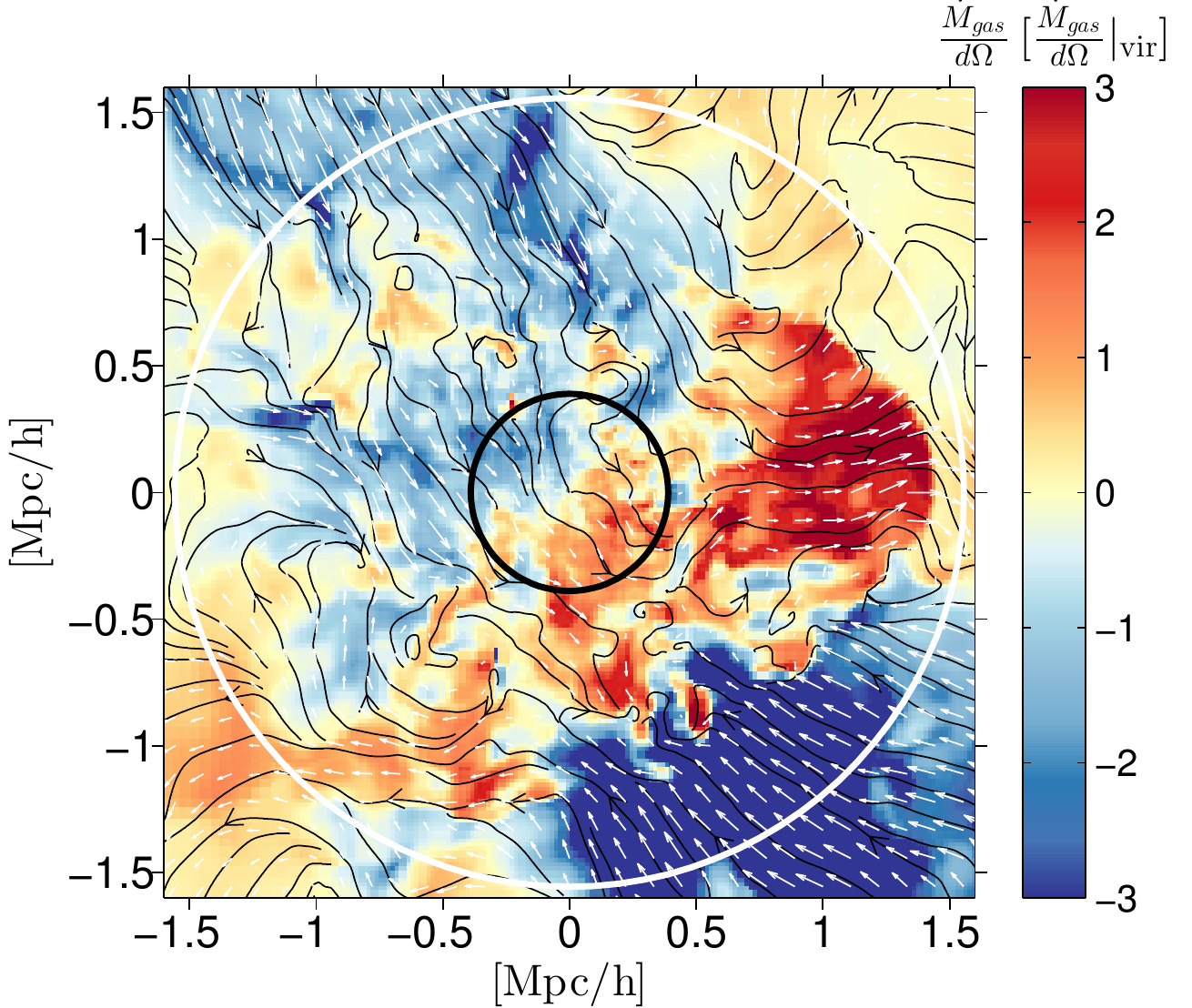}}
  \caption{Maps of mass inflow rate in a thin equatorial slice of the
    cluster CL3 shown at \zeq{0.6} \subrfig{cl3Pen_a06} and at \zeq{0}
    \subrfig{cl3Pen_a1}. White arrows describe the velocity field in
    addition to black streamlines. The virial radius is marked by a
    white circle and the black circle demarks the $0.25\Rv$ radius. At
    the earlier redshift the streams penetrate to the central regions
    while at \zeq{0} the streams are stopped at $r \approx 0.5\Rv$.}
  \label{fig:cl3Pen}
\end{figure*}

\begin{figure*}
  \subfloat[Mass inflow rate at \zeq{0.6}]{\label{fig:cl11Pen_a06}
    \includegraphics[height=7.5cm,keepaspectratio]{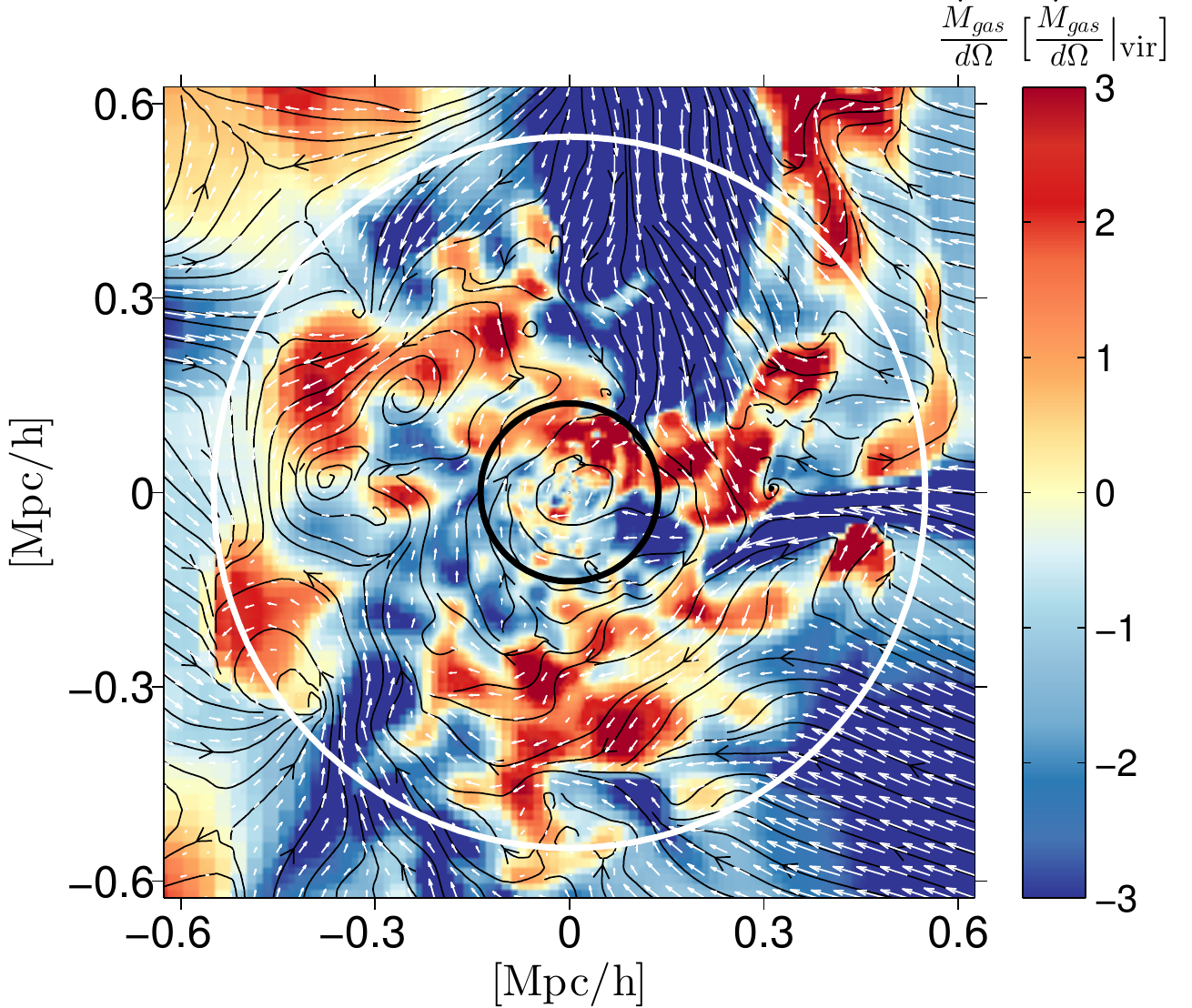}}
  \subfloat[Mass inflow rate at \zeq{0}]{\label{fig:cl11Pen_a1}
    \includegraphics[height=7.5cm,keepaspectratio]{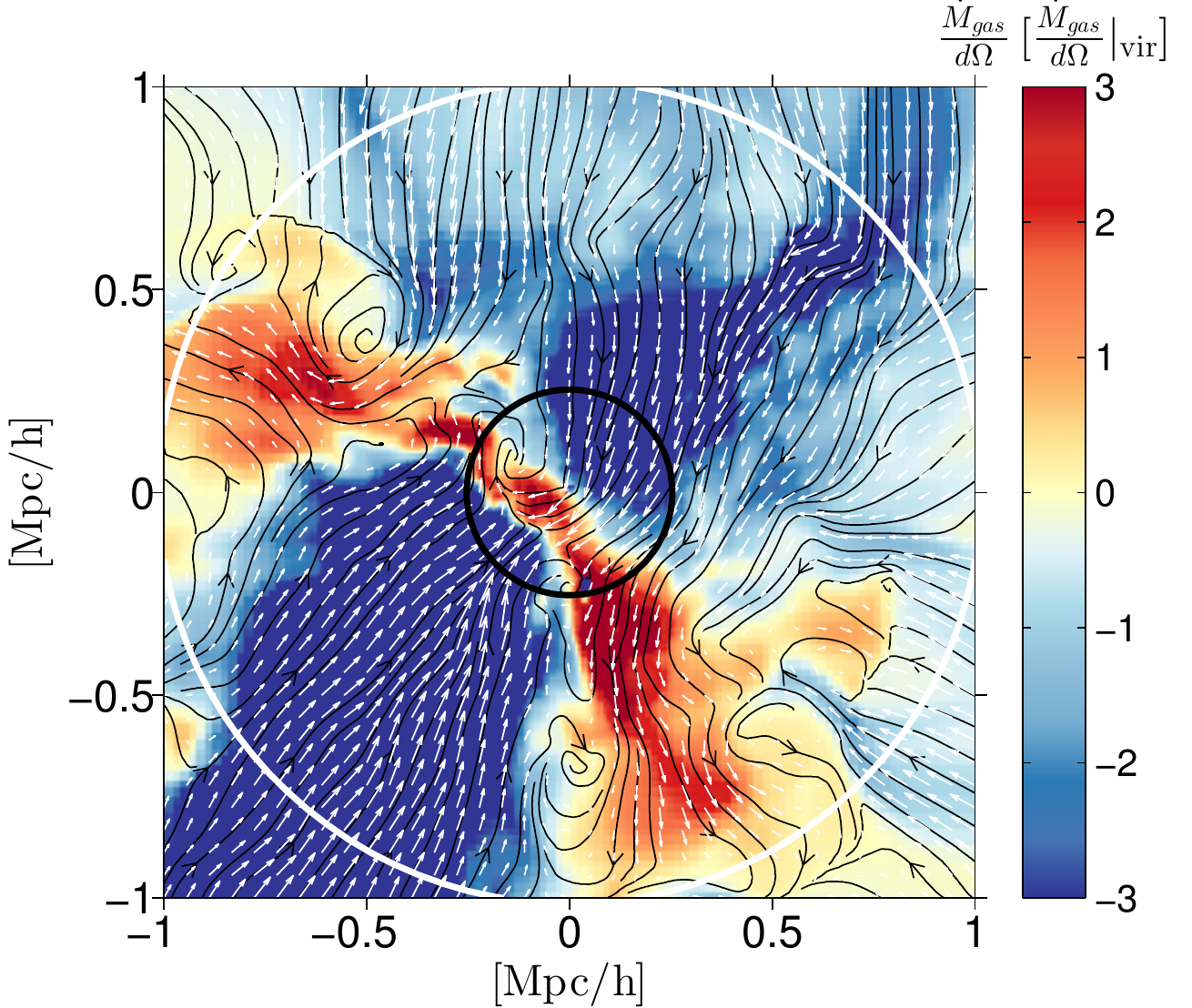}}
  \caption{Maps of mass inflow rate in a thin equatorial slice of the
    cluster CL11 shown at \zeq{0.6} \subrfig{cl11Pen_a06} and at
    \zeq{0} \subrfig{cl11Pen_a1}. At the earlier redshift, the streams
    do not penetrate to the central regions while at \zeq{0} the
    central regions are dominated by deeply penetrating streams.}
  \label{fig:cl11Pen}
\end{figure*}

A deeply penetrating stream, like the stream discussed above, can
funnel energy into the centre of the cluster. As we will see, the
streams typically flow towards the centre till they are stopped and
the ordered flow is dispersed into random motions (e.g.,
\cref{fig:cl6FluxMach,fig:cl14Pen,fig:cl3Pen,fig:cl11Pen}). We can
estimate the energy deposition rate associated with a stream by
following it from outside the system $r_{\mathrm{out}}\sim 2\Rv$ until it
dissipates at an inner radius $r_{\mathrm{in}}$.

 The energy accretion rate is
\begin{equation}\label{eq:streamEnergy}
  \dot{E}_{\mathrm{str}}= \dot{E}_{\mathrm{gain}}
  -\dot{E}_{\mathrm{cool}} = \dot{E}_{\mathrm{grav}}+
  \dot{E}_{\mathrm{kin}}+ \dot{E}_{\mathrm{int}} -
  \dot{E}_{\mathrm{cool}},
\end{equation}
where the first three terms in the second equality account for the
gravitational, kinetic and internal (thermal) energy brought in by the
stream, and the fourth term accounts for the energy lost to radiation
along the flow. The energy deposited in the inner regions of the ICM
will heat the gas or increase the random motions until it is eventually
radiated away.

The gain in gravitational potential energy due to the inflow along the
potential gradient, given the mass accretion rate in a stream
$\dot{M}_{\mathrm{str}}$, can be expressed as
\begin{equation}\label{eq:gravEnergy1}
\dot{E}_{\mathrm{grav}}\approx |\Delta\Phi(r_{\mathrm{in}},r_{\mathrm{out}})|\dot{M}_{\mathrm{str}},
\end{equation}
where the potential gain in a spherically symmetric halo, travelling
from $r_2$ to an inner radius $r_1$ ($r_2>r_1$), can be expressed as
\mbox{$|\Delta\Phi|=\phi \Vv^2$} with 
\begin{equation}\label{eq:potenEnergy}
  \phi(r_1,r_2)= -\frac{V^2(r_2)}{\Vv^2} +\frac{V^2(r_1)}{\Vv^2} +
  \int_{r_1/\Rv}^{r_2/\Rv}
  \frac{3\rho(r^\prime)}{\rho_\mathrm{v}}r^\prime \diff r^\prime,
\end{equation}
where $V^2(r)=GM(r)r^{-1}$ and
\begin{equation}\label{eq:rhovir}
  \rho_{\mathrm{v}}=\frac{3\Mv}{4\mathrm{\pi}\Rv^3}=\delvir\rho_{\mathrm{mean}},
\end{equation}
which follows from \cref{eq:virialDef} \citep{Dekel2008}. The gain in gravitational
energy \cref{eq:gravEnergy1} is therefore
\begin{equation}\label{eq:gravEnergy2}
\dot{E}_{\mathrm{grav}}\approx \phi \Vv^2 \dot{M}_{\mathrm{str}}.
\end{equation}
For an NFW density profile \citep{Navarro1996,Navarro1997} we have
\begin{equation}\label{eq:potNFW}
  \phi(r_1,r_2)=\frac{\ln(1+\cvir x_1)/x_1-\ln(1+\cvir
    x_2)/x_2}{\ln(1+\cvir) -\cvir/(1+\cvir)} ,
\end{equation} 
where $x_i\equiv r_i/ \Rv$ and $\cvir$ is the concentration parameter
of the halo. For cluster-sized systems $\cvir$ is on average
\citep{Bullock2001a}
\begin{equation}\label{eq:cvir}
  \cvir\simeq 4.5 M_{15}^{-0.15}a.
\end{equation}
For a singular isothermal sphere halo density profile we have
\mbox{$\phi(r_1,r_2)=\ln(r_2/r_1)$}.

The potential energy gain of a gas parcel infalling from
$r_{\mathrm{out}}=\Rv,2\Rv\, \&\, 3\Rv$ to a penetration depth of
\mbox{$r_{\mathrm{in}}=0.1\Rv$} is $\phi_{\mathrm{NFW}}\approx 2.3,2.9,3.2$,
respectively. For a penetration depth of \mbox{$r=0.5\Rv$} we have
$\phi_{\mathrm{NFW}}\approx 0.7,1.4,1.7$. For the typical values
quoted above $\phi_{\mathrm{NFW}}$ and $\phi_{\mathrm{Iso}}$ differ by
less than \perc{10}.

\begin{figure*}
  \subfloat[CL6 at \zeq{0}]{\label{fig:flxIsoph_cl6_a1}
    \includegraphics[height=7.5cm,keepaspectratio]{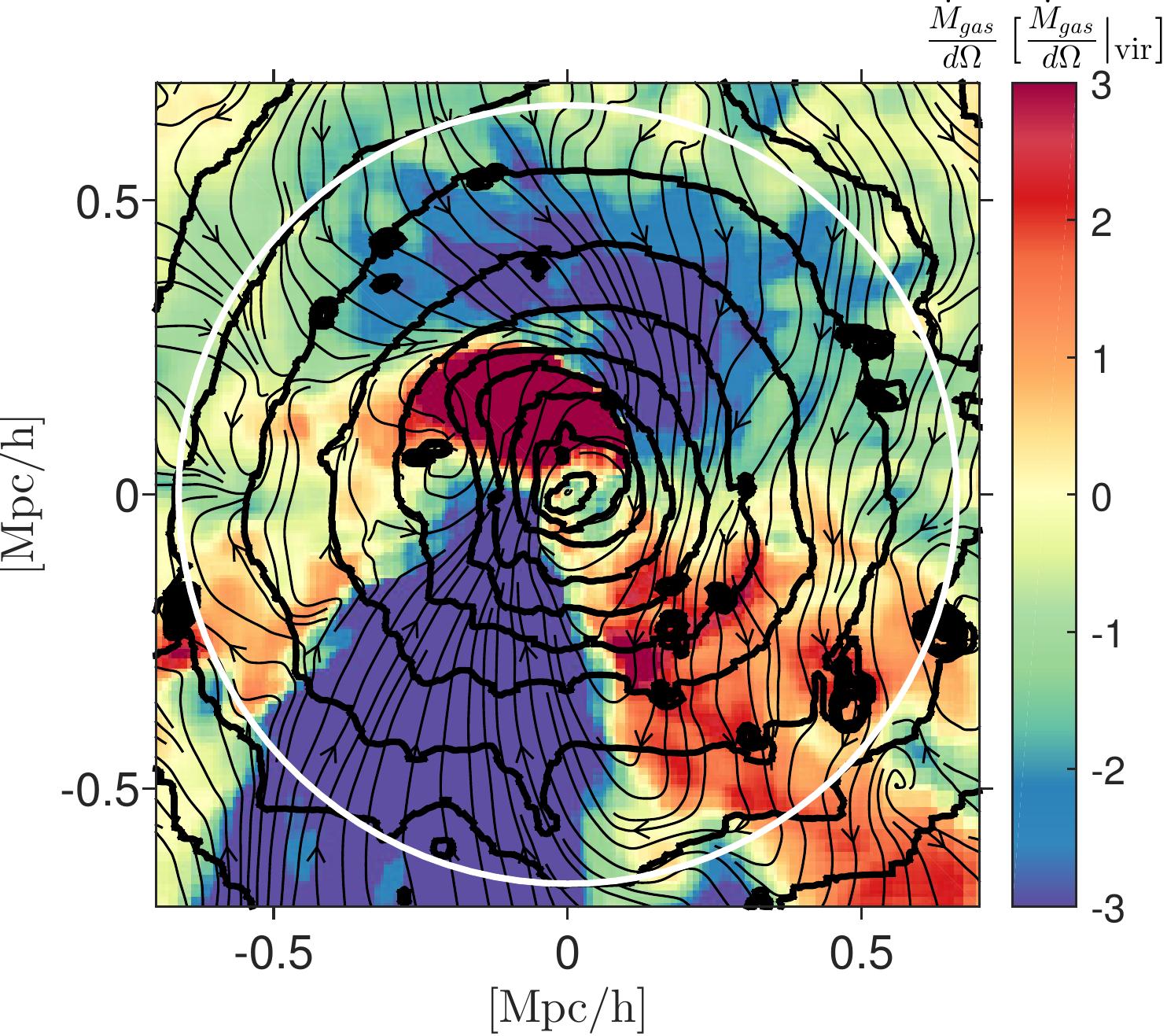}} 
  \subfloat[CL6 at \zeq{0.6}]{\label{fig:flxIsoph_cl6_a06}
    \includegraphics[height=7.5cm,keepaspectratio]{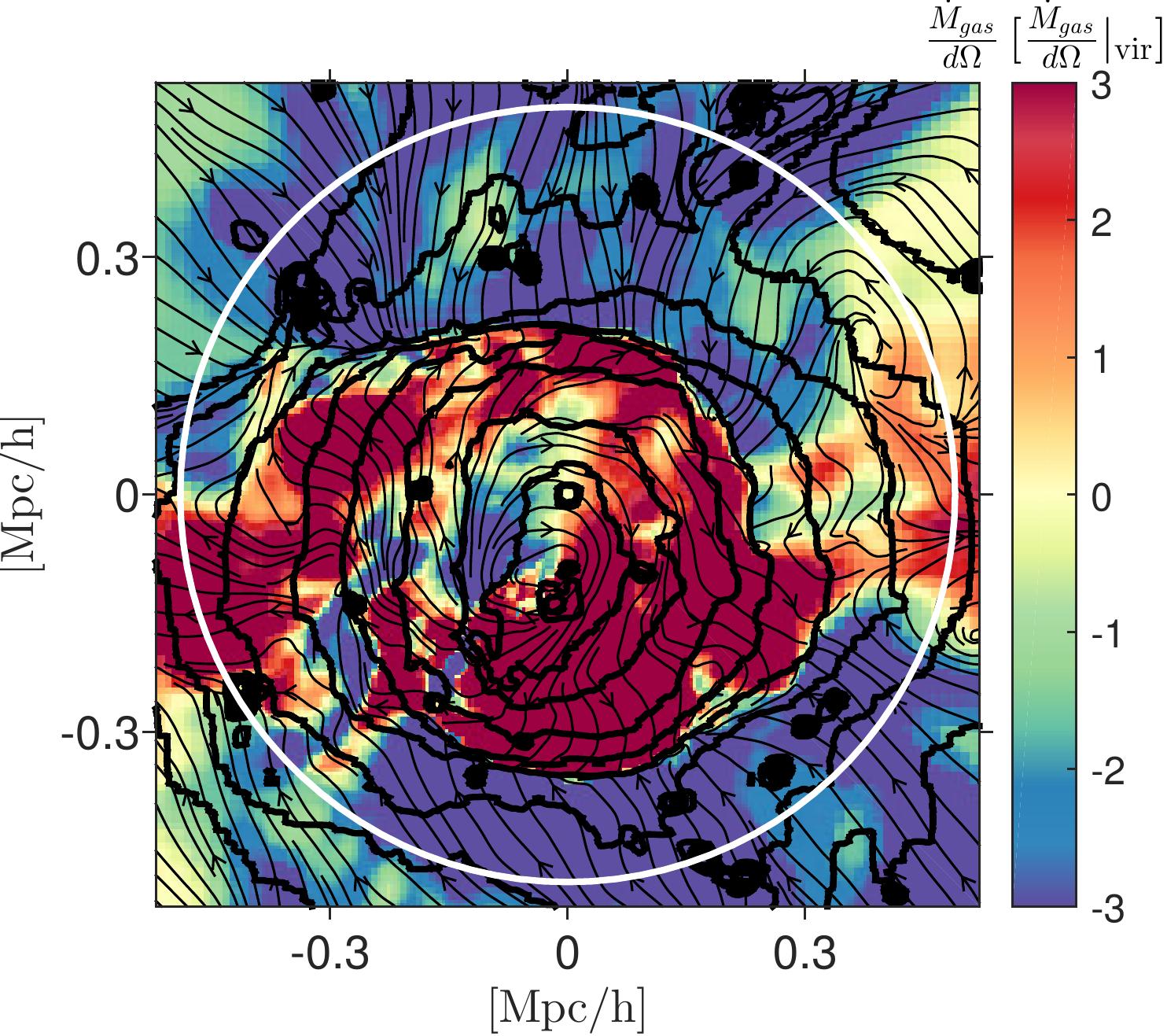}}\\
  \subfloat[CL106 at \zeq{0}]{\label{fig:flxIsoph_cl106_a1}
    \includegraphics[height=7.5cm,keepaspectratio]{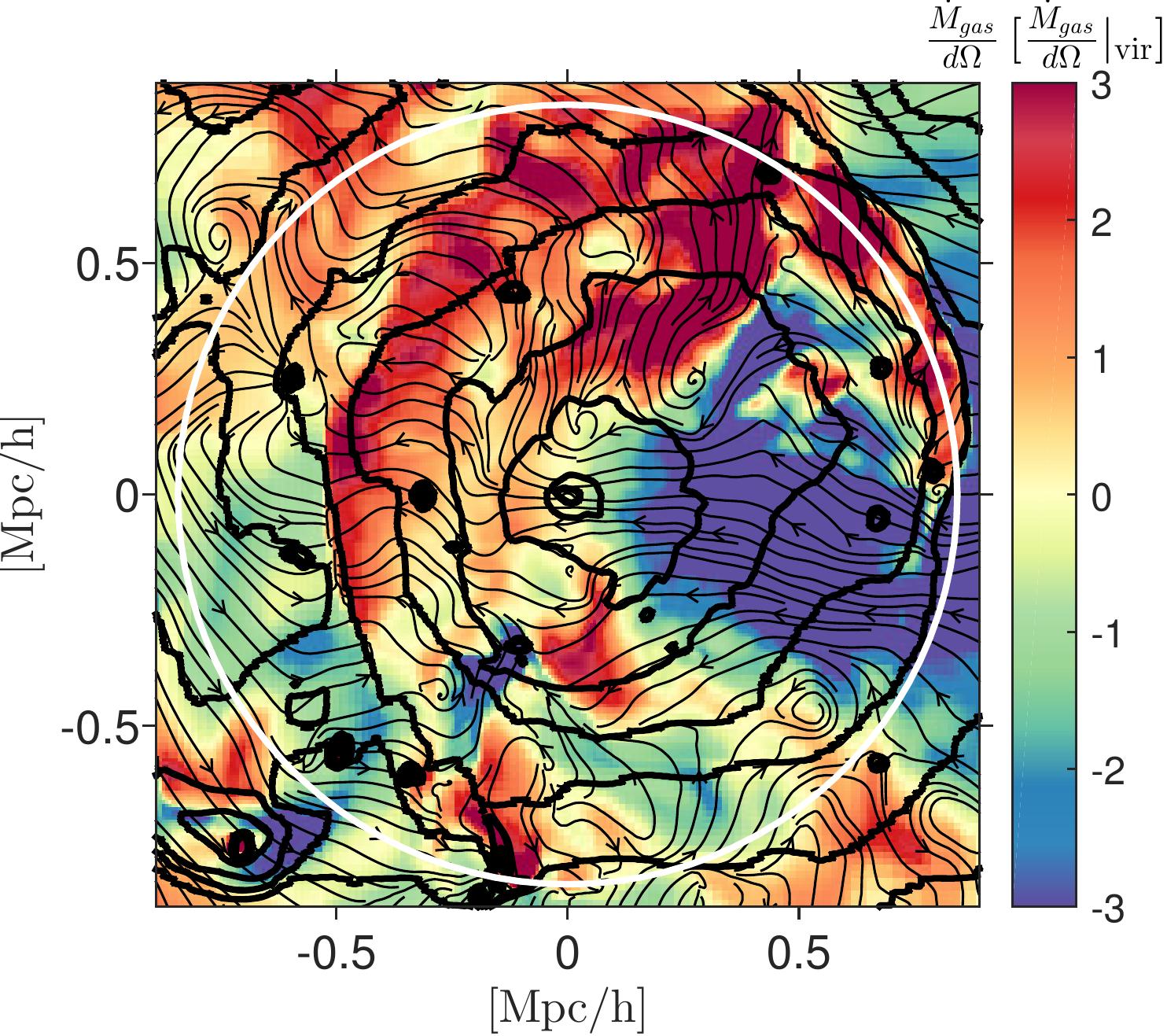}} 
  \subfloat[CL105 at \zeq{0}]{\label{fig:flxIsoph_cl105_a1}
    \includegraphics[height=7.5cm,keepaspectratio]{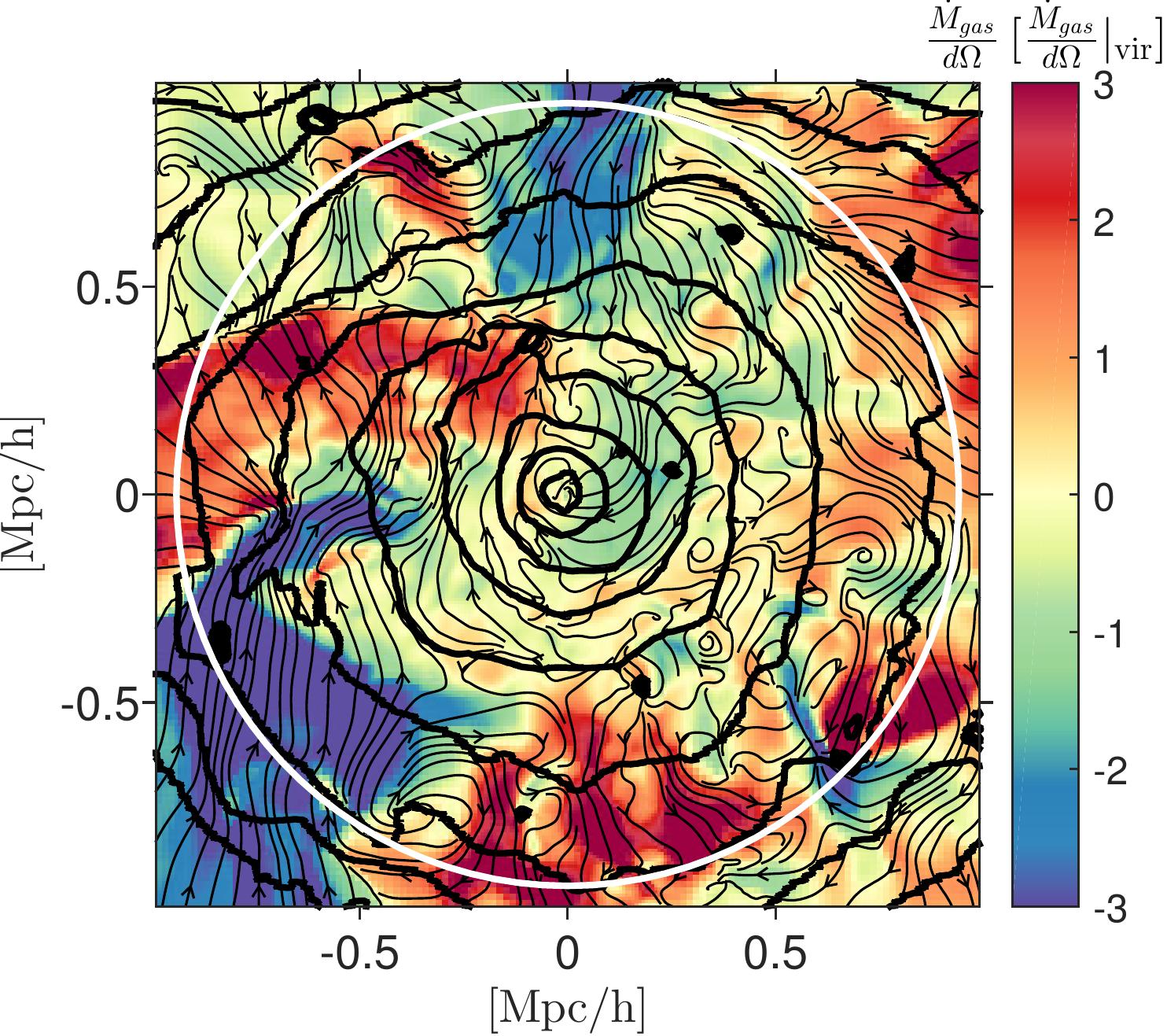}} 
  \caption{The relation between gas streams and relaxedness is
    explored in colour maps of the mass inflow rate overlaid with X-ray
    isophotes (black contours), on the scale of $R_{500,\mathrm{c}}$
    (white circle). The velocity field is shown with black
    streamlines. Four representative NR clusters are shown: CL6 at
    \zeq{0} and \zeq{0.6} \subrfig{flxIsoph_cl6_a1} \&
    \subrfig{flxIsoph_cl6_a06}, CL106 at \zeq{0}
    \subrfig{flxIsoph_cl106_a1} and CL105 at \zeq{0}
    \subrfig{flxIsoph_cl105_a1}.}
  \label{fig:flxIsoph}
\end{figure*}

At the inner radius $r_{\mathrm{in}}$ the stream disperses into random
motions. Shocks, such as the one seen in detail in \cref{fig:cl6CF}
(and several other examples in \cref{fig:flxIsoph}) are a channel in
which the energy in the stream is shared with the ICM. The energy
carried by the stream is subsequently dissipated, mostly into thermal
energy, within this radius. The time-scale of energy dissipation is set
by the largest eddy turnover time both for supersonic and subsonic
random motions \citep{MacLow1998}. The turnover time is set by
$t_{L}\sim\ L /v_L$ where $L\sim r_{\mathrm{in}}$ is the size of the largest
eddy (see \cref{fig:cl14Pen_b1}) and $v_L\sim 500\units{km\,s{-1}}$
is the typical gas velocity. For penetration depths of
$r_{\mathrm{in}}~0.25\Rv$ the eddy turnover time is typically $t_L\lesssim
1\units{Gyr}$ and for $r_{\mathrm{in}}~0.1\Rv$ it is $t_L \lesssim
0.3\units{Gyr}$, both of which are shorter than the cluster crossing
time which is of the order of several giga years, thus it is reasonable to assume
the stream energy is deposited within the central region.

Since we assume the stream is dispersed at the inner radius $r_{\mathrm{in}}$,
its velocity is zero at that point and the kinetic energy gain equal
to the energy brought in by the stream at the outset
\begin{equation}\label{eq:kineticEnergy}
\dot{E}_{\mathrm{kin}} =  \frac{1}{2}  v_{\mathrm{out}}^2  \dot{M}_{\mathrm{str}}=\frac{1}{2}\alpha \Vv^2  \dot{M}_{\mathrm{str}},
\end{equation}
where we have defined $v_{\mathrm{out}}=\sqrt{\alpha}\Vv$ with $\alpha\simeq
1$, since the velocity in the outer regions is of order the virial
velocity $\Vv$.

The internal energy per unit mass of the gas in the stream is
\begin{equation}\label{eq:internalEnergy}
  e_{\mathrm{int}}(T)=\frac{\kboltz T}{\mu m_{\mathrm{p}}}.
\end{equation}
Since the stream has ceased to exist at $r_{\mathrm{in}}$ its final internal
energy is zero, and once again the internal energy contribution is equal
to the internal energy of the stream at the outset
 \begin{equation}\label{eq:internalEnergy1}
\dot{E}_{\mathrm{int}} = \frac{\kboltz T_{\mathrm{out}}}{\mu m_{\mathrm{p}}}\dot{M}_{\mathrm{str}}.
\end{equation}
The temperature in the outer regions is lower than the virial
temperature by a factor of a few at most and we therefore set
$T_{\mathrm{out}}=\beta\Tv$ with $\beta\simeq 0.5$. The gain in internal energy
is therefore
\begin{equation}\label{eq:internalEnergy2}
  \dot{E}_{\mathrm{int}}=\beta \frac{\kboltz \Tv}{\mu
    m_{\mathrm{p}}}\dot{M}_{\mathrm{str}}=\frac{\beta}{2}\Vv^2\dot{M}_{\mathrm{str}}.
\end{equation}

The kinetic and internal energy terms specified above represent the
total amount of kinetic and thermal energy available at the
outset. Obviously, not all of this energy is deposited in the inner
radius since some of it will be lost to the ambient medium. These
energy losses will eventually be radiated away and we therefore
account for them in the energy loss term $\dot{E}_{\mathrm{cool}}$. As
we shall shortly show, the total energy loss of the stream to
radiation is very small compared to the energy gain.

The gain in energy, neglecting the cooling for now, is  
\begin{equation}\label{eq:gainEnergy}
\dot{E}_{\mathrm{gain}}= \dot{E}_{\mathrm{grav}}+
  \dot{E}_{\mathrm{kin}}+ \dot{E}_{\mathrm{int}}=\left(2\phi + \alpha +\beta\right)
  \frac{1}{2} \dot{M}_{\mathrm{str}}\Vv^2,
\end{equation}
where we have used
\cref{eq:gravEnergy2,eq:kineticEnergy,eq:internalEnergy2}. The virial
velocity for clusters can be expressed as
\begin{equation}\label{eq:vvirDef}
  \Vv\simeq
  \begin{cases}
    1290 M_{15}^{1/3}\units{km\,s^{-1}} & z=0 \\ 
    1520 M_{15}^{1/3}\units{km\,s^{-1}} & z=0.6
\end{cases}.
\end{equation}

As we saw in \rfsec{massFlux}, the total gas inflow rate at $\Rv$ is
roughly equal to the analytic estimates of \citet{Birnboim2007},
\cref{eq:analytflux}
\begin{equation}\label{eq:gasInflow}
  \dot{M}_{\mathrm{g}}\simeq 1.5\times10^{13}
  \left(\frac{f_{\mathrm{b}}}{0.14}\right) M_{15}^{1.15} a^{-2.25}
  \units{\msun\,Gyr^{-1}},
\end{equation}
where $f_{\mathrm{b}}$ is the universal baryon fraction. A stream which carries a
fraction $f_{\mathrm{s}}$ of the baryonic inflow will contribute
$\dot{M}_{\mathrm{str}}=f_{\mathrm{s}}\dot{M}_{\mathrm{g}}$.

Thus the power injected by a stream at a dissipation radius of $r_{\mathrm{in}}$ is
\begin{multline}\label{eq:streamEnergy3}
  \dot{E}_{\mathrm{gain}}\simeq8.2\times10^{45} f_{\mathrm{s}} \left(2\phi(r_{\mathrm{in}},r_{\mathrm{out}}) + \alpha +\beta\right)\times \\
  \left(\frac{f_{\mathrm{b}}}{0.14}\right)M_{15}^{1.82}a^{-2.25}
  \units{erg\,s^{-1}}.
\end{multline}
For a $10^{15}\msun$ cluster at \zeq{0}, assuming typical values of
$f_{\mathrm{b}}$, $\alpha=1$ and $\beta=0.5$, the energy injection by a deeply
penetrating stream infalling from $2\Rv$ to $0.1\Rv$ is
\begin{equation}\label{eq:estream15}
  \dot{E}_{\mathrm{gain}}\simeq
  \begin{cases}
    6\times10^{46}f_{\mathrm{s}} \units{erg\, s^{-1}} & r_{\mathrm{in}}=0.1\Rv\\
    3.5\times10^{46}f_{\mathrm{s}} \units{erg\, s^{-1}} & r_{\mathrm{in}}=0.5\Rv\\
  \end{cases}.
\end{equation}
For a $10^{14}\msun$ cluster with the same choice of parameters, the
energy injection rate is
\begin{equation}\label{eq:estream14}
  \dot{E}_{\mathrm{str}}\simeq
  \begin{cases}
    9\times10^{44}f_{\mathrm{s}} \units{erg\, s^{-1}} & r_{\mathrm{in}}=0.1\Rv\\
    5.2\times10^{44}f_{\mathrm{s}} \units{erg\, s^{-1}} & r_{\mathrm{in}}=0.5\Rv\\
  \end{cases}.
\end{equation}

The energy carried in by the streams is sufficiently large to have a
noticeable effect on the inner regions of the cluster. However, it is
not immediately clear how much of the energy is indeed deposited at
the site where the stream dissipates. The potential energy of the gas
stream is converted to other forms of energy by several channels. Some
of the energy is radiated away, some is channeled to heating the gas
and some is converted to kinetic energy (e.g.\@ turbulence) as the gas
velocity increases.

Examining the temperature in the stream we find that it increases by a
factor of $\sim 10$ between $\Rv$ and the inner regions, due to the
gain in gravitational energy (\cref{fig:cl6Map_temp_b8_a1}). In our
analysis, we find that the velocity and velocity dispersion do not
increase by more than a factor of a few. In any case, the changes in
internal or kinetic energy of the stream do not change the amount of
energy in the stream itself (only its form) and thus can still be
shared with the ambient gas at the site of stream dissipation. The
only energy which is truly removed from the stream is that which is
radiated away.

We will now assess the energy losses in the stream and show that they
are negligible compared with the energy gain. We may convince
ourselves that only a small fraction of the stream energy is lost to
radiative cooling along its way by finding an upper limit on the
energy lost to cooling, namely the \emph{total} amount of radiative
cooling \emph{outside} the inner regions of the cluster $r\gtrsim 0.25
\Rv$.

The radiative cooling rate per unit mass is
\begin{equation}\label{eq:coolRate}
  q=\left(\frac{\chi}{\mu m_{\mathrm{p}}}\right)^2\rho_{\mathrm{g}}\Lambda(T,Z),
\end{equation}
where $\chi\simeq 0.52$, $\mu m_{\mathrm{p}}$ is the average particle mass,
$\rho_{\mathrm{g}}$ is the gas density and $\Lambda(T,Z)$ is the atomic radiative
cooling function \citep{Sutherland1993}, which depends on the gas
temperature $T$ and metallicity $Z$. For a spherically symmetric gas
distribution, the total cooling rate in the region between $r_{\mathrm{in}}$
and $r_{\mathrm{out}}$ is
\begin{equation}\label{eq:coolRateTot}
  \dot{E}_{\mathrm{cool}}=4\mathrm{\pi}\int_{r_{\mathrm{in}}}^{r_{\mathrm{out}}}q(r^\prime)\rho_{\mathrm{g}}(r^\prime)^2
      {r^\prime}^2\diff r^\prime.
\end{equation}
Due to the strong density dependence, most of the energy will be
radiated from the centre of the cluster.  We assume the temperature
and metallicity are constant throughout the halo, and scale the gas
density profile
\begin{equation}\label{eq:gasProfScale}
  \hat{\rho}_{\mathrm{g}}(r)\equiv \frac{\rho_{\mathrm{g}}(r)}{f_{\mathrm{b}}\rho_\mathrm{v}},
\end{equation}
where $f_{\mathrm{b}}$ is the gas fraction in the cluster and $\rho_\mathrm{v}$
is defined in \cref{eq:rhovir}.  The cooling rate is now
\begin{multline}\label{eq:coolRateTot2}
  \dot{E}_{\mathrm{cool}}\simeq 3  \times10^{44}\Lambda_{-23}M_{15}f_{0.14}^2 \\
\mathcal{I}\left(r_1/\Rv,r_2/\Rv \right) \units{erg\,s^{-1}},
\end{multline}
where
$\Lambda_{-23}\equiv\Lambda/10^{-23}\units{erg\,cm^3\,s^{-1}}$,
$f_{0.14}\equiv f_{\mathrm{b}}/0.14$ and
\begin{equation}\label{eq:coolInt}
\mathcal{I}(x_1,x_2)=\int_{x_1}^{x_2} \hat{\rho}_{\mathrm{g}}(y)^2 y^2\diff y.
\end{equation}
For typical density profiles the integral converges quickly for
$x_2\to\infty$; thus, an upper limit can be defined as
$\mathcal{I}_{\mathrm{max}}(x)=\mathcal{I}(x,\infty)$. For cooling beyond the
central regions ($r\gtrsim0.25\Rv$) both the NFW and isothermal
profile result in $\mathcal{I}_{\mathrm{max}}(0.25)\simeq 0.45$.

For relevant temperatures of $T>10^6\units{K}$ one can use the
approximation of \citet{Dekel2006} for the cooling function
\begin{equation}\label{eq:lambdaApprox}
  \Lambda_{-23}\simeq 6 Z_{0.3}^{0.7}T_6^{-1}+0.2 T_6^{1/2},
\end{equation} 
where $T_6\equiv 10^6\units{K}$ and $Z_{0.3}\equiv
(Z/0.3)\units{Z_\odot}$. For typical values of the mean metallicity,
$Z_{0.3}\lesssim 1$ \citep{Tozzi2003}, and virial temperatures in
clusters ($10<T_6<100$), we find $\Lambda_{-23}\lesssim 2$.  We can
therefore set $\mathcal{I}_{\mathrm{max}}=0.5$ and $\Lambda_{-23}=2$ to get a
strong upper limit on the total cooling rate from the
\emph{entire} cluster volume in the shell $0.25\Rv < r <2\Rv$,
\begin{equation}\label{eq:coolRateTot3}
  \dot{E}_{\mathrm{cool}}\simeq 3 \times10^{44} M_{15}f_{0.14}^2
  \units{erg\,s^{-1}} .
\end{equation} 
This implies that the cooling rate in the streams account for only a
small fraction of this energy output. Comparing this to the energy
brought in by the streams (e.g.\@ \equnp{estream15}), we see that
radiative cooling in the streams outside $0.25\Rv$ can remove only a
very small fraction of the total energy. We may therefore safely
assume that $\dot{E}_{\mathrm{str}}\approx \dot{E}_{\mathrm{gain}}$,
and that the expression \cref{eq:streamEnergy3} and estimated values
of \cref{eq:estream15,eq:estream14} represent the energy deposition of
the streams.

We find that the energy associated with an inflowing stream, even one
which only accounts for $\lesssim 10$ per cent of the total gas
accretion, is in the same range as the typical X-ray luminosity of the
cluster which is in the range of $10^{43}\textrm{--}10^{46}
\units{erg\, s^{-1}}$ \citep{Peterson2006}.

Most of the energy carried in by the streams is thus transferred into
random motions. The random motions dissipate into heat, but may also
drive the sustained growth of magnetic fields \citep{Iapichino2012}
and the subsequent acceleration of Cosmic Rays (CR). The combined
contribution of turbulent motions, magnetic fields and CR's to
pressure support in the central regions of clusters is of the order of $10$
per cent of the thermal pressure support \citep{Churazov2008}, so it
is safe to assume that the majority of the energy brought in by the
stream is dissipated into heat and eventually radiated away.

The dissipation of random motions is due to the action of
viscosity. While it is true that the viscosity may be affected by the
presence of magnetic fields and CR's, this will only change the
viscous scale where the dissipation takes place but not the rate of
dissipation which is set by the largest eddy turnover time.

As shown above, the largest eddy turnover time is of the order of
$0.1\textrm{--}1\units{Gyr}$ in the central regions of the clusters
and thus the energy dissipation should occur over several turnover
times, on a time-scale of a few giga years. This value is very
similar to those found in observation by \citet{Hallman2011} and
\citet{Rossetti2011}.

Though we have shown that the streams may carry a substantial amount
of energy into the central regions of the cluster in NR clusters, the
current simulations, even those classified as NR still suffer from an
overcooling problem in the very centre of simulated clusters. Though this
may be the result of the numerical limitations of the simulations, it
may indicate that the gas streams on their own cannot offset the
cooling and that additional physical processes, e.g.\@ AGN feedback
and thermal conduction, are still necessary to overcome the
cooling-flow problem perhaps even in NR clusters \citep{Rasia2015}.

\begin{figure*} 
\subfloat[Mass inflow rate]{\label{fig:cl6Hammer_flux}
    \includegraphics[width=9cm,keepaspectratio]{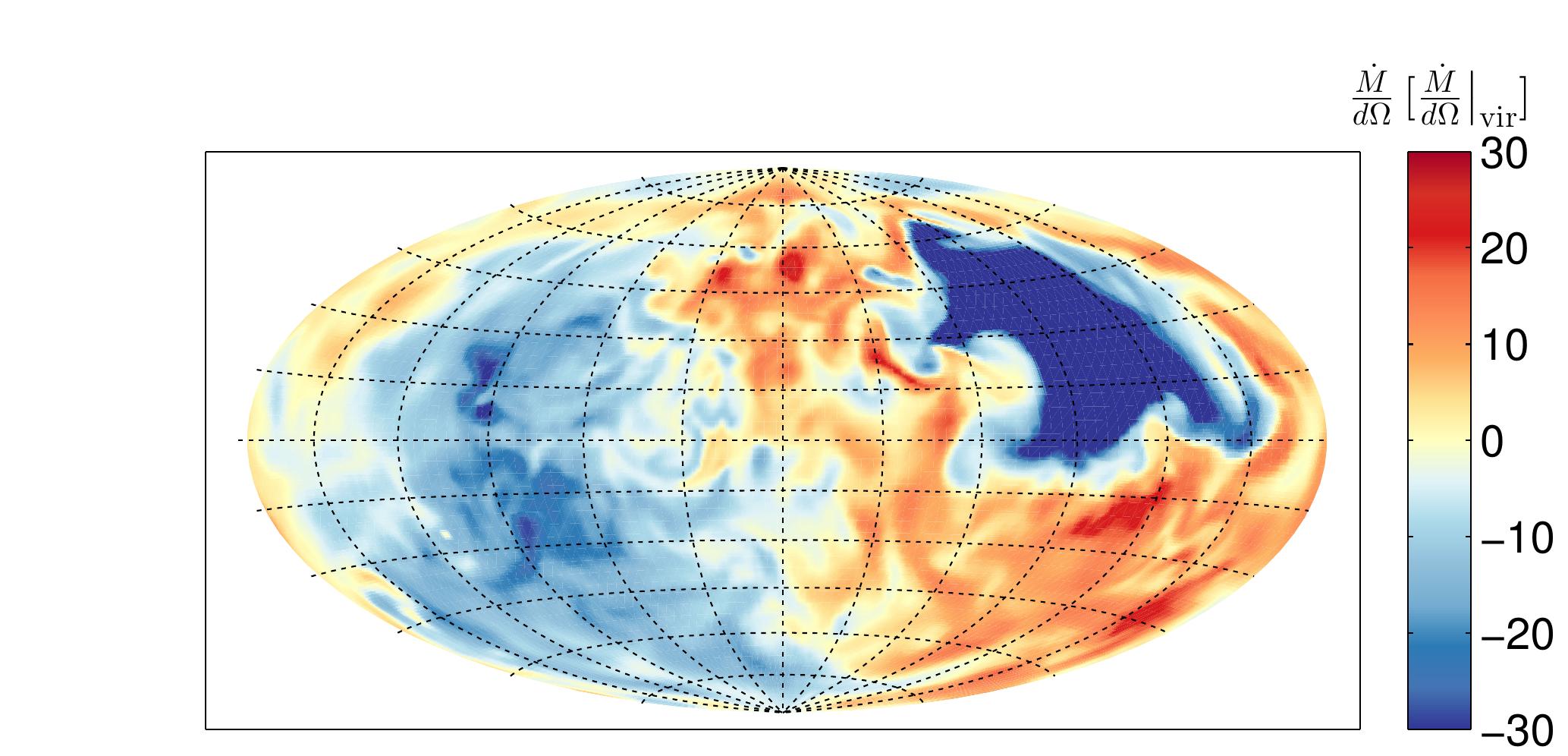}} 
  \subfloat[Entropy]{\label{fig:cl6Hammer_ent}
    \includegraphics[width=9cm,keepaspectratio]{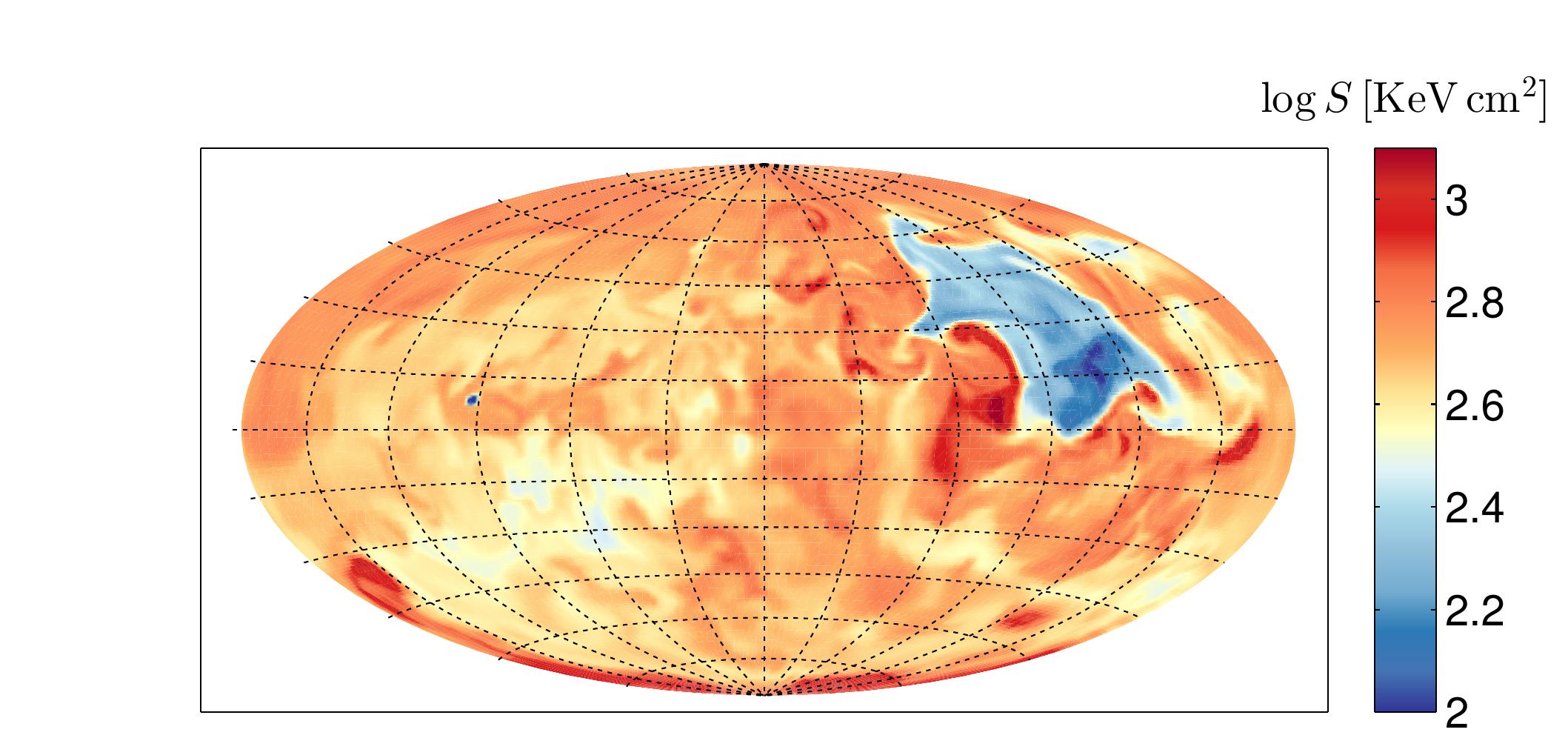}}\\
  \subfloat[Temperature]{\label{fig:cl6Hammer_temp}
    \includegraphics[width=9cm,keepaspectratio]{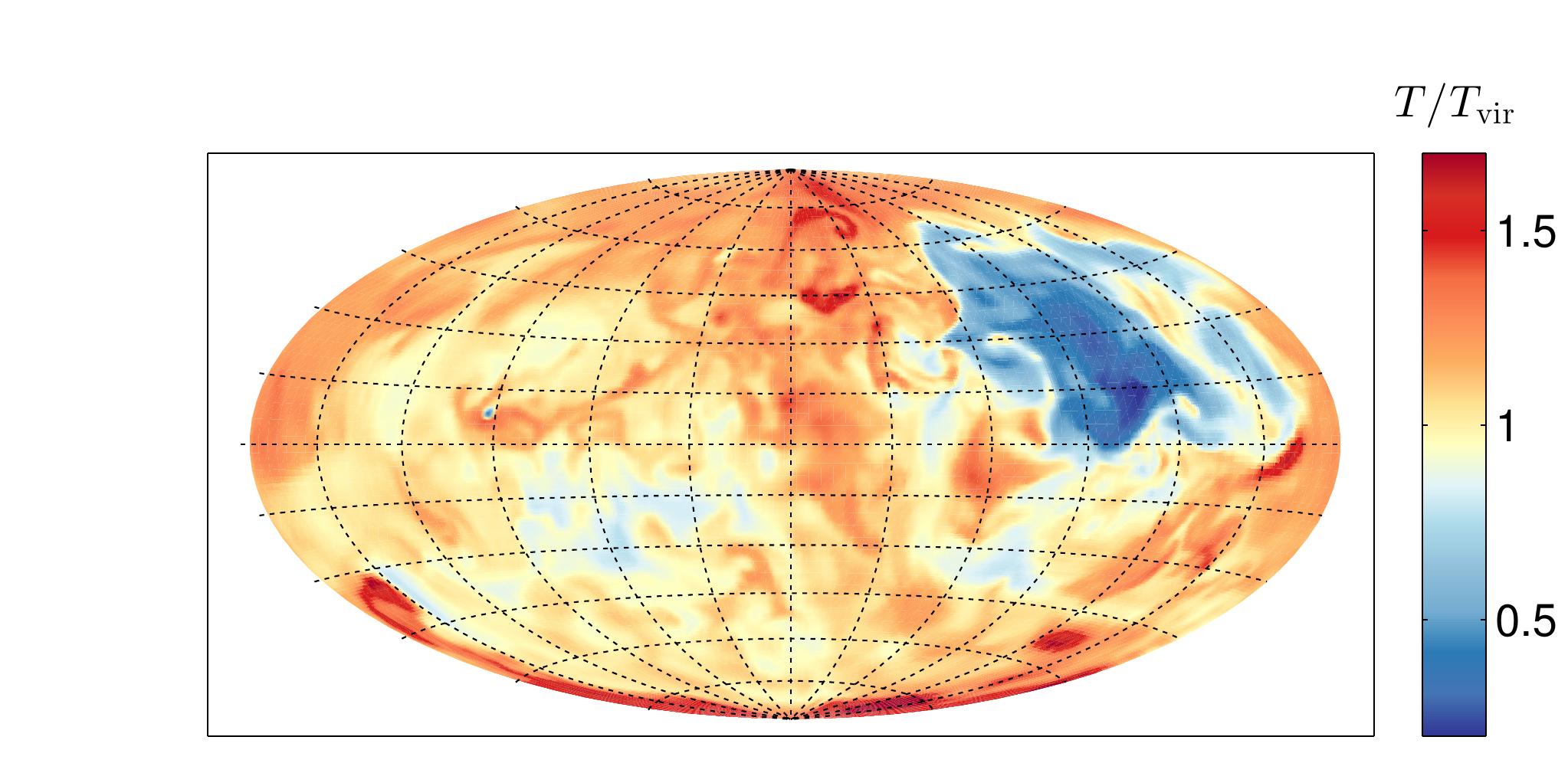}} 
  \subfloat[Gas Density]{\label{fig:cl6Hammer_dens}
    \includegraphics[width=9cm,keepaspectratio]{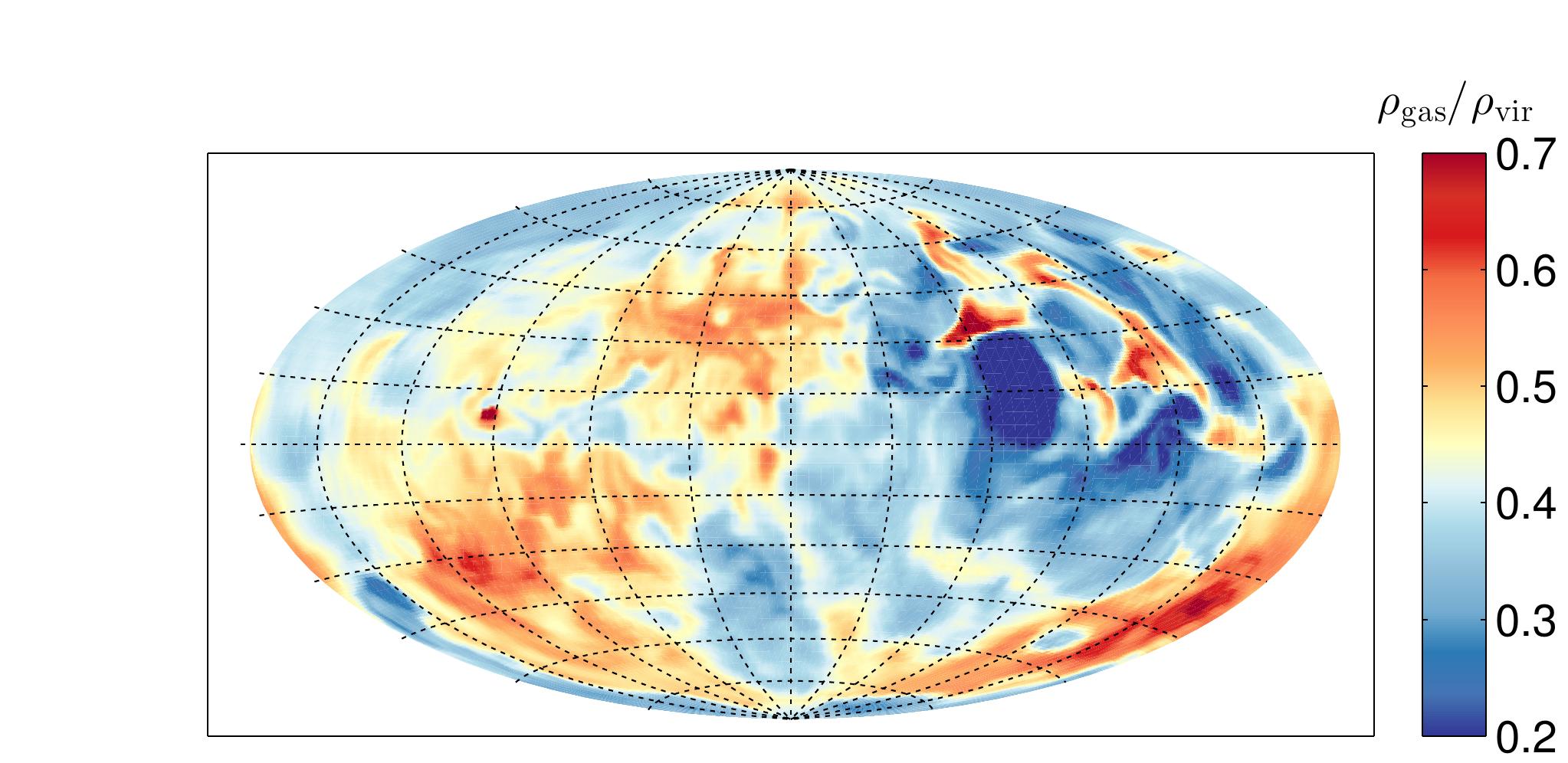}} 
  \caption{Hammer projections of a thin spherical shell of radius $0.3
    \Rv$ of the cluster CL6 shown in \cref{fig:cl6Maps}. Plotted are
    the mass inflow rate \subrfig{cl6Hammer_flux}, entropy
    \subrfig{cl6Hammer_ent}, temperature \subrfig{cl6Hammer_temp} and
    density \subrfig{cl6Hammer_dens}.  A prominent inflowing streams
    is clearly seen in the top-right quadrant, characterized by strong
    inflow and lower temperature and entropy than its surroundings.}
  \label{fig:cl6Hammers}
\end{figure*}

\begin{figure*}
  \subfloat[Mass inflow rate]{\label{fig:cl14Hammer_flux}
    \includegraphics[width=9cm,keepaspectratio]{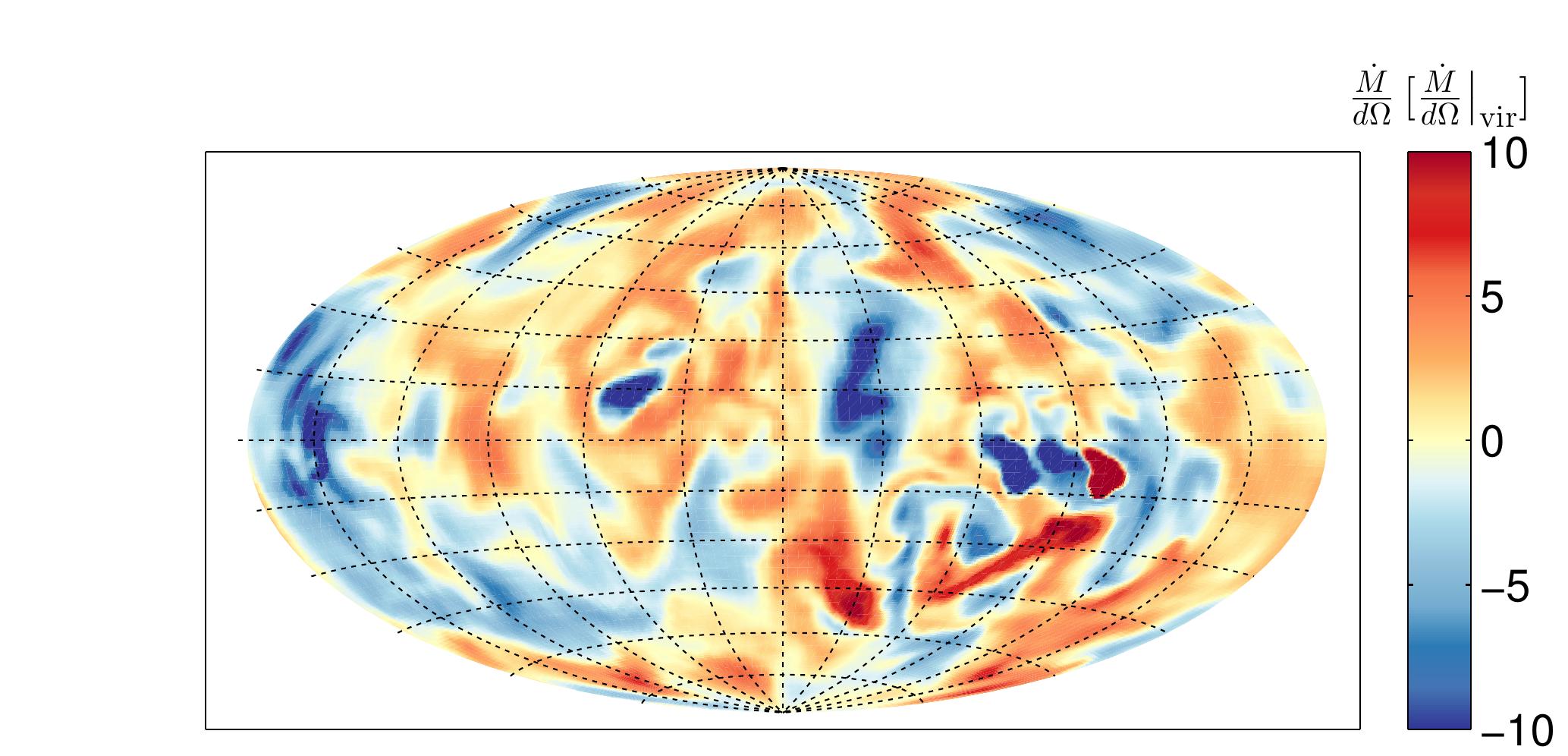}}
  \subfloat[Entropy]{\label{fig:cl14Hammer_ent}
    \includegraphics[width=9cm,keepaspectratio]{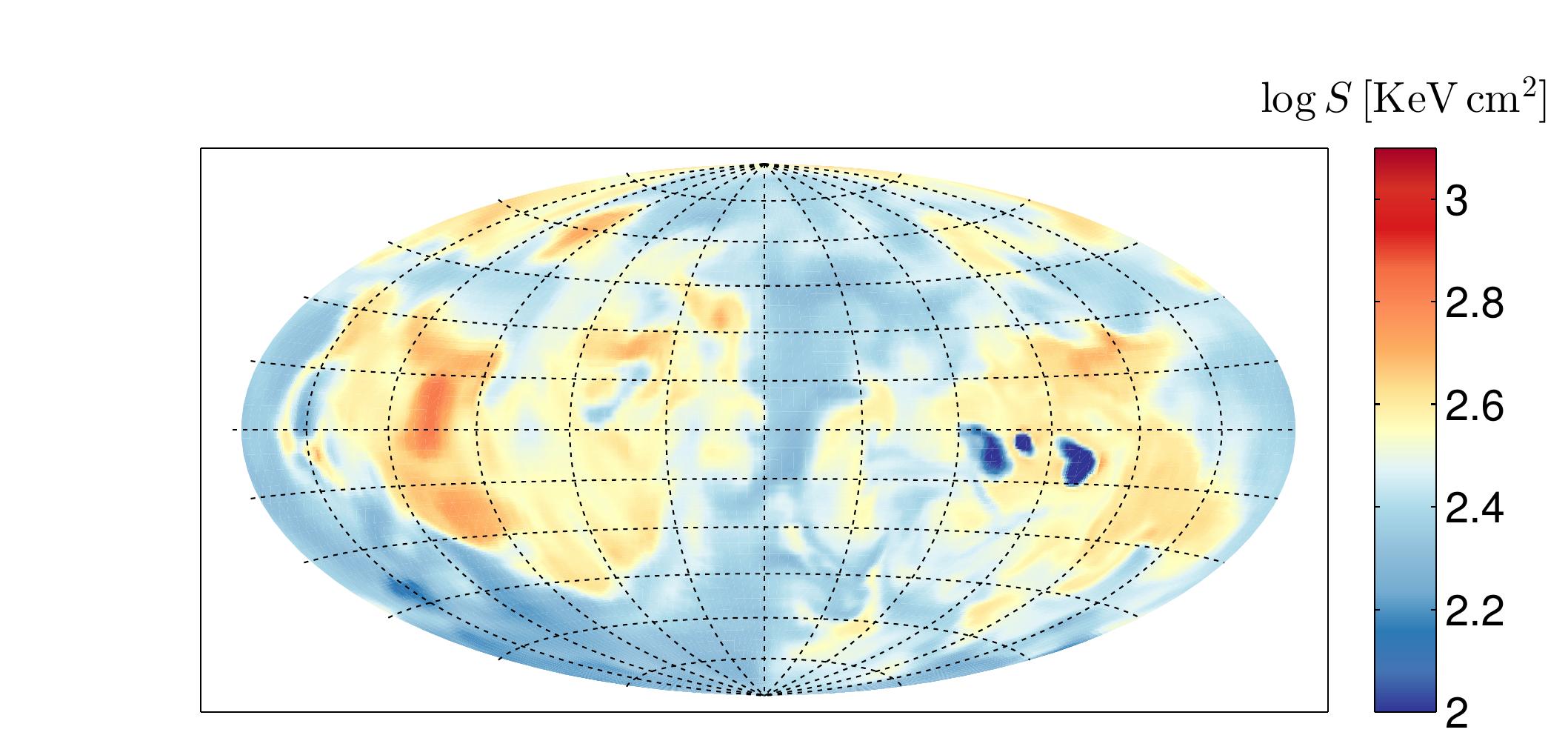}}\\
  \subfloat[Temperature]{\label{fig:cl14Hammer_temp}
    \includegraphics[width=9cm,keepaspectratio]{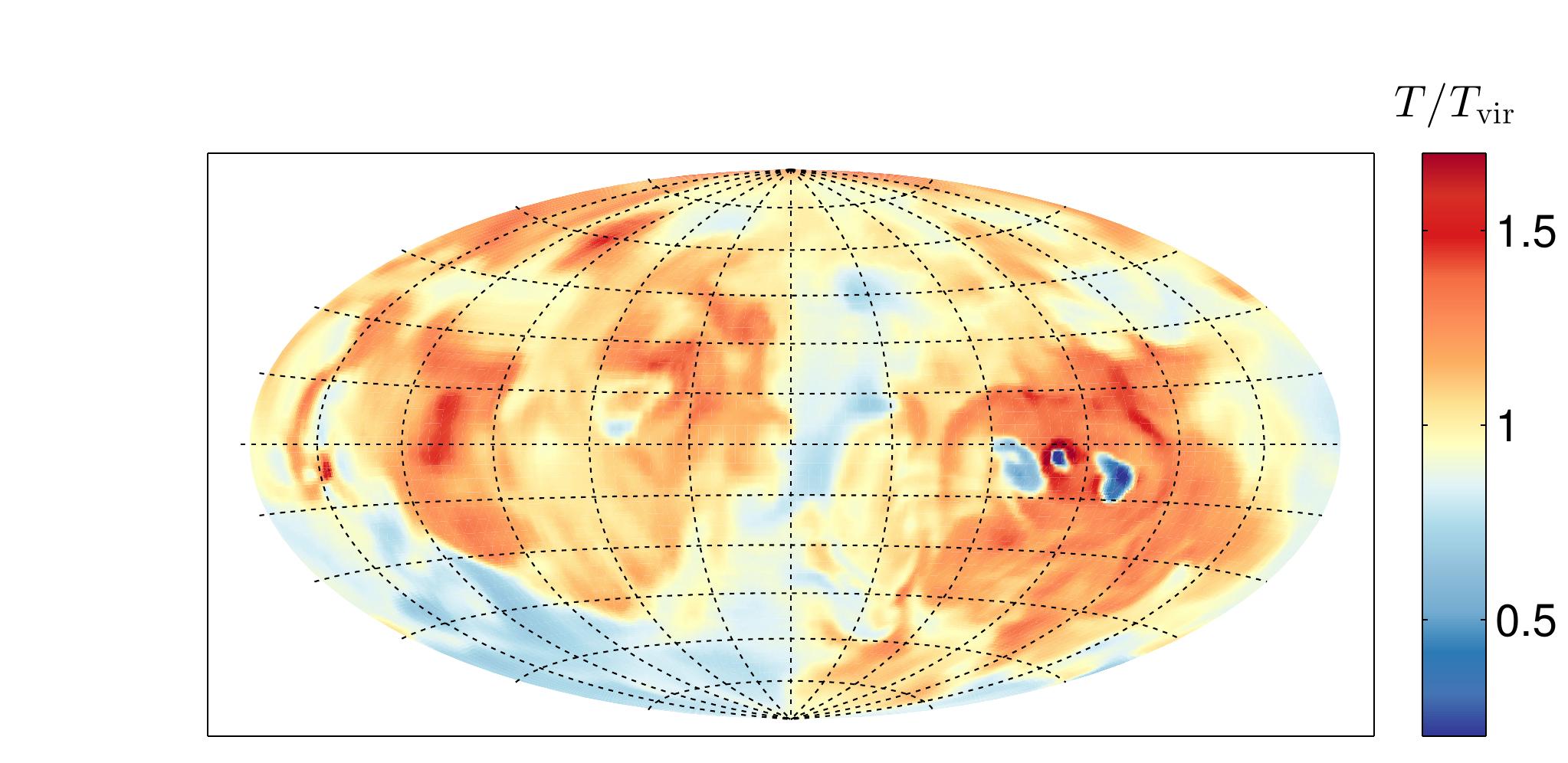}} 
  \subfloat[Gas Density]{\label{fig:cl14Hammer_dens}
    \includegraphics[width=9cm,keepaspectratio]{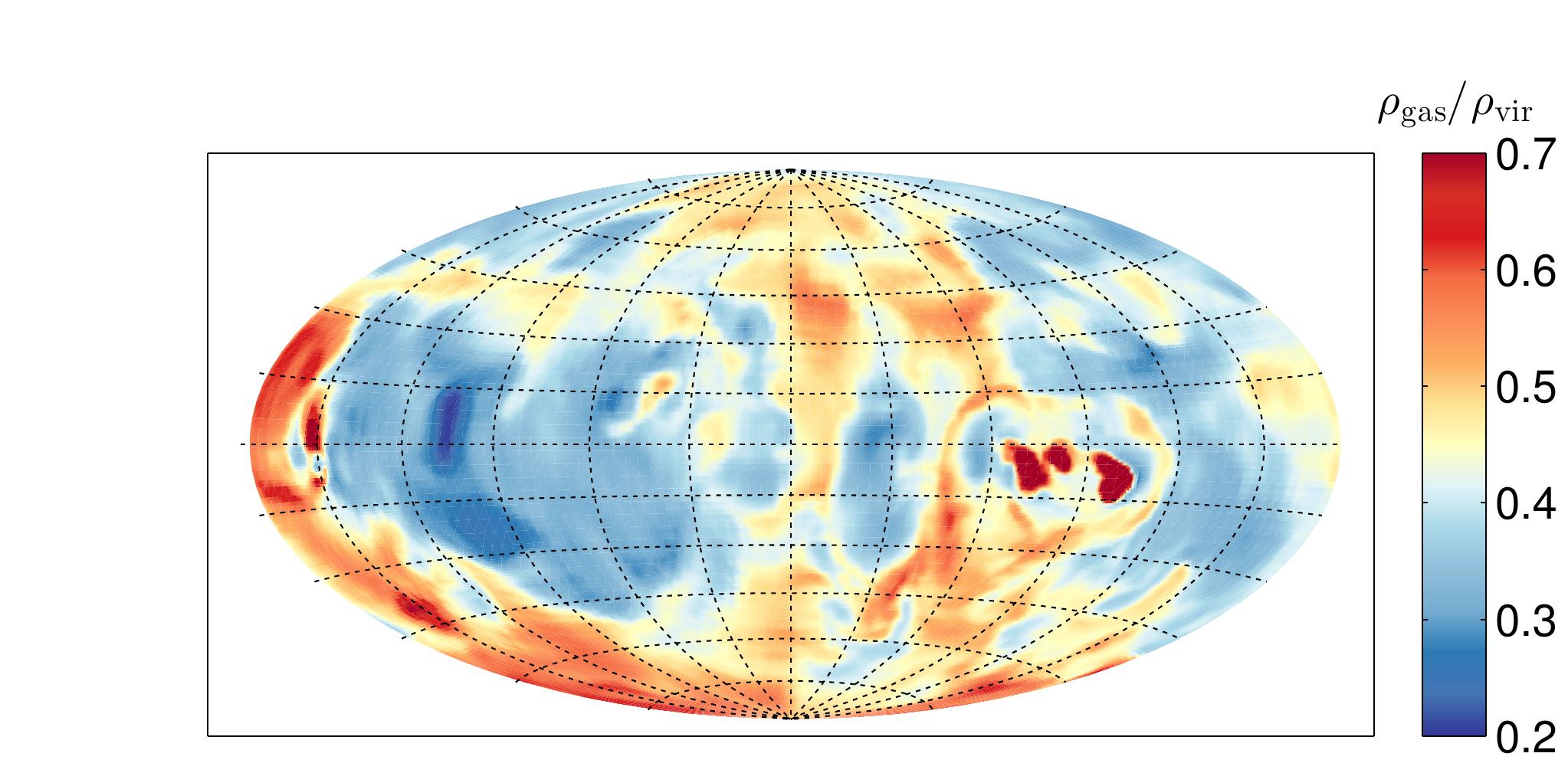}} 
  \caption{Hammer projections of a thin spherical shell of radius $0.3
    \Rv$ of the cluster CL14 shown in \cref{fig:cl14Pen}. Plotted are the
    mass inflow rate \subrfig{cl14Hammer_flux}, entropy
    \subrfig{cl14Hammer_ent}, temperature \subrfig{cl14Hammer_temp}
    and density \subrfig{cl14Hammer_dens}. Mass inflow and outflow
    occur on small areas. Several small high-density cool clumps can
    be seen on the `equator' in the right half of the plots.}
  \label{fig:cl14Hammers}
\end{figure*}

\subsection{Stream Penetration}\label{sec:streamPenen}
As we have shown, the gas streams in the NR cluster CL6 penetrate into
the very centre of the cluster
(\cref{fig:cl6Maps,fig:cl6FluxMach}). The penetrating streams carry
with them a considerable amount of energy (\rfsec{streamEnergy}) which
can be deposited into the central regions. However, examination of
other clusters in the simulation suite shows that not all clusters
possess deeply penetrating gas streams.

In \cref{fig:cl14Pen_b2,fig:cl14Pen_b1}, we examine the cluster CL14 at
\zeq{0}, on the scale of $\Rv$ and focusing on the inner regions. Here
too, several gas streams can be identified entering the system, most
notably from the top-left and bottom-right corners in
\cref{fig:cl14Pen_b2}, but all of them stop abruptly at approximately
$0.5\Rv$. Examination of the central region shows that no coherent
flows are found there, and there is no significant source of mass
inflow. Accordingly, the cluster is classified as R.

Comparing the clusters with their progenitors at earlier times shows
that the degree of stream penetration into the central areas can
change over time. In \cref{fig:cl3Pen_a06}, we examine the cluster CL3 at
\zeq{0.6}, roughly $5.7\units{Gyr}$ ago, where the cluster crossing
time is $2\units{Gyr}$ at \zeq{0} and $1.2\units{Gyr}$ at
\zeq{0.6}. Two prominent streams enter the system, one from the
top-left corner and the other from the bottom-right, and flow unabated
into the very centre of the system. Examining the same cluster at
\zeq{0} in \cref{fig:cl3Pen_a1} shows that the stream from the
bottom-right corner now stops at about $0.5\Rv$, whereas the mass flow
in the stream from the top-left has dwindled considerably and barely
enters the virial radius. The change in stream penetration in the
cluster is accompanied by a change in relaxedness, changing from NR at
\zeq{0.6} to R at \zeq{0} (see
\cref{tab:clusterProperties,tab:clusterProperties06}).

The penetration of streams can also grow over time as can be seen in
\cref{fig:cl11Pen_a06,fig:cl11Pen_a1} where we show the cluster CL11
at \zeq{0.6} and \zeq{0}, respectively. At \zeq{0.6}, the maximal
penetration is found in a stream entering from the top, which
penetrates to just outside the $0.25\Rv$ mark. At \zeq{0}, the streams
can be seen to penetrate deep into the very centre of the cluster.

In order to study the individual stream penetration properties of each
cluster in detail, we employed a two-pronged approach. A simple
stream-identification algorithm to identify the maximal penetration
point of the streams in each cluster was constructed and in addition,
visual identification of the streams was carried out in two
complementary methods. The results of quantitative and visual methods
were compared and combined yielding a robust estimate of the maximal
penetration of the streams in all the clusters.

The stream identification algorithm was based on the high mass inflow
rates and entropy properties of the streams and on the fact that the
streams are supersonic with respect to their surroundings (see e.g.,
\cref{fig:cl6Maps,fig:cl6FluxMach,fig:cl6Hammers}). To ensure that
large-scale streaming features are identified, rather than isolated
clumps and satellites, the simulation data was spatially smoothed over
a scale of $100\units{kpc}$ comoving ($62.5\units{kpc}$ proper at
\zeq{0.6}). 

In the smoothed data, streams were defined by including
simulation cells which simultaneously met the following four conditions: 
\begin{itemize}
\item The mass inflow rate in the cell is at least three times the
  virial accretion rate of \cref{eq:analytflux}. This of course
  entails that the radial velocity in the cell is negative.
\item The entropy in the cell is lower than the average entropy of all
  the cells found at the same radial distance.
\item A ratio of 0.9 or higher between the gas velocity in the cell and the
  typical speed of sound at the cell radius as defined in
  \cref{eq:csonic}.
\item No stellar particles are found in the cell, thus ensuring that
  satellite galaxies are removed.
\end{itemize}

Once all the cells comprising the streams in a given cluster were
identified, the maximal penetration depth can be easily found.

To complement the stream identification algorithm, we carried out a
visual determination of the maximal penetration depth of the
streams. This was done by examining the clusters as a series of
spherical shells by use of Hammer projections\footnotemark \emph{and}
by standard Cartesian projections.\footnotetext{Hammer
  projections are equal-area projections of a spherical surface to a
  2D map.}

We follow the streams in each cluster as they flow into the cluster from 
outside the virial radius until they can no longer be
distinguished from their surroundings. This was done by inspecting
Hammer projections of spherical shells at different radii of the
temperature, density, mass inflow rate and entropy for all the
clusters (examples are shown for CL6 and CL14 at $0.3\Rv$ in
\cref{fig:cl6Hammers,fig:cl14Hammers}).

Gas streams are identified in a Hammer projection of a given radial
shell as areas of which meet the same set of criteria used in the
stream identification algorithm, namely areas of enhanced inflow, at
least three times the virial accretion rate of \cref{eq:analytflux}, in
which the entropy is lower than the mean entropy of the shell. In
addition, the stream signature must be continuous over several
spherical shells ($\sim 300\units{kpc}$), so as not to include
isolated clumps.

Only streams which originate beyond the virial radius at $\gtrsim
4\units{Mpc}$ and persist to within $\Rv$ are considered. The
Hammer projections for a shell at $0.3\Rv$ for a cluster in which the
streams penetrate to the centre (CL6) and for a cluster in which the
streams are stopped at $r\approx 0.5\Rv$ (CL14) are shown in
\cref{fig:cl6Hammers,fig:cl14Hammers}, respectively.

The prominent stream found in CL6 (\cref{fig:cl6Maps,fig:cl6FluxMach})
is clearly seen in the upper-right quadrant of the Hammer plots,
characterized by its high inflow, cooler temperature and elevated
density. No such stream is seen in CL14, but several small clumps
can be seen on the `equator' of the projections on the right half,
characterized by high density and low temperatures.

It is instructive to compare the mass inflow rate maps between the two
clusters, \cref{fig:cl6Hammer_flux,fig:cl14Hammer_flux}, taking note
of the different colour scale of the two plots which highlight the
large difference in the mass accretion rate of the two clusters. In
CL6, the large stream covers a large area in the plot, whereas in CL14
inflowing and outflowing areas subtend a much smaller solid angle and
are also of markedly lower magnitude.

To complement the visual stream penetration analysis, we examined mass inflow
maps of all the clusters in all three Cartesian projections (e.g.\@
\cref{fig:cl6FluxMach,fig:cl14Pen,fig:cl3Pen,fig:cl11Pen}). Once again, gas
streams were identified visually based on the same criteria as before. 
By visually examining the clusters both from the point of view of an outside
observer (Cartesian projection maps) and from the point of view of an
observer at the centre of the cluster (Hammer projections), we achieve
a much more comprehensive estimate of the penetration depth of the
streams.

The error in ascertaining the maximal penetration depth of the streams
is based on the smoothing scale used in the stream identification
algorithm, and the cell size of the simulation ($\sim 5\units{kpc}$
comoving), which also sets the radial spacing of the spherical
shells. In the Cartesian projections, the error is of order several
times the cell size. In the Hammer projections, the shell thickness is
twice the cell size and once again the error is of order several times
the shell thickness. As a result, we estimate that the error in
determining the maximal penetration to be no more than $\pm 50
\units{kpc}$ comoving.

It stands to reason that in a cluster with deeply penetrating streams,
the high mass and energy inflow rate into the central regions will
have an important effect on the properties of the ICM. The energy
associated with a strong inflow of matter can heat its surroundings,
drive random motions and thus create a system which is dynamically NR.

\begin{figure*}
  \includegraphics[width=12cm,keepaspectratio]{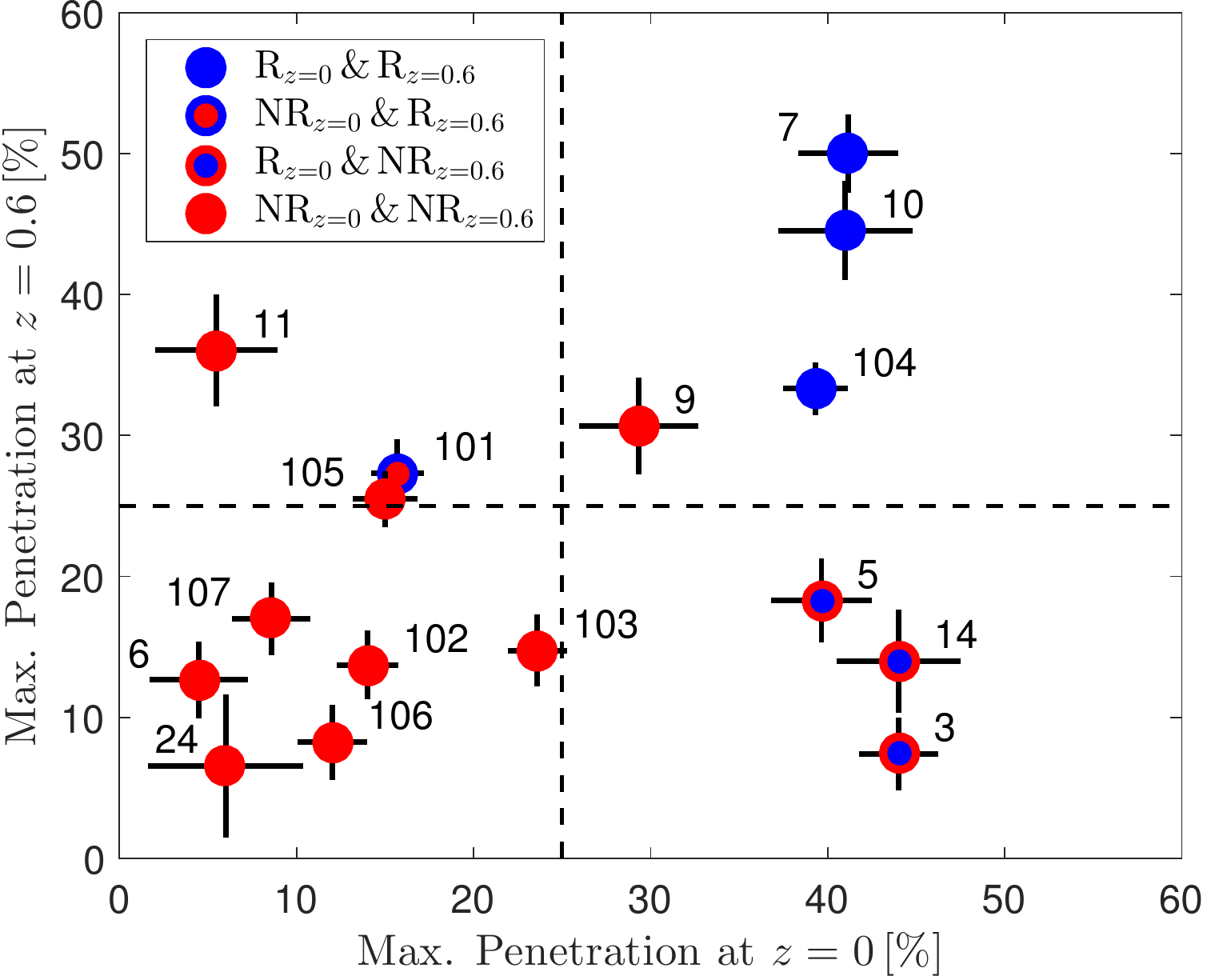}
  \caption{The link between the relaxedness and the maximal
    penetration of streams in the simulated cluster suite. The axes
    show the maximal penetration of streams into the clusters at
    \zeq{0} and \zeq{0.6} ($X$ and $Y$ axes respectively), in per cent of
    the virial radius $\Rv$. The colours mark the dynamical state of
    the clusters. Clusters which are R at both epochs are marked in
    blue, while clusters which are NR at those times are marked in
    red. Clusters which were NR at \zeq{0.6} but are R at \zeq{0} are
    shown as blue circles with red outlines, while clusters which were
    R at \zeq{0.6} but are NR at \zeq{0} are shown as red circles with
    blue outlines. The uncertainties in determining the penetration
    depth are shown by the black horizontal and vertical lines. The
    black dashed lines mark the $0.25\Rv$ region. The link between
    stream penetration and relaxedness is evident -- deeply
    penetrating streams are found only in NR clusters and all but two of
    the NR clusters host deeply penetrating streams. Of particular
    interest are the three clusters found in the bottom-right
    quadrant,which host deeply penetrating streams at \zeq{0.6} but
    not at \zeq{0}, and whose dynamical classification changes from NR
    to R accordingly. CL101 found in the top-left quadrant, also
    changed from R at \zeq{0} to NR at \zeq{0} in conjunction with an
    increase in stream penetration between the two epochs. }
  \label{fig:penen_relax}
\end{figure*}

Our findings concerning the correlation between stream penetration and
relaxedness are summarized in \cref{fig:penen_relax}, which shows the
maximal stream penetration into the clusters in the simulations suite
at \zeq{0} and at \zeq{0.6}. Each cluster is represented by a coloured
circle on the plane, with a label identifying the cluster, such that
its $X$ coordinate shows the maximal penetration into the cluster at
\zeq{0} and its $Y$ coordinate shows the penetration at \zeq{0.6}. The
uncertainty in the determination of the penetration depth is marked by
the black lines. The units of penetration are in per cent of $\Rv$ for
each cluster, thus the streams in cluster CL3 penetrate to $\sim 5$ per cent
of $\Rv$ at \zeq{0.6} (\cref{fig:cl3Pen_a06}) but at \zeq{0} penetrate
only to $\sim 36$ per cent of $\Rv$ (\cref{fig:cl3Pen_a1}).

The colours of the circles indicate the dynamical state of the cluster
(R versus NR) at the two epochs: clusters which are R at both epochs
are represented by blue circles and clusters which are NR at both
epochs are marked as red circles. Clusters which are NR at \zeq{0.6}
but R at \zeq{0} are marked as blue circles with red outlines and
clusters which are R at \zeq{0.6} and NR at \zeq{0} are shown as red
circles with blue outlines.

It is immediately apparent that there exists a correlation between
penetration and relaxedness. Of the $10$ NR clusters at \zeq{0} all
but one are characterized by streams which penetrate to within
$0.25\Rv$ and four of them host streams which penetrate to within $0.1
\Rv$. Meanwhile, none of the six R clusters at \zeq{0} host a stream
which penetrates deeper than $0.4\Rv$. By and large this relationship
holds at \zeq{0.6}. Of the four R clusters at that epoch none have streams 
which penetrate into the inner regions. Most of the NR clusters, $10$ out of
$12$, host streams which penetrate to at least $0.25\Rv$ and half of
those penetrate to within $0.15\Rv$.

Especially interesting are the clusters CL3, CL5 and CL14 (found in
the bottom-right quadrant of \cref{fig:penen_relax}), which are
characterized by deeply penetrating streams at \zeq{0.6} which no
longer penetrate into the central regions at \zeq{0}. For these
clusters, the transition between penetrating to non-penetrating
streams is accompanied with a change of classification from NR to R,
strengthening the relation between stream penetration and
unrelaxedness. In the cluster CL101 the opposite trend is observed,
with the cluster transitioning from R at \zeq{0.6} to NR at \zeq{0}
with a corresponding increase in stream penetration between the two
epochs.

At \zeq{0.6}, two exceptions are found, CL9 and CL11, which are
classified as NR but do not exhibit strongly penetrating streams. A
possible explanation for these clusters, besides the possibility that
they are mis-classified as NR, is that a process other than smooth
stream penetration is responsible for the NR status of the cluster,
such as an ongoing merger event. Examination of these systems reveals
that at \zeq{0.6} both CL11 and CL9 host multiple X-ray peaks in their
centres which can account for the NR classification.  In the next
section, we address the role of gas streams versus satellite haloes in
determining the dynamical state of the clusters.

The cluster CL9 is classified as NR at \zeq{0}, but hosts streams
which only penetrated to within $0.29\Rv$. Since this is the only case
in which the penetration is somewhats at odds with the relaxedness we
consider it to be an isolated anomaly.

\section{Gas Streams versus Mergers}\label{sec:mergers}
In the previous section, we found a strong correlation between the
penetration of gas streams into the cluster and the dynamical state of
the system, which suggests that the strong inflows of mass into the
central regions of clusters drive random gas motions and lead to
dynamically unrelaxed systems. A question naturally arises regarding
the importance of gas streams in determining the classification into R
and NR, compared to other factors such as satellites and mergers,
which are known causes for NR clusters \citep{Burns2008}. Since the
gas streams originate in the large-scale filaments of the cosmic web,
and indicate the preferred directions along which mass accretion takes
place, it is reasonable to assume that satellites will be found
preferentially along the streams, especially on their way in for the
first time. This begs the question: could both the relaxedness and
stream penetration properties be correlated with the presence of
satellites or other substructures in the cluster without being linked
directly?

As described in \rfsec{sims}, the hallmarks of an NR cluster are the
existence of secondary luminosity peaks in the X-ray emission,
significant shifts of the centres of the isophotes and filamentary
X-ray structures. R clusters, on the other hand, are characterized by
isophotes with only small deviations from elliptical symmetry. The
mere existence of some substructure within the cluster is of course
not necessarily a sign of unrelaxedness, otherwise all clusters would
be classified as NR, seeing as all clusters contain substructures to a
certain degree. Identification of these features in mock X-ray images,
which may suffer from projection effects, and is carried out
subjectively, as is also the case in observations.

The existence of a secondary luminosity peak in the central regions is
a clear mark of a merger event (except in the case of projection
effects). X-ray filamentary structure can occur either as a result of
stripped gas forming an extended emission area behind (or ahead) of
the sub-clump/galaxy or as a result of several separate substructures
forming an apparent filament due to projection effects. In our
analysis, we found that the large-scale gas streams themselves do not
register as filamentary X-ray structures. Isophotal centroid shifts
and deviations from elliptical symmetry may be the result of
substructure, but as we shall see, can also occur due to the inflowing
streams.

\begin{figure}
  \includegraphics[width=\columnwidth,keepaspectratio,keepaspectratio]{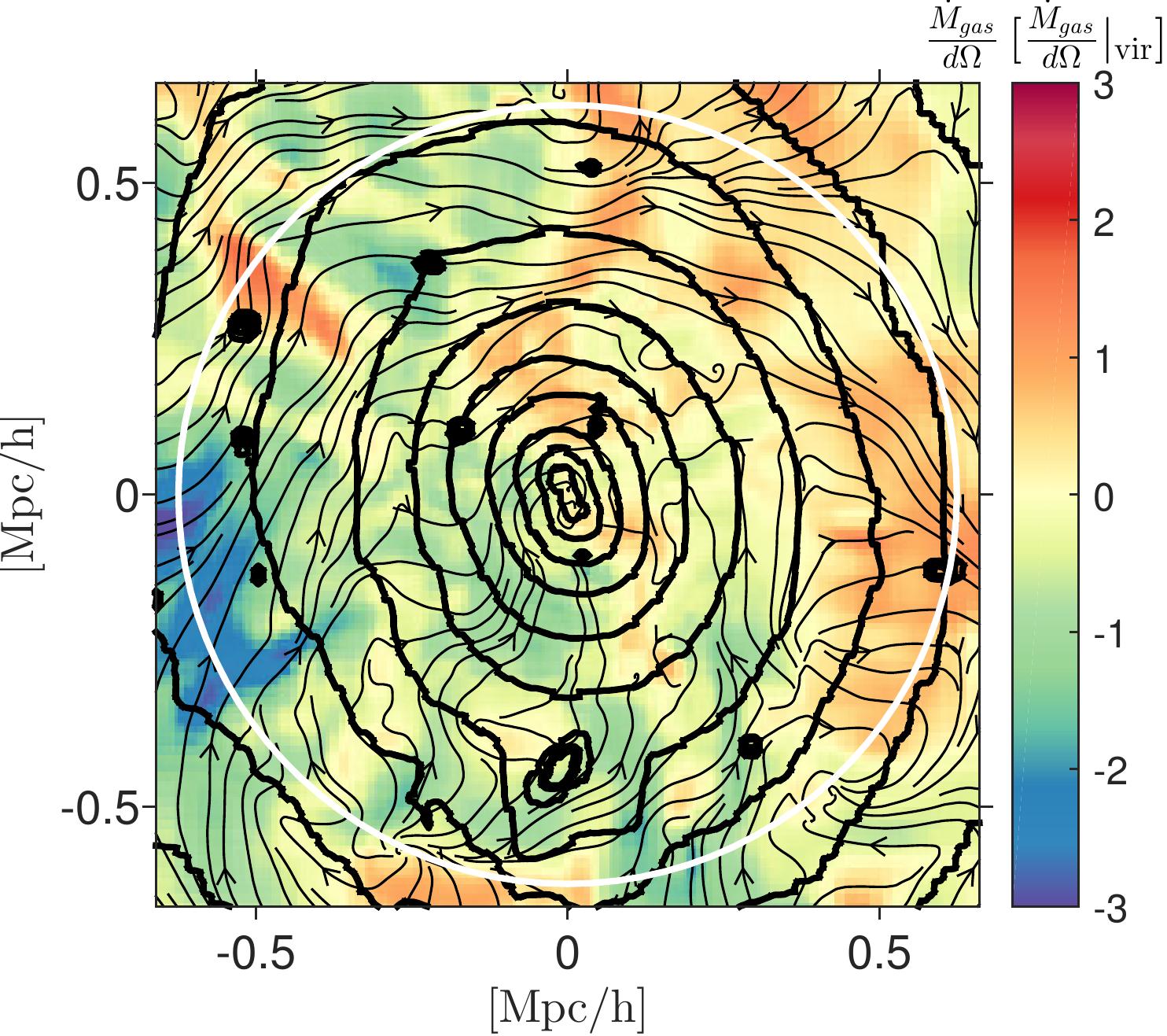}
  \caption{An example of a R cluster, CL7 at \zeq{0}. Colour maps of
    the mass inflow rate overlaid with X-ray isophotes (black
    contours), on the scale of $R_{500,\mathrm{c}}$ (white circle). The
    velocity field is shown with black streamlines. Substructure
    within $R_{500,\mathrm{c}}$ does not preclude a dynamically relaxed state.}
  \label{fig:flxIsoph_cl7}
\end{figure}

We explore the relation between the smooth streams and the
observational signatures of relaxedness by examining maps of the mass
inflow rate in the clusters overlaid with X-ray isophotes.  We examine
the cluster on the scale of $R_{500,\mathrm{c}}$, that is the radius at which
the mean density is $500$ the \emph{critical} density of the
universe. $R_{500,\mathrm{c}}\simeq 0.45 \Rv$ based on
\cref{eq:virialDependence}.  Several representative examples of NR
clusters are shown in \cref{fig:flxIsoph}, and an example of an R
cluster is shown in \cref{fig:flxIsoph_cl7}. 

It is important to stress that the images shown here were \emph{not}
used to determine the dynamical classification. As noted earlier, the
dynamical classification to R and NR was based on mock \emph{Chandra}
observation of the clusters which took into account instrument
sensitivity, exposure times and other observational considerations
\citet{Nagai2007a,Nagai2007}, which we do not address here.

The X-ray emission per unit volume from a gas element is calculated by 
\begin{equation}\label{eq:xrayLum}
  L_{\mathrm{X}}=\left(\frac{\chi}{\mu m_{\mathrm{p}}}\right)^2\rho_{\mathrm{g}}^2\Lambda(T,Z).
\end{equation}
The X-ray isophotes are determined by summing the X-ray luminosity
along the line of sight for a given projection.

In \cref{fig:flxIsoph_cl6_a1}, the inflowing smooth stream from the
bottom, overshoots the centre and shocks against gas coming in from
the top (\cref{fig:cl6CF}). The isophotes in the centre are distorted
in the region of the inflowing stream and the plume shaped outflow,
which results in an NR classification, despite the relative dearth of
significant substructure in the cluster, seen as small blobs in the
isophotes. The distortions in the isophotes also do not correlate
with the positions of the substructures seen in the cluster.  A
similar effect can be seen in \cref{fig:flxIsoph_cl105_a1} where the
isophotes are distorted due to the incoming smooth stream on the
bottom left leading to a significant deviation from elliptical
symmetry. Here too the small amount and size of the substructure, as
well as their positions, cannot account for the distortions in the
isophotes which attest to the unrelaxedness of the cluster. In
\cref{fig:flxIsoph_cl106_a1}, the isophotes are also skewed due to the
incoming smooth stream with small signatures of substructures
found in the cluster.

In \cref{fig:flxIsoph_cl6_a06}, two luminosity peaks can be found in
the central region, one at a position of $(0,0)$ and the other at a
position of roughly $(-0.1,0)$, indicating a merger in the system. The
cluster is classified as NR, due to the combined effect of the merger
and the deeply penetrating stream it hosts. Examining the isophotes
beyond the merger event, we find that they trace the shock fronts which
formed at the collision sites between the inflowing smooth streams
coming from the top and the bottom of the plot. The density and
temperature jump associated with the shock lead to an enhancement of
the X-ray luminosity accounting for this feature (this can also be
seen in \cref{fig:flxIsoph_cl6_a1,fig:flxIsoph_cl106_a1}). Thus, even
if the cluster was not in the throes of a merger event, the distorted
isophotes which trace the shock fronts would still lead to an NR
classification.

The combined sample of clusters at both redshifts contains 22 NR
clusters: 10 at \zeq{0} and 12 at \zeq{0.6}. Of these, only in nine
clusters can the dynamical state be unequivocally linked to mergers
and substructures, by dint of the existence of secondary luminosity
peaks or filamentary structure (e.g.\@
\cref{fig:flxIsoph_cl6_a06}). In nine additional clusters, the
contribution of substructure to the non-relaxedness is seen to be of
minor importance in light of the small number and size of the observed
substructure, and the unrelaxedness is due primarily to the smooth
streams (e.g.\@
\cref{fig:flxIsoph_cl6_a1,fig:flxIsoph_cl106_a1,fig:flxIsoph_cl105_a1}). In
the remaining four systems it is unclear what is the relative
contribution of the streams and substructure to the NR state of the
cluster. In all likelihood, it is the combined effect of the streams
and the substructures which lead to the unrelaxedness.

In light of these findings, we conclude that there is a real connection
between smooth stream penetration and the relaxedness of the systems.
The correlation found in \cref{fig:penen_relax} is not an artefact of the
correlation between the streams and substructures and the
unrelaxedness of a cluster is not caused solely by mergers.
 
\section{Discussion}\label{sec:discuss}
Galaxy clusters are classified as R or NR based on observational
signatures of dynamical activity. We utilized a suite of simulated
cluster systems, each classified as R or NR based on the same
observational parameters employed for observed clusters, to study the
source of this classification.

Our starting point in the analysis was to relate the classification,
based on mock observations, to the actual dynamical forces at play in
the system. We employed a form of the Jeans equation for the radial
velocity component, modified for the Eulerian simulation scheme, which
resulted in an equation of motion for spherical gas shells. In the NR
clusters, we find that a higher fraction of the pressure support
against gravity originates in non-thermal processes, mostly due to
random gas motions. As a result, these clusters are not in
steady-state equilibrium. In R clusters, the non-thermal contributions
are lower by a factor $\sim 2$ compared to the NR clusters, except
within the very centre of the system ($r \lesssim 0.05\Rv$).

The increased level of gas motions in the NR cluster population is
matched by higher mass inflow rates in these clusters, compared to the
R cluster population. The picture that emerges is that the higher
inflow rates bring in more energy and momentum, which allow deeper
penetration of the streams which can then drive random motions in the
cluster, stirring it up and leading to a dynamically disturbed system.

The significant role the build-up of turbulence plays in determining
the properties of the ICM has been well established
\citep{Parrish2010,Hallman2011,Nelson2012,Lau2013,Nelson2014}.  Other
studies have also shown that the turbulent pressure created by the
random motions of the gas is important in stabilizing the system and
in the central regions may be the dominant force fighting the pull of
gravity \citep{Lau2009,Nelson2014a}. \citet{Zhuravleva2014} have shown
that the heating rate due to the dissipation of turbulence may be
sufficient to offset the radiative cooling in the cluster centres. The
random motions can also lead to the dispersal of metals over larger
volumes in the central regions, as is seen in observations
\citep{DeGrandi2002a,Rebusco2005,Rebusco2006,Simionescu2015}.

While treating the clusters as spherical systems can yield important
insights, a full understanding of the structure and dynamics of the
ICM can only be achieved by studying the full 3D structure of the
system. The main reason for this is the asymmetric nature of gas
accretion into the cluster which occurs along large-scale gas streams
which originate from the cosmic web. These streams play an important
role in determining the behaviour of the system, especially in the 
central regions.

The gas streams are a prominent feature in the clusters, especially in
terms of entropy and inflow rate. Mass inflow rate into the system
occurs almost solely along the gas streams. The gas within the stream
has undergone early shock heating when it was accreted to the stream long
before entering the cluster and is heated further via weak shocks
within the stream. As a result, the stream temperatures are $\gtrsim
10^6\units{K}$ before entering the cluster, earning them the name
`warm streams'. After crossing the virial radius the streams are
heated to $\sim \Tv$, and are not significantly cooler than their
surroundings.

The inflowing streams carry a significant amount of energy with them,
not much lower than the entire radiative output of the cluster, only a
fraction of which is lost to radiation as they flow towards the
centre. The energy is deposited in the region where the stream
disrupts and dissipates. In our study, we find some gas streams which
penetrate into the centre of the clusters while others were stopped at
larger radii. Examining the clusters at two epochs, \zeq{0} and
\zeq{0.6}, reveals that the penetration depth of streams may evolve
over time. In some cases, streams which penetrated into the centre in
early times have been `pushed out' and dissipate at larger radii at
\zeq{0}, while in other cases the opposite occurs with streams which
increase their penetration depth over time.  The cause of the possible
disruption of streams at a given radius is an interesting open issue,
one which we plan to address in future studies (Mandelker et al., in
preparation; Padnos et al., in preparation).

A deeply penetrating stream, depositing its energy in the very centre
of the cluster, may have a profound effect on the dynamics of the ICM
in that region. Examining our simulated clusters, we find a strong
connection between the dynamic state of a cluster and the presence of
a deeply penetrating stream. All but two of the 22 NR clusters,
comprised of clusters at both epochs, possess at least one stream
which penetrates to $r\lesssim 0.25\Rv$ whereas the penetration into R
clusters does not exceed $0.35\Rv$, with one exception.

Especially interesting are three clusters which possessed deeply
penetrating streams at \zeq{0.6} but by \zeq{0} no streams were found
in their central regions. In conjunction, the dynamic classification
of these clusters changed from NR to R. In another cluster, the
penetration of the streams increased between the two epochs and the
cluster transitioned from R to NR accordingly. This strengthens the
link established between the two properties.

An important result of this paper is that the build-up of random
motions can occur not only due to episodic merger events
\citep{Vazza2011,Nelson2012}, but also by the continuously inflowing
gas streams originating from the large-scale filamentary structure of
the cosmic web. We have taken special care to ensure that there is
indeed a direct link between deeply penetrating gas streams and the NR
dynamical classification in the relevant clusters, besides the effect
of merging substructure, which is known to induce dynamical
disturbance in the ICM.

Clusters which are R are often also CC clusters
\citep{Sanderson2009,Walker2012} whereas the NCC clusters are often
associated with NR clusters. The common wisdom is that a CC cannot
easily form in a dynamically disturbed environment. In this manner, it
may be that the dichotomous population of CC versus NCC clusters is a
direct and natural result of stream penetration in clusters. This
scenario complements the work of \citet{Burns2008} in the sense that
the key factor in preventing the formation of CC is a strong mass (and
energy) inflow into the core of the system, whether it be in the form
of smooth streams or mergers. \citet{Birnboim2011} have also shown
that infalling gas clumps can be an effective form of heating the ICM.

We note that our simulation suite, which reproduces the R--NR
dichotomous population, does not explicitly reproduce the associated
CC--NCC dichotomy population. Rather, all the clusters, both R and NR,
suffer from central overcooling and thus possess a density peak
reminiscent of a CC cluster, over a region which is only barely resolved by
the simulation mesh. It is likely that this is due to the numerical
limitations of the simulation but it is also possible that key
physical process is missing, such as AGN feedback and thermal conduction.

If the latter is the case, we find that the energy injected by the
streams is not sufficient on its own to offset the runaway cooling in
CC clusters. However, the random gas motions stirred by the streams 
may increase the effectiveness of the AGN heating. Examination of new 
simulations with higher resolution will be able to shed light on whether 
the gas streams on their own can avert the overcooling or whether they are
only one of a number of players which lead to the formation of NCC
clusters.

Since the streams are the principal channels of mass accretion into
the clusters, sub-haloes, gas clumps and satellites are also expected
to be found along the streams. This makes it difficult to disentangle
the contribution of the smooth versus clumpy accretion to the heating
and build-up of random gas motions in the central regions of the
ICM. One may argue that for some purposes such a distinction is
superfluous and that accretion is accretion, regardless of its
form. 

However, since the relevant physical processes, such as ram-pressure
stripping, hydrodynamical drag, tidal stripping etc.\@, act
differently on clumps and streams, it may be necessary in some cases
to differentiate between the different forms of accretion. Tidal
interactions are sensitive to the geometrical shape of the
interacting bodies and thus the forces acting on a clump and stream in
the tidal field of the host cluster will be very different. Dynamical
friction is another physical process which may affect clumps and
streams in a different manner.

Another key difference is that clumps of gas, if they are not
disrupted, may travel through the centre of the cluster and affect the
ICM for much longer periods \citep{Birnboim2011} and over larger areas
than streams which are dispersed once they reach the centre.

As we have seen, the penetration depth of a stream can change in a
given cluster over time and so does the dynamical state of the
cluster. Thus, a cluster may evolve from being an NR cluster to a R
cluster, and vice-versa. In our simulation suite, we have found three
examples of NR to R transitions, and one of R to NR all of which are
linked to changes in the stream penetration
depth. \citet{Rossetti2010} found that most observed NCC clusters in
their observed sample host regions which have some characteristics of
CC, such as low-entropy gas and high metal abundances, supporting the
notion that clusters alternate between the CC and NCC state.

In our scenario, transition from an NR to R cluster involves the
dissipation of the random gas motions in the central regions which we
estimate to take of the order of several giga years. This is in agreement with
\citet{Hallman2011} who find the turbulence dissipation time-scale to
be $\sim 1 \units{Gyr}$. \citet{Rossetti2011} estimate that in
observed clusters, the typical transition time-scale in which an NCC
relaxes into a CC is $\sim 3\units{Gyr}$.

An interesting issue worthy of follow-up research is the interplay
between the effects of gas streams and AGN on the ICM. Turbulent
motions have been suggested as a mechanism for spreading the energy
supplied by an AGN through jets over large volumes
\citep{McNamara2007}. If the central regions of the cluster have
already been stirred up by the infalling gas streams, the
effectiveness of the AGN in heating the ICM may be bolstered. A study
of this issue would require simulations which incorporate both AGN
models and sufficient resolution to model the gas motions generated by
the streams in the central regions.

The analysis of the effects of streams on the structure of cluster,
particularly in relation to the formation of CC versus NCC systems,
would be incomplete without a detailed examination of the heating and
cooling processes in the inner regions.  This will allow one to
assess whether the energy brought in by the streams is effective in
heating the central regions or not.

\section{Summary}\label{sec:summary}
In this paper, we have revealed the existence of large-scale gas
streams in clusters and explored how they affect their host systems,
specifically in relation to the dynamical state of the cluster. We
summarize our main conclusions below.
\begin{itemize}
\item NR clusters are characterized by higher levels of non-thermal
  pressure support against gravity, mostly due to random gas motion,
  compared to R clusters. This is due to the higher mass inflow rates
  found in these clusters.
\item The mass inflow occurs predominantly along gas streams which
  originate in the cosmic web and flow into the cluster. These streams
  are not significantly cooler than the surrounding ICM (`warm
  streams'). Some streams are found to penetrate into the very centre
  of the cluster, while others are stopped abruptly before reaching
  the centre.
\item A strong connection was established between the dynamical state
  of the cluster (R versus NR) and the presence of at least one deeply
  penetrating stream. The degree of penetration was found to change
  over time coupled with a corresponding change in dynamical state. In
  this manner, the observed R versus NR dichotomous population, was
  found to be a natural result of the degree of stream penetration
  into the cluster.
\item The deeply penetrating streams can carry large amounts of energy
  and momentum into the very centre of the cluster. These can drive
  the random gas motions in the cluster centre which account for the
  NR state of the cluster.
\end{itemize}

The study of these objects is still in its early stages and we are
confident that with better simulations and more advanced observational
instruments, much more can be learned about the gas streams which may
help answer some of the important open questions which pertain to the
understanding the structure and evolution of galaxy clusters.

\section*{Acknowledegments}
We acknowledge stimulating discussions with Nir Shaviv and Alexey
Vikhlinin. This work was partly supported by the grants ISF 124/12,
I-CORE Program of the PBC/ISF 1829/12, BSF 2014-273, and NSF
AST-1405962.

\bibliographystyle{mnras}

\bibliography{zinger_clusterStreams.bib}

\bsp

\label{lastpage}

\end{document}